\RequirePackage{fix-cm}
\documentclass[smallextended]{svjour3}       
\smartqed  
\usepackage{graphicx}
\usepackage{xcolor}
\usepackage{cite} 
\usepackage{hyperref}
\usepackage{amsmath}    
\usepackage{tikz}
\usepackage{pgfplots}
\usetikzlibrary{arrows,shapes,positioning,calc}

\pgfplotsset{compat=1.16} 
\begin{document}

\tableofcontents

\title{Studies of the equation-of-state of nuclear matter by heavy-ion collisions at intermediate energy in the multi-messenger era 
}
\subtitle{A review focused on GSI results}

\titlerunning{Studies of the equation-of-state of nuclear matter...}        

\author{P.~Russotto \and
        M.D.~Cozma \and
        E.~De~Filippo \and
        A.~Le~F\`{e}vre \and
        Y.~Leifels \and
        J.~\L{}ukasik
}


\institute{P. Russotto \at
              INFN-Laboratori Nazionali del Sud, Catania, Italy \\
              Tel.: +39-095-542275\\
              \email{russotto@lns.infn.it}           
           \and
            M.D.~Cozma \at
              IFIN-HH, Bucharest, Romania
           \and
            E.~De~Filippo \at
              INFN-Sezione di Catania, Italy
            \and  
            A.~Le~F\`{e}vre \at
            GSI, Darmstadt, Germany
            \and
            Y.~Leifels \at
            GSI, Darmstadt, Germany
            \and
            J.~\L{}ukasik \at
            IFJ PAN, Krak\'{o}w, Poland; 
}

\date{Received: date / Accepted: date}

\maketitle

\begin{abstract}
The study of the nuclear matter equation-of-state (EoS) is a relevant topic of modern nuclear physics. 
It governs the behaviour of nuclear matter away from the normal conditions found in nuclei interiors and plays a major role in heavy-ion collisions, in determining neutron skin thicknesses of neutron rich nuclei and the mass-radius relation of neutron stars, and in modelizations of supernovae explosions. Its uncertain knowledge is related to difficulties in solving the many-body problem with realistic nuclear interactions. 
In the last decades several studies, from both theoretical and experimental sides, have allowed relevant progress in the description of the EoS, both for the isospin-symmetric matter and for the isospin-asymmetric matter, the so called symmetry energy, especially at densities below the saturation point. 
In this paper we review some of the studies on the high-density behavior of the EoS, obtained by studying heavy-ion collisions with incident energies between several hundred MeV up to about 2 GeV per nucleon, with a focus on those carried out at the GSI laboratory in Darmstadt (Germany) by using the SIS18 accelerator beams. 
Constraints on the isospin-symmetric matter EoS, based on studies of kaon and pion production and collective flows, are reviewed. 
Regarding the symmetry energy, results based on charged pion ratios and neutron-to-charged particles elliptic flow ratios are discussed. A brief overview on the heavy-ion collisions studies sensitive to densities below the nuclear saturation density is presented, but the main emphasis of the review is on the density dependence of the symmetry energy above saturation density, which is also the region which is particularly important for astrophysics. 
Estimates of neutron star radii, as deduced by the results of heavy-ion collisions discussed in this review, are compared to astrophysics results, including the recent ones made possible by gravitational waves detection and X-ray satellite based observation.
A multiple source analysis, using theoretical calculations, astrophysics and heavy-ion collision results, in constraining neutron star radii, is discussed as an illustrative example of the role played by heavy-ion collision results in the multi-messenger astronomy era.
Finally, some future perspectives and experimental possibilities are outlined.
\end{abstract}

\section{Introduction}
\label{intro}

The study of the nuclear matter equation-of-state (EoS) constitutes one of the central topics in contemporary nuclear physics; a good example of the breadth of this research field is the topical issue on the nuclear symmetry energy which appeared in Eur. Phys. Jour. A in 2014 \cite{refId0}.\\
The EoS describes the relation between energy, density, pressure, temperature and the isospin asymmetry $\delta = (\rho_n -\rho_p) / \rho$ of nuclear matter, where $\rho_n$, $\rho_p$, and $\rho$ are the neutron, proton and nuclear matter densities, respectively. It is conventionally split into a symmetric matter part, independent of $\delta$, and an isospin dependent term, expressed as a product of the symmetry energy\footnote{In the whole document we use the term "symmetry energy", instead of the more appropriate term ``asymmetry energy'', for consistency with what is presently commonly used by the scientific community and in the literature.}, $E_{sym}(\rho)$, and $\delta^2$ \cite{BARAN2005335,Li:2008gp,Burgio:2020fom}. For the case of T=0: 
\begin{equation}
 E(\rho, \delta)=E(\rho,
0)+E_{sym}(\rho)\delta^2+o(\delta^4)+....
\label{eq01}
\end{equation}
\\
Astrophysical investigations of neutron star properties has been another source of information for the high density behaviour of the EoS. Recently, this has produced pioneering results after the gravitational waves discovery and recent neutron star investigations by satellite based X-ray observation. The new possibility of studying some astrophysical phenomena, such as binary neutron star merger events or supernovae, by the combined observation and interpretation of different signals (gravitational waves, electromagnetic radiation and, eventually, neutrinos and cosmic rays) has been named ``multi-messenger'' astronomy \cite{Piekarewicz:2018gtd}. And the topics related to the EoS of neutron-rich matter, especially in the high-density domain, have become of even more importance and interest with the opening of the multi-messenger astronomy era.\\

The zero-order term of the EoS, $E(\rho, 0)$, provides the dominant contribution to the EoS and represents the energy per nucleon of (symmetric) nuclear matter. It is commonly expressed as an expansion around the saturation density $\rho_0$:
\begin{equation}
 E(\rho, 0)=E(\rho_0)+\frac{K_0}{18}\frac{(\rho-\rho_0)^2}{\rho_0}
\label{eq02}
\end{equation}

\begin{figure}
\centering
\includegraphics[width=0.49\textwidth]{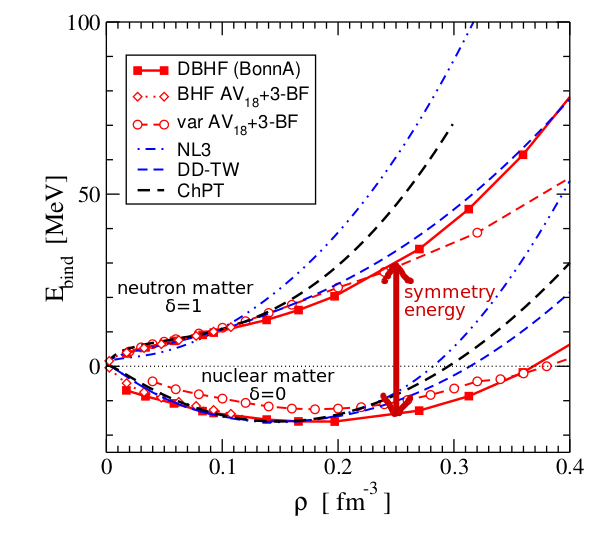}
\includegraphics[width=0.49\textwidth]{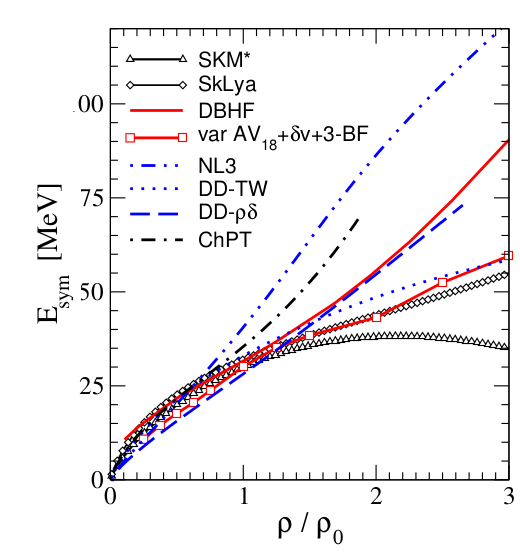}
\caption{(adapted from Ref. \cite{Fuchs:2005yn} under permission) Left panel: lower curves are predictions for EoS of symmetric nuclear matter as given by different theoretical models using various approaches; upper curves show predictions of the same models but for EoS of neutron matter. The arrow indicates that the differences between the upper and lower curves, given by the same model, corresponds to the predictions for the density dependence of the symmetry energy. Right panel: corresponding predictions for the symmetry energy. See \cite{Fuchs:2005yn} for more details. }
\label{fig:eos}
\end{figure}

The $K_{0}$ value fixes the compressibility of nuclear matter in the vicinity of the saturation point, determining how much energy is needed to compress/decompress symmetric nuclear matter starting from normal saturation density; in fact, for many years isoscalar giant monopole resonances served as an important source of information on the nuclear incompressibility \cite{BLAIZOT1980171,Youngblood-PRL.82.691}. The density evolution, in terms of maximum densities reached, in a typical heavy-ion nuclear reaction is essentially driven by this part of the EoS. 
Compressibility of the nuclear matter can be considered as a collective manifestation of the properties of the, still not well known, nuclear interaction. Theoretical approaches based on first principles, such as ab-initio calculations or variational methods, realistic nucleon-nucleon interactions and effective field theory have been extensively used to study the EoS, resulting in a large variety of predictions \cite{Fuchs:2005yn,BARAN2005335}. As an example, the left panel of Fig. \ref{fig:eos} shows predictions for the EoS of symmetric matter (lower group of curves) as given by diverse theoretical models, adapted from Ref. \cite{Fuchs:2005yn}. We notice significant divergences among the model predictions, especially when going in the high density region.\\ 
The study of heavy-ion collisions (HICs) in the intermediate energy regime has proven to be very effective in constraining the isospin-independent term of EoS. Observables connected to pion and kaon production and the study of flows of proton and light clusters have produced important results. In one of the first works \cite{Dani02Esymm} on this subject, Danielewicz \textit{et al.} compared the directed transverse flow measured in Au+Au collisions, at incident energies ranging from about 0.15 to 10 GeV/nucleon, to BUU transport model simulations, ruling out, both, very repulsive and very soft EoSs. The elliptic flow of protons was found to be less sensitive, relative to the directed flow, to the stiffness of EOS. However, a relevant uncertainty was the poor knowledge of the isospin-dependent term of EoS (see below), indicating the importance of studying the symmetry energy.\\ 


Moving to the $2^{nd}$ order term of eq. (\ref{eq01}), and neglecting
higher order terms, the fact that $E_{sym}(\rho)=E(\rho,\delta=1)-E(\rho,\delta=0)$ shows that $E_{sym}$ can be interpreted as the difference of energy between pure neutronic ($\delta$=1) and symmetric ($\delta$=0) matter, hence it corresponds to the energy needed to convert symmetric matter into neutronic one. This is also sketched in the left panel of Fig. \ref{fig:eos} where the upper group of curves show model predictions for the EoS of pure neutronic matter; therefore, $E_{sym}$ predictions, shown in the right panel of that figure, are just the difference between the the corresponding upper and low curves of left panel.\\ 
The density dependence of $E_{sym}$ is an important ingredient for the determination of the drip lines, masses, surface or halo densities, and collective excitations of neutron-rich nuclei \cite{Brown:2000pd, Roca-Maza:2011qcr}, for flows and multi-fragmentation in HIC at intermediate energies \cite{Li:2008gp, PhysRevLett.102.122701}, but also for astrophysical phenomena like supernovae and neutron stars \cite{STEINER2005325}, where the knowledge on the high-density dependence of $E_{sym}$ is of crucial importance.\\
Our current knowledge of $E_{sym}$ is poorer with respect to one we have for the isospin-symmetric contribution to the EoS. In fact, when trying to study the symmetry energy, one has to consider that this portion of the EoS plays a minor role in the HIC dynamics, hard to be evidenced, with respect to the one played by the isospin-symmetric part. Assuming an $E_{sym}(\rho_{0})$ value of 32 MeV and a neutron-rich Au nucleus, the contribution to the energy per nucleon due to the symmetry energy amounts to about 1.25 MeV,  $\sim 8\%$ with respect to the contribution due to the isospin-symmetric part of the EoS. But, since the effect of the symmetry energy is just to produce different potentials/forces for neutrons and protons moving in asymmetric nuclear matter, it follows that its tiny role can be better evidenced by looking at ``relative'' observables for isospin pairs. For that, the ratio or difference of neutron and proton, or light isobar nuclei, observables, or strongly related to it, the relative $\pi^{+}-\pi^{-}$ production yield in the pion production energy regime, have been suggested by the transport models (see below) used in this field. The use of relative observables gives also the  advantage of decreasing, and to some degree canceling, the effects of other ingredients, such as the stiffness of isospin-symmetric EoS, acting in the same way on both neutrons and protons.\\
The symmetry energy represents a key quantity to be accounted for when dealing with ``highly'' asymmetric-nuclear matter, and becoming of main relevance in the case of neutron-rich matter, as in a neutron star. In fact, neutron star radii, are mainly governed by the symmetry energy at about twice  nuclear matter saturation density \cite{Fattoyev-PRL.120.172702}.\\
Different density dependencies of $E_{sym}(\rho)$ can be characterized by expanding  $E_{sym}$ around the normal nuclear matter density, $\rho_o$, leading to the following expression:
\begin{equation}
E_{sym}(\rho) = E_{sym,0}
+ \frac{L}{3}\left( \frac{\rho-\rho_o}{\rho_o} \right)
+ \frac{K_{sym}}{18}\left( \frac{\rho-\rho_o}{\rho_o} \right)^2 + ... 
\label{equ02}
\end{equation}

\noindent with the value of the symmetry energy at normal density $E_{sym,0}
\equiv E_{sym} ( \rho = \rho_o)$, the slope parameter $L \equiv 3 \rho_o \;
\frac{\partial E_{sym}(\rho)}{\partial \rho} \Big |_{\rho = \rho_o}$, and the
curvature parameter (symmetry compressibility) $K_{sym} \equiv 9 \rho_{o}^{2}\;
\frac{\partial^{2} E_{sym}(\rho)}{\partial \rho^{2}}\Big |_{\rho = \rho_{o}}$.\\

A theoretical determination of the nuclear EoS from first principles making use of microscopic models is challenging and a subject of intense scientific research since several decades \cite{Fuchs:2005yn}. In fact, microscopic calculations of the density functional of nuclear matter employing different forms to the nucleon-nucleon interaction predict rather different EoSs, with very different behaviours of the density dependence of $E_{sym}$. 
As shown in right panel of Fig. \ref{fig:eos}, most calculations coincide at or slightly below normal nuclear matter density, which demonstrates that constraints from finite nuclei are active for an average density smaller than $\rho_o$ and surface effects play a role. In contrast to that, extrapolations to supra-normal densities diverge dramatically, calling for more tight experimental constraints in this region. In addition, it is worth noting that in the case of isospin-symmetric EoS a rather precise reference value at $\rho_{0}$ exists, as extracted by the volume term of the Bethe-Weizs\"acker semi-empirical mass formula. In contrast, the symmetry term of the Bethe-Weizs\"acker formula includes surface effects. It describes nuclear matter at an effective density of about $2/3~\rho_{0}$, and the extrapolation to $\rho_{0}$ leads to a correlation between $E_{sym,0}$ and $L$ \cite{Horowitz:2014bja}.\\
In the last years a tendency to produce increasingly more accurate predictions has been observed, in particular for calculations based on chiral effective field theory ($\chi EFT$), combined with advanced statistical methods \cite{Drischler:2020yad}. However, experimental investigations are always needed in order to  validate theoretical findings and, eventually, reduce their uncertainties, especially when moving away from normal nuclear conditions.


\begin{figure}
\centering
\includegraphics[width=0.49\textwidth]{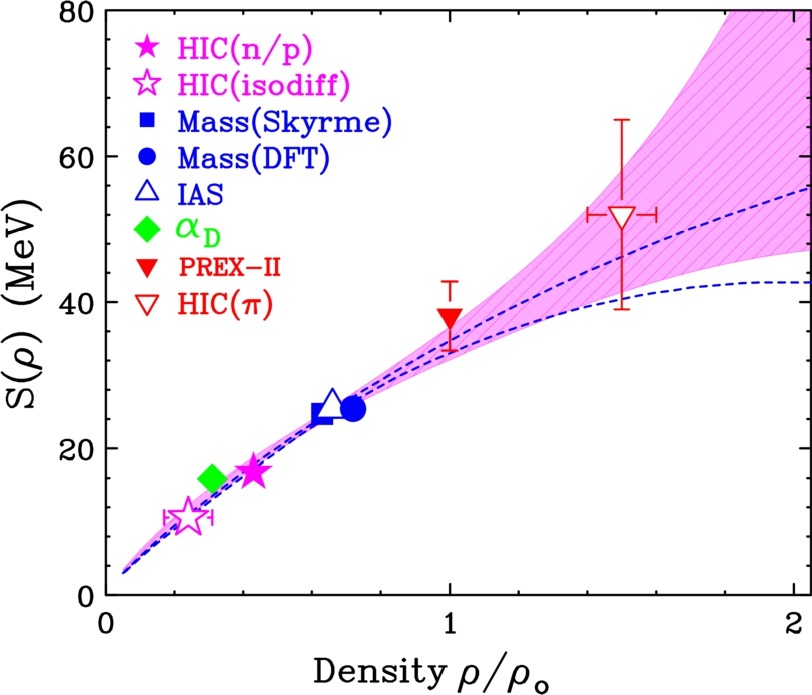}
\caption{(reprinted from Ref. \cite{lyn21} under permission) Symmetry energy as a function of nuclear density. Symbols are the constraints taken into account in the Lynch and Tsang work \cite{lyn21}, placed at the specific sensitive density. The shaded region is the 1$\sigma$ allowed region as resulting from a fit, including a cubic term, of the used constraints. The dashed curves indicate the 1$\sigma$ boundaries from a fit without the cubic term. See Ref. \cite{lyn21} for more details.}
\label{fig:lynch21}
\end{figure}

The study of the symmetry energy has received a remarkable attention by the scientific community in the last few decades, but has gained even more interest than before after gravitational waves (GW) discovery in binary neutron star merger events, like the one observed in the GW17082017 \cite{LIGOScientific:2017vwq}. In these events, one of the key observables is the so-called tidal polarizability $\Lambda$, being highly sensitive to the neutron star radius ($\Lambda \sim R^{5}$), a quantity that is notoriously difficult to measure. A second event of a binary neutron star merger, GW190425, has been recently reported in \cite{LIGOScientific:2020aai}. 
In addition, the recently commissioned NICER (Neutron star Interior Composition Explorer) instrumentation \cite{Riley:2019yda,Miller:2019cac} on-board of the International Space Station (ISS), devoted to the study of neutron stars through soft X-ray timing, has made available new data, enlarging even more the possibilities to investigate the EoS of neutron-rich matter. A special feature of NICER is the possibility of simultaneous estimation of both mass and radius of neutron stars. This allows to fix a region of the neutron star mass-radius relation, constituting a stringent constraint for each proposed EoS. All these new opportunities have made the study of $E_{sym}$ at high density even more intriguing than in the past, and opened the way to works where results from astrophysical observations are compared with the ones obtained in terrestrial laboratories and with predictions of theoretical models \cite{Fattoyev-PRL.120.172702,Ghosh:2021bvw,Zhang:2018bwq,Huth:2021bsp}.\\
From the terrestrial laboratory point of view, several observables have been proposed and effectively used as sensitive probes of the symmetry energy in heavy-ion studies. In the low density regime, isospin diffusion \cite{PhysRevLett.102.122701} and migration \cite{DeFilippo:2012qd}, properties of isobaric analogue states \cite{Danielewicz:2013upa}, double magic nuclei \cite{Brown:2013mga} and isotope binding energy differences \cite{Zhang:2013wna}, thickness of neutron skins in neutron rich nuclei \cite{Roca-Maza:2011qcr}, electric dipole polarizability in giant dipole resonances \cite{Tamii-PhysRevLett.107.062502,Pieka-PhysRevC.85.041302}, isoscalar monopole resonances along isotopic chains \cite{Li-PhysRevLett.99.162503,Li-PhysRevC.81.034309}, competition between fusion-like and binary reactions in HIC at 25 MeV/nucleon \cite{Amorini:2008sm} have been extensively used, producing quite consistent results.\\Recently, Lynch and Tsang \cite{lyn21} have emphasized the importance of carefully evaluating which density region is effectively probed by a given observable. This limits the obtained constraint on the EoS to the density region effectively probed, avoiding unreliable extrapolations to other density regions.
In detail, they studied the density region effectively probed by analyses of nuclear masses, isobaric analog states, isospin diffusion, electric dipole polarizability of $^{208}Pb$ and double magic nuclei (see Ref. \cite{lyn21} for more details). The specific sensitive densities probed by these analyses were found to range from 0.25 to 0.7 $\rho_0$ and the corresponding $E_{sym}$ constraints at those densities were calculated, as presented here in Fig. \ref{fig:lynch21}. They found a highly consistent description of $E_{sym}$ at low densities, solving some of the discrepancies that occur when $E_{sym,0}$ and $L$ values from different works are compared, being in some cases just an extrapolation to saturation density, that is, outside the probed density region. In addition, also constraints at saturation density, from recent PREX-II result on $^{208}Pb$ neutron skin thickness \cite{adh21}, and around 1.5 $\rho_0$, from pion studies at RIKEN \cite{SRIT:2021gcy} and ASY-EOS experiment \cite{Russotto:2016ucm} (see below), were taken into account. By fitting all these constraints together, magnitude ($S$) and slope ($L$) values at $\rho_{01}=0.1$ fm$^{-3}$ were obtained ($S_{01}=24.0\pm0.5$ MeV and  $L_{01}=54\pm6$ MeV). As a final result, the 1$\sigma$ obtained region for the $E_{sym}$ is shown in Fig. \ref{fig:lynch21}. The inclusion of PREX-II and pion results led to a stiffening of $E_{sym}$, relative to what was obtained by using only low densities results.
\\

For studying $E_{sym}$ at high densities, observables related to the production of isospin pairs, such as the ratio of negative to positive pions \cite{Xiao:2008vm}, kaon multiplicity ratio $K^{+}/K^{0}$ \cite{Ferini:2006je} as well as yields and flows of neutrons and protons or light clusters \cite{Russotto:2011hq} have been proposed. Compared to the case of the isoscalar part of the EoS, the situation was, and still it is, less clear.
The double kaon multiplicity ratio, while promising when used for infinite matter calculations, showed a reduced sensitivity  to the symmetry energy for realistic HIC \cite{FOPI:2007gvb}.
Attempts to study the symmetry energy using the single pion ratio have been faced with a strong model dependence of this observable \cite{Xiao:2009zza,Feng:2009am,Xie:2013np,Hong:2013yva} related to the differences in the treatment of in-medium pion production, propagation and, eventually, re-absorption in the available transport models, as discussed in the next sections. A collaborative effort on the theoretical side is on-going to clarify the situation \cite{Xu:2016lue,Zhang:2017esm,Ono:2019ndq,Colonna:2021xuh,TMEP:2022xjg}. At the same time, additional experimental measurements of this observable, from the S$\pi$IRIT Collaboration, have recently become available \cite{SpRIT:2020blg,SRIT:2021gcy}. 
Observables related to the neutron-proton ratios, such as the differential directed flow ratio proposed by Li \cite{Li:2002qx} proved to be hard to measure, especially because of the difficulties connected to a precise measurement of neutron multiplicity over a large solid angle range. Instead, observables related to flows have been successfully used to get $E_{sym}$ constraints in the high-density regime \cite{Russotto:2011hq,Cozma:2013sja,Russotto:2014EPJA,Russotto:2016ucm,MDCozEPJA18}. These results appear to be more robust, relative to the pion ones, even if, also in this case, some dependence on model ingredients has been reported.\\
However, the current status presents still few, compared to the low density regime, HIC measurements and results explicitly devoted to the study of $E_{sym}$ at high densities. In the case of neutron-to-proton observables, the lack of experiments has to be ascribed to the difficulties of measuring the neutron component, in addition to that of protons (or charged particles in general), requiring the use of neutron detectors covering an important fraction of the solid angle. Unfortunately, just few detectors \cite{LAND:1991ffr,R:2021lxa,Nakamura:2015phw} suited for this task are operative worldwide. Also the pion measurements are not straightforward, requiring the use of very accurate and precise pion detection tracking based systems. Hence, the need of new and more accurate measurements to confirm what has been observed so far, improve the obtained accuracy and extend the results toward higher densities.\\ 

The interpretation of HIC experiments in the intermediate energy range is usually performed using semi-classical transport models. These fall essentially into two large classes based on the method used to treat fluctuations. Firstly, Boltzmann-Uehling-Uhlenbeck (BUU) models~\cite{Bertsch:1988ik,Bonasera:1994zz} solve the Boltzmann-Vlasov equation, supplemented by a two-body collision term to account for the short-range residual interaction ~\cite{Carruthers:1982fa}, to determine one-body phase-space distribution functions of each particle specie. Fluctuations can be accounted for by extending the collision term to include a Langevin stochastic contribution~\cite{Abe:1995yw,Chomaz:2003dz}. Secondly, Quantum Molecular Dynamics (QMD) models~\cite{Aichelin:1991xy,Feldmeier:1989st,Ono:1998yd,Papa:2000ef,Hartnack:1997ez} solve the n-body Schr\"odinger equation in the Hartree approximation, supplemented also in this case by a collision term. The \textit{ansatz} for the total wave function as a product of fixed-width Gaussian wave packets, together with an event-by-event solution of the equations, ensures that fluctuations, albeit of classical nature, are preserved during the evolution of the simulated system. The quantum nature of nuclear systems is more realistically described within the framework of off-shell transport models~\cite{Cassing:2009vt,Buss:2011mx}, which naturally account for the finite width of nucleons in dense nuclear matter and the effect of short-range correlations~\cite{Hen:2014nza,Hen:2016kwk}. All results concerning the EoS described in this review have been obtained, where appropriate, using either BUU or QMD type of transport models. Similar analyses employing off-shell transport models would be extremely important, in view of the recent claims regarding the relevance of short-range correlations~\cite{Hen:2014yfa} and threshold effects~\cite{Song:2015hua} on particle production and nucleonic observables.\\
In general, past studies of the EoS employing HICs have revealed that theoretical predictions employing transport models often bear significant model dependence. Understanding the origin of these differences and their eventual resolution has been the goal of several past and present collaborations among transport theorists~\cite{Kolomeitsev:2004np,TMEP:2022xjg,Reichert:2021ljd}.\\
Given the small contribution of the symmetry energy to the full reaction pattern, it is hard to isolate its effects and to extract observables sensitive to it. In fact, the effects of other transport model ingredients, such as stiffness of isospin-symmetric EOS,  momentum dependence of the mean-field potential, momentum and/or density dependence of the in-medium N-N elastic cross section, etc., and even strategies adopted to implement in the code basic physical principles, like the Pauli-blocking in the collision term, can be bigger than the one due to the symmetry energy. As said before, this issue can be partially solved by using n-p relative observables. Also difficulties in getting reliable ``asymptotic'' simulated quantities, to be compared to experimental ones, starting from the configuration produced by the transport model at the times the calculations are stopped, exist. For that, more or less advanced clusterization algorithms have been developed, from simple approaches, based on minimum-spanning-tree algorithms the make use of the distance of nucleons in the coordinate and momentum spaces, to more sophisticated treatments, taking into account also excitation energy, isotopic composition and, in general, stability of the built-in cluster \cite{LeFevre19}. Hence, transport model final results also depend on the strategy adopted in the clusterization method, and on the parameters used for it. A way to circumvent this dependence is the use of coalescence invariant observables ~\cite{Famiano:2006rb,Coupland-PhysRevC.94.011601,morf19} or the  use of transport models that propagate explicitly the cluster degrees of freedom ~\cite{Danielewicz:1991dh,Ono:2013aaa}.\\ 


Presenting an extensive overview of the above mentioned studies is well beyond the aim of the current paper. We will limit the presentation here to a review on a restricted part of the experimental investigation of the EoS, relevant to studies of the supra-saturation density regime, i.e. by using heavy-ion reactions in the intermediate energy regime around 1 GeV/nucleon beam energy. Many of these studies have been carried out at the GSI laboratory at Darmstadt (Germany) by using heavy-ion beams delivered by the SIS18 accelerator system. But we will also discuss relevant recent studies on pion production carried out at the RIKEN laboratory (Japan) that complement the former ones from GSI. When discussing the symmetry energy, space will be also devoted to provide a sample of the studies carried out at the Fermi energy, probing the sub-saturation regime; this will allow to present a more complete description of the current results on the global density dependence of symmetry energy, giving a view on the consistency between studies performed above and below the saturation density.
In the next sections, for both isospin-symmetric EoS and symmetry energy studies we will discuss, first, results based on particle production (pion and kaon) and, after, those based on flow of nucleons and light clusters. This choice reflects somehow the chronological order in which these results were released. However, it is worth to state that, in this energy region, the primary effect of the EoS is to affect the evolution of the collision, hence, dynamical observables like stopping of the collision and flow and momentum distributions of nucleons and light clusters, also with respect to the isospin asymmetry degree of freedom. The produced particles then act as tracers of this evolution, thus this can be seen as a secondary, but sensitive, probes of the EoS.

As said before, astrophysical observations have resulted to be a rich source of information on the symmetry energy at high-density in the last few years. It has thus become possible to compare laboratory and astrophysical constraints. In sect. \ref{sec:2323} estimations for neutron star radii from astrophysical sources, such as low-mass X-ray binary systems, millisecond pulsars, gravitational waves and multiple source analyses, will be presented and compared to the ones obtained by using elliptic flow measurement in HIC, allowing to discuss the evolution of results in the field and the role played by HIC results. Moreover, results for pressure of neutron matter and $E_{sym}$  as given by elliptic flow studies will be discussed and compared to those based on multi-messenger astronomy analyses.

However, despite the new and relevant results coming from astrophysical observations and the increasingly more accurate results coming from theory, new and more precise laboratory results are still needed. In fact, astrophysical observations are still affected by important uncertainties, in part related to model and analysis procedures, thus independent laboratory confirmation is needed. Moreover they do not probe effectively the region just above the nuclear matter saturation densities, being mainly sensitive above $\sim~2~\rho_0$ \cite{Fattoyev-PRL.120.172702,Huth:2021bsp}. Being the nuclear structure and Fermi-energy HIC studies sensitive to densities below $\rho_0$, the intermediate-energy HIC studies are thus needed for filling the gap in density in the $1-2.~\rho_0$ region, and completing the astrophysical observations at higher densities. In addition, advanced theoretical calculations, as the one given by $\chi EFT$, in most of the cases, have been shown to be very accurate in the low density regime, but tend to be affected by larger uncertainties in the supra-saturation region \cite{Huth:2021bsp} and, also in this case, terrestrial laboratory confirmation is desirable. The uncertainties at high densities result mainly from uncertainties in the treatment of many-body (mainly many-neutron) interactions.
Advanced analysis using theory, astrophysical and heavy-ion reaction results jointly, appears to be a promising way to produce precise and reliable constraints for the symmetry energy at high densities. This guarantees also the needed cross check of results coming from different fields. As a relevant example, this was  done in the Ref. \cite{Huth:2021bsp} that will be presented in more detail in sect. \ref{sec:2323}.  

Finally, some perspectives, as triggered by the past studies of the EoS at high densities presented in the following, will be given in sect. \ref{sec:233}. \\

\section{Experimental investigations of the EoS of isospin-symmetric nuclear matter above saturation density}
\label{sec:1}
 
During a HIC at incident energies above 100 MeV/nucleon hot and dense matter is created in the overlap zone 
of the two nuclei. Densities up to several times normal nuclear matter density $\rho_0$ are reached for relatively long times 
in the course of such reactions. Therefore, HICs are a unique tool to probe the nuclear matter EoS in terrestrial laboratories. The high pressure and density reached in HICs at relativistic incident energy in the compressed overlapping region -- dubbed "fireball" or "participant", at "mid-rapidity" between projectile and target rapidity --- give rise to a multitude of different observables, which have been described in the literature \cite{Reisdorf:1997fx, Herrmann:1999axy, Friman:2011zz}: formation of clusters even in the most central collisions, longitudinal and transverse rapidity distributions and stopping, transverse, directed, and elliptic flows, and particle creation even at energies below their production threshold in nucleon-nucleon collisions. 
After the first pioneering experiments in Berkeley with the Plastic Ball, where directed and elliptic flows were discovered \cite{Gustafsson:1984ka, Gutbrod:1989gh}, and the Streamer Chamber, which allowed to study the production of strange particles below production thresholds for the first time \cite{Schnetzer:1982}, various aspects of HICs in this energy regime have been studied with various experimental set-ups at the SIS18 synchrotron at GSI, Darmstadt. Fragmentation of the projectile spectator and the characteristics of the mid-rapidity source have been studied with the ALADiN spectrometer together with various detectors of large angular coverage (e.g. the INDRA detector) \cite{Hubele:1991ss}. Equipped with a highly selective trigger and particle identification by magnetic rigidity the KaoS spectrometer was used to study strangeness production and flows within a small acceptance \cite{SENGER1993393}. With the TAPS array of BaF2 crystals \cite{Novotny:1991}, it was possible to measure the production of neutral particles decaying into $\gamma$'s, and the large acceptance spectrometer FOPI \cite{GOBBI1993156} allowed to study the emission pattern of all charged particles in reactions starting from Ca+Ca to Au+Au covering the whole SIS18 energy regime. These experiments are not operational any more. Currently, the HADES experiment \cite{Agakichiev_2009} is used to study heavy-ion reactions at SIS18 energies. The HADES set-up is capable to record events with the highest rates achieved so far for HICs in this energy regime. Thus, very rare probes, e.g. dileptons resulting from decays of vector mesons, or production of the double strange $\Xi^-$, are accessible.\\ 
It was early recognized that heavy-ion reactions in this energy regime are essentially dynamical processes, that is, the close-to-equilibrium situation assumed by the hydrodynamical approaches (for a recent study see Ref. \cite{Inghirami:2022afu}) is not reached.
Therefore,  microscopic transport theory which does not require the assumption of local equilibrium was necessary for the interpretation of the observables.
Many dynamical models suitable for HICs have been developed in the recent years and are continuously upgraded (see \cite{TMEP:2022xjg} for a review). Consequently, transport models are necessary  to assess the properties of nuclear matter, and precise and systematic measurements of HICs varying system size and beam energy are necessary to constrain the various input parameters to the models, which cannot be deduced from first principles. 
In order to confine the EoS of symmetric matter, observables have to be selected which are either sensitive to the density or the pressure reached during a HIC. Density and pressure are quantities which are closely connected to the nuclear matter EoS. If the nuclear matter EoS is rather soft, i.e. the nuclear matter is easily compressible, the densities which are reached during a HIC are higher than in the case of a hard EoS. On the other hand, a hard EoS results in a higher maximum pressure. In the following, the most relevant observables which have been used to constrain the EoS of symmetric matter are introduced and the results are discussed.

\subsection{Studies on particle production}
\label{sec:11}

In elementary nucleon-nucleon (NN) reactions, particle production is possible if the energy in the elementary collision exceeds  the production threshold of the particle under consideration. In the dense, hot fireball of a HIC various processes occur, such that particle production below the respective production threshold is observed. In particular, at low energies the Fermi energy of nucleons inside the nuclei contributes to the available energy and enhances the production rates close to threshold.  In dense and extended systems of many colliding nucleons, energy may be stored in resonances.  In a first collision a nucleon could be excited into a nucleonic resonance, e.g. a $\Delta$, which collides in the next step with another nucleon. In the latter collision the production threshold for a particle, e.g. kaon, is significantly reduced.  The modifications of hadron characteristics in dense matter~\cite{Kaiser:2001bx,Korpa:2003bc,Korpa:2004ae} influence the production of particles as well: The kaon-nucleon (KN) interaction leads to a slightly  heavier mass for the K$^+$ at normal nuclear matter densities (+ 10-20 MeV); whereas the attractive K$^-N$ interaction leads to reduction of the K$^-$ mass  of up to  ~100 MeV \cite{HARTNACK2012119}.

\subsubsection{Results of studies on kaon production}

The production of positively charged kaons in HICs at sub-threshold energies ( $E^{thr}_{NN} =$ 1.6~GeV) is a multistep process where intermediate  $\Delta$ resonances and pions are used as energy reservoirs, hence, it is strongly density dependent. The number of collisions is rising with the maximum density reached during the HIC, and consequently  the production probability grows with the density. The production threshold energy for K$^-$ is much higher, $E^{thr}_{NN} = 2.1$ GeV, because of its different quark content, ($\bar{\mbox{u}}\mbox{s}$) in contrast to  ($\mbox{u}\bar{\mbox{s}}$)  for the K$^+$. Negative kaons are mainly produced in collisions of pions with intermediate hyperons at sub-threshold energies. Hyperons, denoted with Y, are created together with K$^+$ in order to conserve strangeness in reactions like  $\pi$~+~N~$\rightarrow$ ~Y ~+~K$^+$. Therefore, K$^-$ production is closely linked to the production of K$^+$. However, the absorption cross section of K$^-$ in the nuclear medium is rather high, whereas K$^+$ leave the reaction zone almost undisturbed. Hence, K$^+$ could be an ideal messenger for the density reached in a HIC. This was confirmed by predictions of microscopic transport models \cite{AICHKO}.  It should be noted, that the (in-medium) cross sections of the reactions involved in kaon production, which are needed in the microscopic transport codes,  are not all experimentally accessible and have been deduced from theoretical models. 

\begin{figure}
\centering
\includegraphics[width=0.6\textwidth]{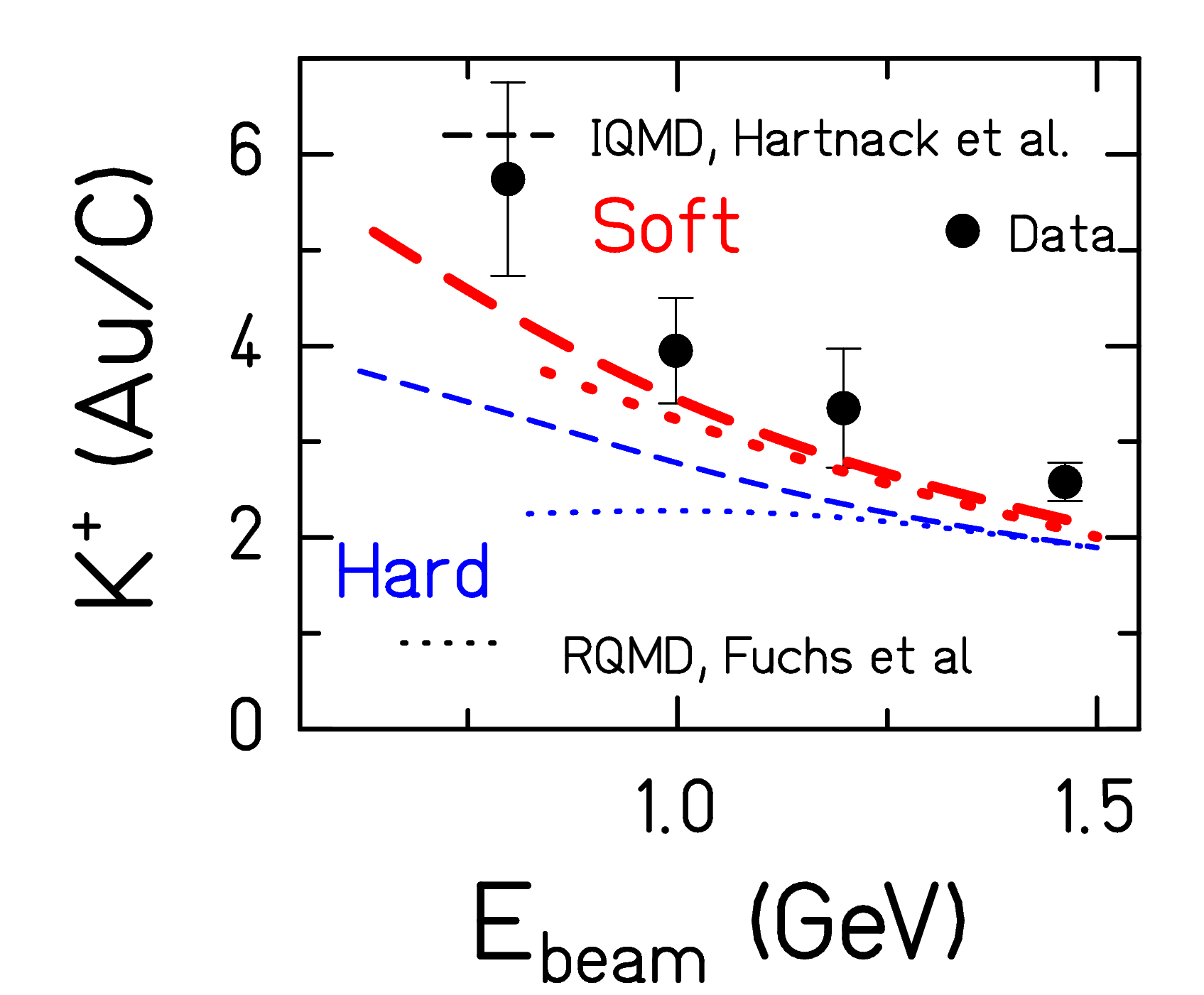}

\caption{Experimental data on the ratio of K$^+$ production in Au+Au collisions to the one in C+C as function of energy are shown as black dots. Predictions from the IQMD model are shown as long dashed lines, predictions from RQMD as short dashed lines. Predictions utilizing a soft(hard) EoS are shown in red(blue). Data points and model predictions are taken from Ref. \cite{HARTNACK2012119}}.
\label{fig:kaons_hartnack}
\end{figure}

The KaoS collaboration measured $K^+$ production in Au+Au collisions from 600 MeV/nucleon up to 1.5 GeV/nucleon.  Data from C+C collisions were utilized as normalization to reduce the influence of the partially unknown elementary cross sections when comparing to the predictions of theoretical models. The ratio of K$^+$ production in Au+Au  to C+C collision as a function of energy is shown as black circles in Fig.~\ref{fig:kaons_hartnack}. The experimental data is compared to the predictions of two different microscopic transport codes, IQMD (long dashed lines) \cite{Hartnack:1997ez} and RQMD (short dashed lines) \cite{Fuchs:2001gv}. For both models, the data is best described by the predictions utilizing a soft EoS, shown in red in Fig.~\ref{fig:kaons_hartnack}. In addition, it has been shown that this conclusion is not altered, when changing the elementary cross sections, employing in-medium effects on the kaons or changing the momentum dependence of the NN potential in the transport code \cite{Hartnack:2006}. For that, this result can be considered quite robust.

\label{sec:111}

\subsubsection{Results of studies on pion production}
\label{sec:112}
Pions are the most abundant newly created particles in HICs at higher energies ($E^{thr}_{NN}=280~$MeV). Most of them are produced via creation and subsequent decay of $\Delta(1232)$, which is reflected in the two-slope shape of pion spectra characteristic for heavier colliding systems \cite{Wagner:2000ak,FOPI:2005tyo}. The steeper low momentum part results from the decay of the $\Delta(1232)$. At higher energies,  N$^*$ resonances are excited and start contributing to pion production. The first resonance in the pion-resonance chain is created during the high density stage of the reaction~\cite{Li:2004cq}. Initially, this raised hopes that one may deduce the stiffness of the EoS by using deficits in pion production relative to expectations based on a scenario which assumes that the total available energy is shared between compressional and thermal energy~\cite{Stock:1985xe}, which may be linked to pion production. However, transport model simulations have shown that the influence of the EoS on total pion yield is rather weak, mainly because pion production is a chain of creation and absorption reactions before  pions finally leave the dense medium. Nevertheless, highly energetic pions, which were shown to be emitted at an early stage of the reaction~\cite{Bass:1994af,Wagner:2000ak}, still retain some of the sought-after information on the EoS. 

The sensitivity of pion yields to the compressibility modulus of symmetric nuclear matter is accompanied by an even stronger dependence on the effective nucleon mass~\cite{Schonhofen:1989pt}, which complicates the interpretation of experimental data. Energy and momentum dependencies of the interaction are the consequence of non-localities in time and spatial coordinates and can be described equivalently by a non-relativistic effective mass~\cite{Jaminon:1989wj}. Changing the effective mass modifies the attraction/repulsion of the interaction in the momentum range probed in a particular heavy-ion reaction. This in turn affects the maximum density reached during the compression stage of the collision, which plays a crucial role in determining the total pion yield via the dependence of collision rate on density. The importance of the effective mass has re-emerged more recently during the attempt to simultaneously describe pion multiplicities and elliptic flow of protons~\cite{Hong:2013yva}.


Pion production in intermediate energy HICs has been extensively studied experimentally at GSI by several collaborations: TAPS~\cite{Schwalb:1994zz}, KaoS~\cite{Brill:1993xh,Wagner:2000ak}, FOPI~\cite{FOPI:2005tyo,FOPI:2006ifg} and HADES~\cite{HADES:2009mtt,HADES:2020ver}. The EoS of symmetric nuclear matter has been addressed mainly in connection with FOPI collaboration data, most notably in the works of Hong $et\,al.$ and Reisdorf $et\,al.$ mentioned above. A comparison with transport model simulations has revealed systematic discrepancies for yields and their rapidity or transverse mass spectra that affect even the most recent versions of essentially all widely used transport models~\cite{HADES:2020ver} at all studied impact energies. A more careful treatment of the momentum dependent part of the interaction and associated threshold shift effects for particle production in a dense medium appears as a possible part of the solution to this problem~\cite{Cozma:2021tfu}.  

In the work of Hong $et\,al.$~\cite{FOPI:2005tyo} experimental results for transverse momentum and rapidity spectra of charged pions emitted in central Ru+Ru collisions at impact energies of 0.4 and 1.5 GeV/nucleon have been reported. Investigation of the symmetric matter EoS was performed by comparing IQMD model predictions, using momentum dependent interactions, for rapidity spectra to experiment. At lower beam energies the sensitivity to the nuclear incompressibility modulus $K_0$ is negligible, while at 1.5 GeV/nucleon a stiffer EoS is closer to measured values. However, both soft and stiff results over-predict data by an amount comparable to the soft-stiff difference. This discrepancy is seen to increase towards lower impact energies.

In Ref.~\cite{FOPI:2006ifg} an exhaustive set of results for pion related observables due to the FOPI collaboration have been presented. Data for several systems (Ca+Ca, Ru+Ru, Zr+Zr and Au+Au) and impact energy in the range 0.4-1.5 GeV/nucleon are available. The EoS of symmetric nuclear matter has been studied using stopping, polar anisotropies and elliptic flow. As before, the IQMD model has been used for theoretical predictions. Pion stopping in Au+Au collisions is seen to be sensitive to $K_0$, a soft value (210 MeV) being favored at higher impact energies. Close to threshold, both soft and stiff EoSs overpredict experimental data, which is due to discrepancies in longitudinal rapidity distributions. Polar anisotropies show dependence to the EoS at the higher end of impact energy. Sensitivity can be further increased by selecting high transverse momentum pions. The soft EoS is closer to experimental data, but model simulations underpredict data by sizable margins particularly at lower impact energies. For the case of elliptic flows of both $\pi^-$ and $\pi^+$ the stiff EoS is favored by a slight margin over the soft choice.\\ 

Summarizing, only few works have addressed the study of the symmetric part of the EoS with pions. The results are rather qualitative and often inconsistent for different observables.


\subsection{Results from flow studies}
\label{sec:121}

Over the last two decades, major experimental efforts have been devoted to measuring the nuclear EoS with HIC experiments performed at relativistic incident energies \cite{Dani02Esymm,Fuchs:2005yn,Zhang:2020dvn}.
These collisions of atomic nuclei form a hot, dense fireball of hadronic matter in the overlapping region, which expands in time and reaches the surrounding detectors as baryons and mesons. 
The phase-space distribution of particles flowing from the fireball during the expansion phase is strongly dictated by the compression achieved in the colliding region and is, therefore, sensitive to the EoS of the hot and dense nuclear matter created in the collision. 

The so-called elliptic flow ($v_2$) of particles emerging out of the reaction plane is the main observable, which has been used to experimentally constrain symmetric nuclear matter at supranuclear densities with HICs. 
It is given by the second moment of the Fourier expansion of the distribution of azimuthal angle $\Phi$ of the emitted particles with respect to that of the reaction plane $\Phi_{\rm RP}$,
\begin{equation}
\begin{aligned}
    \frac{d\sigma(y,p_{t})}{d\Phi}= &C(1+2v_{1}(y,p_t)\cos(\Phi-\Phi_{\rm RP})\\
    &+2v_{2}(y,p_t)\cos~2(\Phi-\Phi_{\rm RP})+...)\,,
\end{aligned}
\end{equation}
where all expansion coefficients $v_{n}$ are functions of longitudinal rapidity $y = \frac{1}{2}\ln\left(\frac{E+p_z}{E-p_z}\right)$, with $p_z$ being the momentum along the beam axis and $E$ the total energy, and of transverse momentum $p_{t}=\sqrt{p_x^2+p_y^2}$ of the particle, with $p_x$ and $p_y$ denoting the momentum components perpendicular to the beam axis, where $p_x$ is the component into the reaction plane, i.e., the plane containing the beam axis and the impact parameter direction. 

In experiments, the orientation of the reaction plane is event-wise reconstructed from the azimuthal distribution of particles recorded in the forward and backward hemispheres, and the Fourier coefficients are corrected for the resolution due to the finite number of products emitted and instrumental accuracy \cite{Andronic:2006ra}.
A positive elliptic flow $v_2$ indicates a preferred emission in the reaction plane whereas a negative flow indicates an emission out of the reaction plane.

It has been shown that the negative elliptic flow $v_{2}$ of protons and light isotopes emitted at mid-rapidity in HICs at incident energies of about 100 MeV/nucleon up to about 2 GeV/nucleon (so-called "intermediate energies") offers the strongest sensitivity to the nuclear EoS \cite{LeFevre:2015paj,Dani02Esymm,FOPI:2010xrt}, as evident from calculations made with various transport models. 
This dependence on the nuclear EoS is predicted by QMD \cite{LeFevre:2015paj,Hartnack:1997ez, FOPI:2011aa, Wang:2018hsw} and Boltzmann-Uehling-Uhlenbeck \cite{Dani02Esymm} models. The origin of the out-of-plane flow phenomenon at mid-rapidity has been investigated in detail by Le F\`evre \textit{et al.} \cite{LeFevre:2016vpp}, within the framework of the  quantum molecular dynamics (QMD) model. There, at intermediate incident energies, due to the relative slowness of the escape of the target/projectile spectators compared to the expansion of the fireball, absorptions and rescatterings by the spectator matter of in-plane emitted fireball particles occur, dubbed "shadowing", which cause an excess of mid-rapidity particle emission in out-of-plane directions. Furthermore, the strength of this anisotropy depends strongly on the EoS by the different density gradients (and therefore different forces) in the direction of the impact parameter (x direction) as compared to the direction perpendicular to the reaction plan (y direction), caused by the presence of the spectator matter. The stronger density gradient in the y direction, connected to the shadowing of the spectators, accelerates the particles more and creates therefore a negative $v_2$. This useful phenomenon is vanishing around 2 GeV/nucleon incident energy, where mid-rapidity emissions tend to become isotropic ($v_{2}$ tends to zero) because spectators escape too fast to interact any longer with the expanding fireball.   

As shown by Danielewicz \textit{et al.} \cite{Dani02Esymm}, at higher beam energies between 1 and 10 GeV/nucleon, the sensitivity of the directed flow $v_1$ to the stiffness of the EoS of symmetric nuclear matter becomes comparable to that of $v_2$ when not stronger. The EoS constraints from Danielewicz \textit{et al.} \cite{Dani02Esymm} for symmetric nuclear matter have been deduced from HIC experiments at the Bevalac accelerator at Lawrence Berkeley National Laboratory (LBL) and the Alternating Gradient Synchrotron (AGS) at Brookhaven National Laboratory (BNL) where Au nuclei were collided. 
The information from this series of HIC experiments allows us to further constrain the EoS in a density range ($3-4\rho_0$) where, however, 
the interpretation of experimental data using presently available transport models becomes less robust due to the strong production of baryonic resonances and mesons together with the high temperature reached. Therefore, such experimental data above intermediate energies should be revisited in a near future.
Overall, from HICs performed at incident beam energies of a few hundred MeV/nucleon up to around 10 GeV/nucleon, the flow data indicate an EoS for symmetric nuclear matter with an incompressibility $K_0$ below 260~MeV, using a soft momentum-dependent EoS. It turned out that the momentum-dependence is required to reproduce as well the rapidity distributions of protons in semi-central collisions \cite{FOPI:2011aa}  and the yields of pions in central collisions, in particular at near or sub-threshold incident energies \cite{FOPI:2006ifg}.
 
Using FOPI data on the elliptic flow in Au+Au collisions between 0.4 and 1.5 GeV/nucleon, thanks to the broad acceptance of the detector, an enhanced precision in the determination of the EoS could be achieved. Including the full rapidity of the elliptic flow of protons and heavier isotopes \cite{LeFevre:2015paj} in the analysis with the Isospin-QMD (IQMD) transport model, the incompressibility was determined as $K_0=190\pm30$~MeV. This result has been obtained using the following method. 

\begin{figure}
\centering
\includegraphics[width=0.6\textwidth]{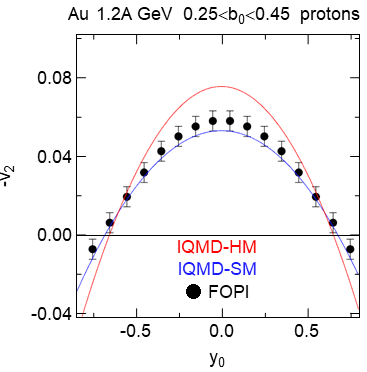}

\caption{(reprinted from Ref. \cite{LeFevre:2015paj} under permission) Proton elliptic flow data from mid-central collisions of Au+Au at 1.2 GeV/nucleon incident energy measured by the FOPI set-up: -v2(y0) (black dots), and IQMD-SM/HM simulations (blue/red curves, respectively). See text and source in Ref. \cite{LeFevre:2015paj} for further explanations.}
\label{fig:FOPI_IQMD_v2_y0}
\end{figure}

In Fig. ~\ref{fig:FOPI_IQMD_v2_y0}, extracted from Ref. ~\cite{LeFevre:2015paj}, we show a sample of proton 'elliptic', $-v_2$, flow data from the FOPI Collaboration (black dots with error bars) together with simulations using the IQMD transport model \cite{Hartnack:1997ez} with a stiff version of the EoS (HM, red) and a soft version (SM, blue) with momentum dependent interactions -- which is compulsory to describe other observables as already mentioned --, as a function of the rapidity in the centre of the colliding system scaled to that of the projectile $y_0 = y / y_{proj}$. The reaction is Au + Au at an incident energy of 1.2 GeV/nucleon. In Ref.~\cite{FOPI:2011aa} many more such data spanning the incident beam energy range 0.15–1.5 GeV/nucleon and varying centrality and system composition and size are shown. By plotting $-v_2$ (rather than $+v_2$) one sees in this energy regime an enhancement of strength around mid-rapidity, i.e. predominantly out-of-plane emission, which was given the suggestive name ‘squeeze-out’ in the pioneering, and predictive, theoretical work \cite{Stoecker:1981pg} on the subject. Nowadays this phenomenon is named elliptic flow which includes both effects of shadowing and density gradients. At these energies the elliptic flow changes sign at high $|y_0|$ becoming predominantly in-plane.

Taking a closer look at $-v_2(y_0)$ we see that the predicted shape is sensitive to the EoS in the full rapidity range. To take advantage of this feature, a quantity named $v_{2n}$ has been introduced defined by $v_{2n} = |v_{20}| + |v_{22}|$ where the parameters are fixed by a fit to the flow data using $v_2(y_0) = v_{20} + v_{22} \cdot y_0^2$ in the scaled rapidity range $|y_0| < 0.8$. Low momenta are cut off both in the data and the simulations. The cutting out of low transverse momenta (see \cite{LeFevre:2015paj} for details), originally forced by apparatus limitations, actually turns out to raise the sensitivity of $v_2$ to the EoS as flow is generally converging to zero at low momenta (see e.g. \cite{FOPI:2011aa}).

\begin{figure}
\centering
\includegraphics[width=0.6\textwidth]{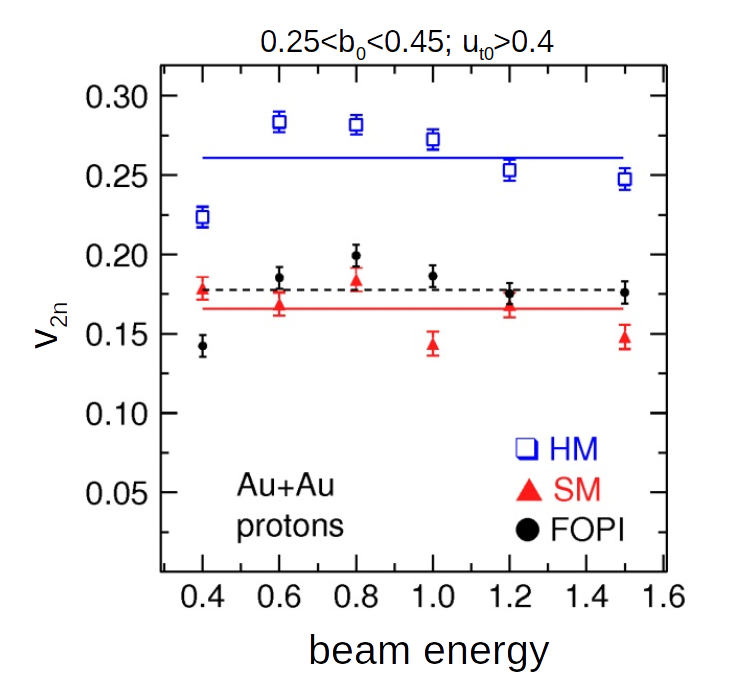}
\caption{(reprinted from Ref. \cite{LeFevre:2015paj} under permission) Elliptic flow $v_{2n}$ for protons as function of incident beam energy (in GeV) from mid-central collisions of Au+Au measured by the FOPI set-up. See text and source in Ref. \cite{LeFevre:2015paj} for further explanations.}
\label{fig:FOPI_IQMD_v2n_Einc}
\end{figure}

The $v_{2n}$ obtained for Au + Au between 0.4A and 1.5 GeV/nucleon with FOPI are shown in Fig. ~\ref{fig:FOPI_IQMD_v2n_Einc} for protons, from Ref. ~\cite{LeFevre:2015paj}. As the beam energy dependencies are rather weak, the average behavior is indicated by straight lines. The comparison of the data for $v_{2n}$ with the calculations shows a rather convincing preference for SM (soft with momentum dependence) EoS. In this work, the same conclusions have been drawn using elliptic flows of deuterons, tritons and $^3$He isotopes. The sensitivity is large: overall, there is a factor $1.63 \pm 0.06$ between HM and SM, a difference significantly exceeding the indicated experimental error bars. If we compare this factor with the fluctuations of the experimental data points around the average experimental values (dashed line) we can estimate the uncertainty of the deduced EoS. From the information using four different isotopes, the authors of Ref. \cite{LeFevre:2015paj} obtained in comparison with IQMD predictions a weighted average of the nuclear incompressibility modulus $K_0 = 190 \pm 30$ MeV. 

\begin{figure}
\centering
\includegraphics[width=0.49\textwidth]{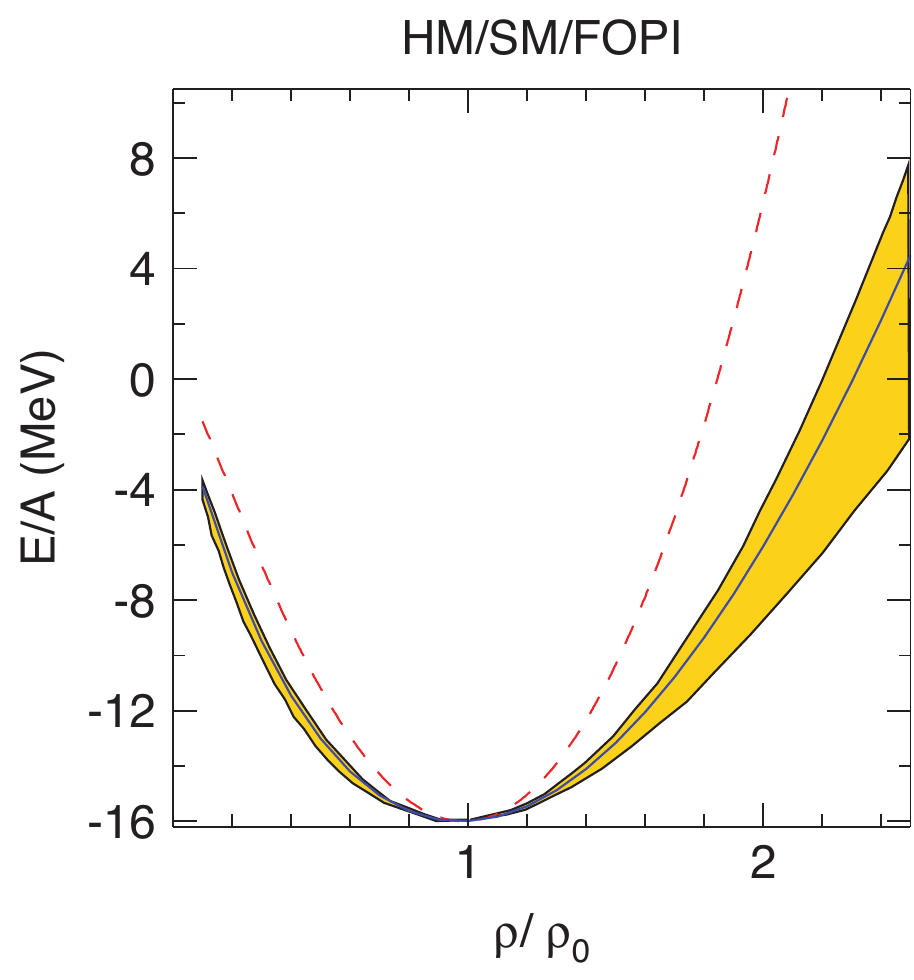}
\includegraphics[width=0.49\textwidth]{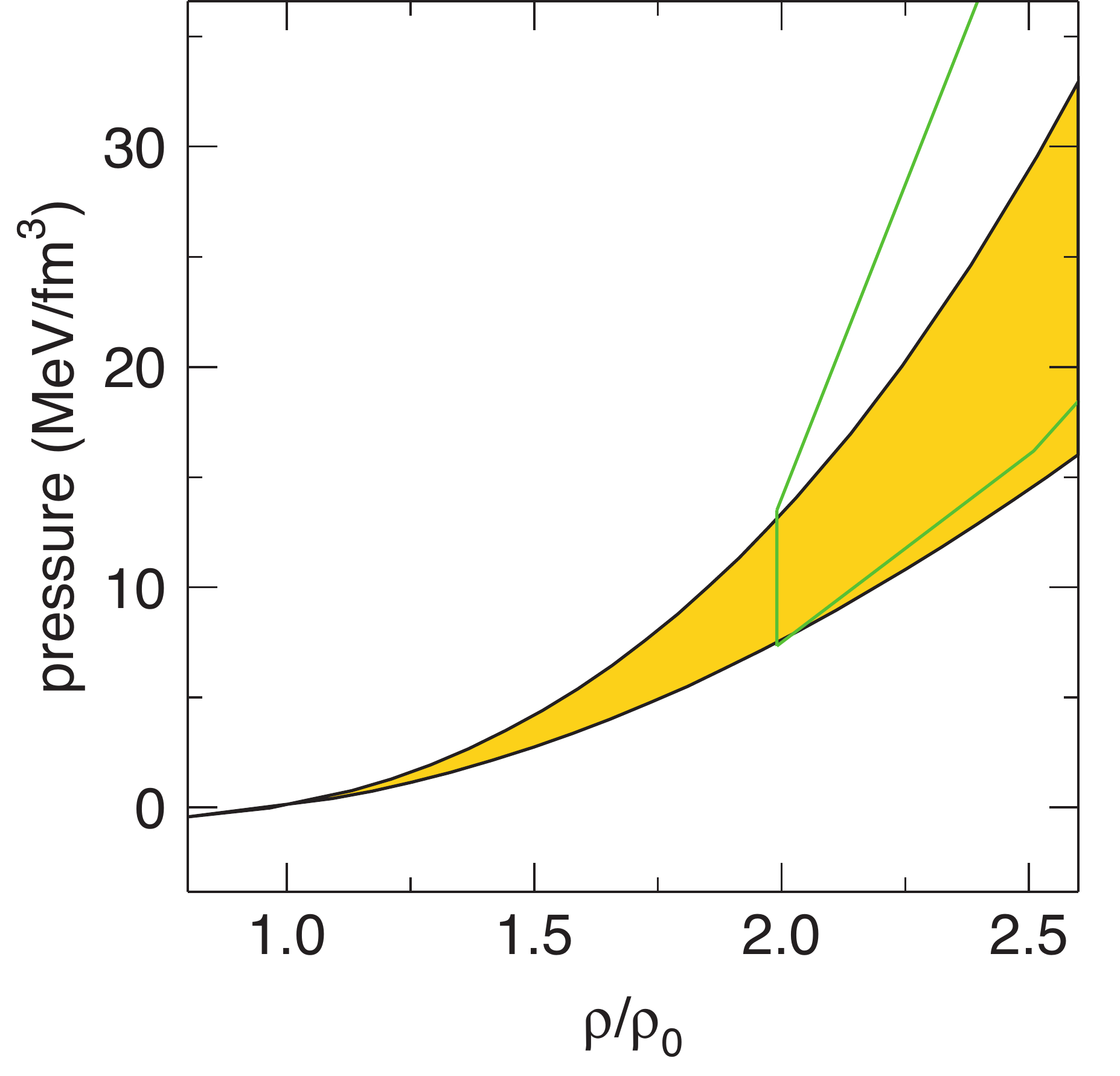}

\caption{(adapted from Ref.~\cite{LeFevre:2015paj} under permission) Left panel: Various symmetric nuclear matter EoS. Dashed (red) curve, HM; full (blue) curve, SM. The yellow band delimits the FOPI EoS constraints $K_0 = 190 \pm 30$ MeV. Right panel: Nuclear EoS in terms of pressure versus density as obtained from FOPI data (yellow band). 
The area framed with green lines originates from the analysis of \cite{Dani02Esymm}.
}
\label{fig:FOPI_EOS}
\end{figure}

The resulting EoS with its uncertainty band is plotted in Fig. ~\ref{fig:FOPI_EOS} (left panel). For comparison the ‘trial’ EoSs, HM and SM, used in the IQMD simulations are shown. Note that the phenomenological EoS HM and SM include the saturation point at $\rho/\rho_0=1$, $E/A$ = -16 MeV by construction. This fixes the absolute position of the curves: the heavy-ion data are only sensitive to the shape, i.e. the pressure which is essentially the derivative. Therefore the uncertainty of this fundamental point, about 0.5 MeV for the binding energy and 0.01 fm$^{-3}$
for $\rho_0$, is not included in the uncertainty band. However, one can conclude, in complete agreement with Ref. \cite{Dani02Esymm}, that a stiff EoS, characterized by $K_0 = 380$ MeV is not in agreement with flow data in the incident energy range (0.4–1.5) GeV/nucleon. For 0.4 GeV/nucleon this had also been suggested in \cite{FOPI:2004hyz}.
The corresponding EoS constraint in terms of pressure is shown in Fig. ~\ref{fig:FOPI_EOS} (right panel), along with its extension at larger densities as deduced from \cite{Dani02Esymm}.

\begin{figure}
\centering
\includegraphics[width=0.6\textwidth]{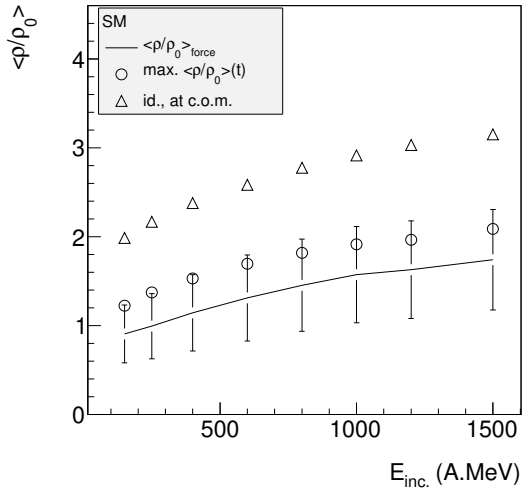}
\caption{(reprinted from Ref.~\cite{LeFevre:2015paj} under permission) Density probed by the elliptic flow in semi-central Au+Au collisions measured by FOPI from Ref. ~\cite{LeFevre:2015paj} as a function of the incident energy. Open circles and triangles: maximum of the density $<\rho(t)/\rho_0 >_{max}$ reached in the central volume of the collision and at the centre-of-mass of the system respectively. Line: “force-weighted” mean value of the reduced density (see text) seen by protons in their final configuration averaged until the passing time of the colliding system. The error bars represent the standard deviations of the distributions.}
\label{fig:FOPI_density}
\end{figure}

In Ref.~\cite{LeFevre:2015paj}, the sensitive density range, using the same IQMD predictions, has been found to span between around 0.7 and 3 times the saturation density, over all incident energies studied.
This conclusion has been obtained from the observation that the elliptic flow signal is dominated by the acting of the mean-field potential. This led authors of this work to determine, from IQMD simulations fitting FOPI data -- therefore with a SM EoS --, the density range which is connected to the elliptic flow of protons. It is obtained by weighting the mean value of the density seen by all protons building the flow (with the same low momentum and rapidity cut-off as in the flow analysis) by the strength of the force due to the local mean field. This quantity integrated up to the full passing time of the colliding system is displayed by the line of Fig. ~\ref{fig:FOPI_density} as a function of bombarding energy. The error bars represent the variances of the “force-weighted” densities. Using such a method, one is not limited to some small most central volume of the system, a procedure which then naturally yields significantly larger densities, as depicted by triangles in Fig. ~\ref{fig:FOPI_density} corresponding to the innermost region of the collision. There, densities up to $3.5\rho_0$ are reached at the highest beam energies measured by the FOPI Collaboration. The maximum density experienced by all protons measured in the region around mid-rapidity $|y_0| < 0.8$ (circles) is substantially lower on average (up to $2\rho_0$).

This result was confirmed later by interpreting the same data with three Skyrme energy-density functionals introduced into the ultrarelativistic QMD (UrQMD) transport model \cite{Wang:2018hsw}, leading to $K_{0}=220\pm40$~MeV. The interval of confidence used in \cite{Huth:2021bsp} $K_0=200\pm25$~MeV reflects both predictions. 
Note that the constraints deduced from the analysis of elliptic flow are compatible with earlier findings of the kaon studies \cite{Sturm:2000dm,Fuchs:2001gv} already discussed in Sect. \ref{sec:111}.

\section{Experimental investigation of the equation-of-state of isospin-asymmetric matter at low-density}


\label{sec:211}

In the Fermi energy domain, from about 20 to 100 MeV/nucleon, HICs can create 
nuclear matter with density, excitation energy, isospin asymmetries and temperature different from the initial 
state at saturation density $\rho_0$. Nuclear matter produced in such collisions has as main decay modes the emission 
of fragments, clusters, light charged particles and neutrons and this is the main source of experimental information 
upon the nuclear symmetry energy in heavy ion collisions. In order 
to reach this goal measurements with projectiles and targets with the most possible 
large isospin asymmetries (by using stable or radioactive beams) are used in order to 
identify observables that are sensitive to the symmetry energy. 
As an example, in central most violent collisions the nuclei, initially 
compressed, can expand to densities as low as 0.3$\rho_0$. These nuclei are expected to 
undergo a liquid-gas phase transition whose main characteristic should be 
the production of a high multiplicity of intermediate mass fragments (IMF) and 
light particles in the so-called nuclear multifragmentation phenomenon \cite{bor08}. 
Note that the onset of multifragmentation, characterized by excitation energies below the binding energy 
of nuclei, temperatures around 3-7 MeV and sub-saturation densities, can be reached 
in collisions at Fermi energies \cite{bor08}
as well as in the spectator break-up at relativistic energies \cite{poc95},
independently from the beam energy.
In fact, results at different beam energies can be compared, in terms of caloric curves,  
within the same systematic \cite{nat02}.

Indeed, light-ion clustering happening 
at extreme low densities ($\le$ 0.2 $\rho_0$) has been observed \cite{wad12,qin12} by 
looking at the expansion and cooling of an intermediate source in semi-central collisions. 
It has been shown \cite{wad12} that cluster formation increases the 
symmetry energy at very low densities and mimics conditions of temperatures and densities that are relevant in 
studying stellar evolution and core-collapse supernovae as in the supernova neutrino 
sphere \cite{rob12, hag16, carl17}. Going from central to semi-peripheral 
collisions several experimental observables have been used to constrain the density 
dependence of the symmetry energy (see reviews \cite{Li:2008gp,hor14,tsa12,mcin19,col20} for an exhaustive 
description and summary). 

\begin{figure}[t]
\centering
\includegraphics[width=0.8\textwidth]{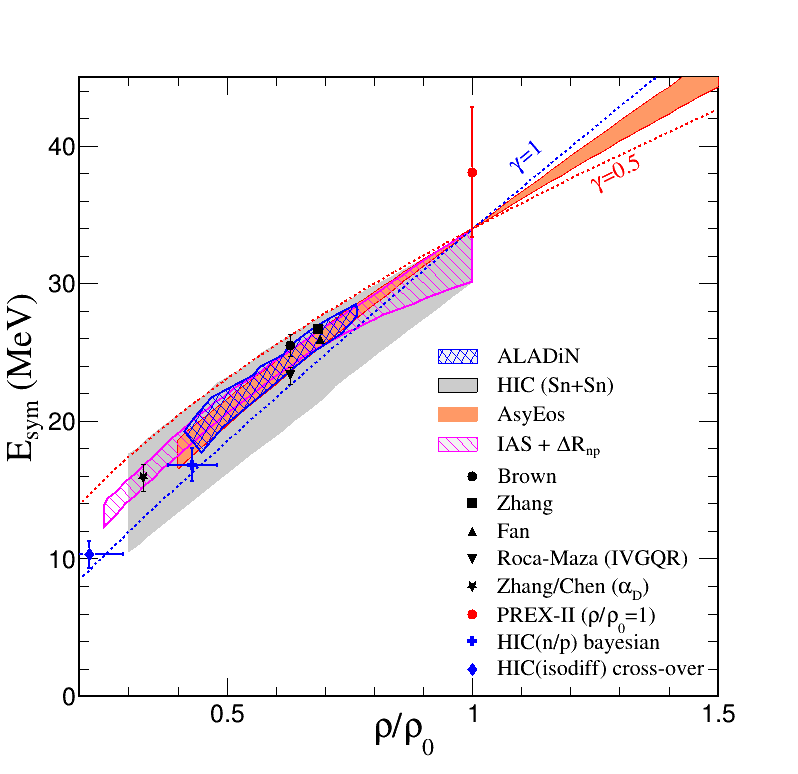}
\caption{Symmetry energy E$_{sym}$ as a function of baryonic density normalized to the 
nuclear saturation density $\rho_0$. The grey area corresponds to isospin diffusion data on Sn+Sn 
collisions \cite{PhysRevLett.102.122701}. The blue hatched area shows results of the S254 ALADiN experiment at GSI (see text)
\cite{LeFevre22}. The orange area displays ASY-EOS Au+Au data extrapolated at low density \cite{Russotto:2016ucm}. 
The magenta hatched area corresponds to Isobaric Analog States (IAS) analysis and neutron skin data from \cite{Danielewicz:2013upa}. 
For comparison two different parameterizations of the symmetry energy (see text) 
corresponding to different values of the $\gamma$ parameter (soft=0.5, stiff=1) in eq. (\ref{urqmd-par}) are also indicated. Black symbols correspond to nuclear-structure observables. Red symbol is from PREX-II experiment \cite{reed21}. Blue symbols are from Refs. \cite{lyn21} and \cite{morf19} (see text for a detailed description).}
\label{figlowdens}
\end{figure}

Constraints from HICs in the Fermi energy regime are obtained in central collisions, 
for example from multifragmentation and isoscaling studies \cite{she07}, or the neutron-proton yield ratio 
\cite{Coupland-PhysRevC.94.011601}, and in semi-peripheral collisions focused on isospin transport dynamics: 
isospin diffusion \cite{PhysRevLett.102.122701,sun10,cam21} and migration through a transient ``neck'' region \cite{DeFilippo:2012qd,hud12,pia21}. 
The basic idea is, for isospin-asymmetric nuclei in non-central collisions, 
that the isospin transport (diffusion) of nucleons 
(protons and neutrons) handles the degree of the isospin equilibration and is driven by 
the symmetry energy strength at low density \cite{riz08}. In addition, in semi-peripheral 
events, the formation of a low density neutron rich ``neck'' region is expected by transport model calculations, 
due to a larger neutron flow (isospin migration) from the two reaction partners of the collision 
(projectile-like and target-like fragment) towards the neck. This flow is ruled by the density derivative 
of the symmetry energy. In this case light IMFs that are formed from the neck fragmentation are expected to be 
more neutron rich depending on the stiffness of the symmetry energy at low density \cite{BARAN2005335} as 
effectively has been experimentally probed \cite{DeFilippo:2013ipa,pag20}. 
Indeed, a rich physics connected to the isospin properties of the emitted fragments 
is related to the deformation
and fission of the projectile-like fragment in the dynamical stage of the reaction \cite{rusfis20,jedel17} 
that seems promising to bring new knowledge into the EoS of asymmetric matter around or below $\rho_0$.  

Fig. \ref{figlowdens} shows a summary of constraints on the symmetry energy 
as a function of the baryonic density obtained from different experimental data from heavy ion 
collisions and nuclear structure observables. In particular, the grey band shows experimental data from 
isospin diffusion in mid-peripheral Sn+Sn (50 MeV/nucleon) heavy ion collisions \cite{PhysRevLett.102.122701} 
analyzed with the quantum molecular dynamics (ImQMD) transport model \cite{zhaplb08}. The blue hatched area 
corresponds to the analysis of the projectile spectator decay in ALADiN S254 experiment \cite{LeFevre22} 
(Au, $^{124}$Sn, $^{107}$Sn and La beams at 600 MeV/nucleon \cite{sfi09}), with the FRIGA+IQMD \cite{LeFevre19} transport model predictions; 
the main observable exploited in those data is the dependence on charge (Z) of the width of mass ($\sigma A$) distributions of clusters emitted in the projectile spectator break-up in semi-central collisions. This width probes the density dependence of the symmetry energy at sub-saturation densities at values between 0.4 and 0.8 $\rho_0$.  
The symmetry energy function used in the model is of the form  $E_{sym}(\rho) = 9.0(\rho/\rho_0)^{2/3}+23.2(\rho/\rho_0)^\gamma$ being $\gamma$ the stiffness parameter describing the potential term of the symmetry energy. These values
correspond to E$_{sym}=32.2$ MeV at $\rho_0$. 
The data analysis is performed over a broad range of clusters (from Z=3 to Z=26). Depending on cluster size the value of the most probable $\gamma$ value and uncertainties, the probed density and the E$_{sym}$ values are determined and reported as the blue hatched area in Fig. \ref{figlowdens}. 

The orange area corresponds to ASY-EOS, Au+Au data extrapolated at low density \cite{Russotto:2016ucm} that will be 
fully described in chapter \ref{sec:232} where flow data are compared with the UrQMD model. The symmetry 
energy functional used in the model (see eq. \ref{urqmd-par}) is plotted for reference as a line for two 
different indicated values of the $\gamma$ parameter (0.5 and 1).    

The plot shows 
comparisons (black points) with those obtained from nuclear structure data and calculations: 
isobaric analog state (IAS) and analysis on neutron skin data (hatched pink area) \cite{Danielewicz:2013upa}; 
electric dipole polarizability related to the Giant Dipole Resonance in nuclei (GDR) 
\cite{zha15,Tamii-PhysRevLett.107.062502} (labeled Zhang/Chen $(\alpha_{D})$) and determination of symmetry energy related to properties of the 
isovector giant quadrupole resonance (IVGQR) in $^{208}$Pb \cite{roc13}; 
data from nuclear masses of neutron rich nuclei (labels Brown, Zhang and Fan in the figure) 
\cite{Zhang:2013wna, Brown:2013mga, fan14}. In all these cases the value of the symmetry energy S($\rho$) 
is plotted at the sub-saturation cross density $\rho$ values at which it is constrained, as reported from the 
respective publications. For example, Zhang and Chen \cite{Zhang:2013wna} calculate, using a Skyrme–Hartree–Fock approach, the symmetry energy 
E$_{sym}$=26.65$\pm$0.20 and $L$=46$\pm$4.5 MeV at $\rho_{c}\approx$0.11 fm$^-3$ (as indicated in the figure by a filled square), by using isotope binding energy difference and neutron skin thickness, and then, also, extrapolate these values at $\rho_0$. 

The slope parameter $L$ of $E_{sym}$ around the saturation density is also 
correlated linearly with the neutron skin
thickness of heavy nuclei \cite{Roca-Maza:2011qcr}. Notable are the recent measurements of the neutron skin thickness 
in $^{48}Ca$ (CREX data) and $^{208}Pb$ (PREX-II data) based on the parity-violating asymmetry in the elastic scattering of polarized 
electrons from $^{48}Ca$ \cite{adh22} and $^{208}Pb$ \cite{adh21,reed21} respectively. Analysis of PREX-II data leads to a value of $E_{sym}=38.1\pm4.7$ MeV 
and $L=106\pm 37$ MeV, when the slope $L$ is calculated at $\rho_0$ (indicated as a red circle in the figure), 
which is larger (stiffer density dependence of the $E_{sym}$) 
than previous average values from both theoretical approaches and experimental measurements, 
but closer to the systematic \cite{lyn21} ($L_{01}=71.5\pm 22.6$) 
if the slope is calculated at $2/3\rho_0 \approx 0.1 fm^{-3}$.  This new approach, recently presented in Ref. \cite{lyn21}, is based on the most probable density range that a given observable probes, extracting the values of $L$, $E_{sym}$ or pressure relative to the explored density rather than at the saturation density. It has been already summarized in the introduction and results are shown in Fig. \ref{fig:lynch21}. Finally the blue points show results from a Bayesian analysis on neutron/proton ratio observable in central Sn+Sn collisions at 120 MeV/A \cite{morf19} and the cross-over analysis for isospin diffusion in Sn+Sn heavy ion collisions  \cite{PhysRevLett.102.122701}.

Thus, results in Fig. \ref{figlowdens} suggest a remarkable coherence between HIC and nuclear structure results at low densities. Indeed data from nuclear structure 
have to be extrapolated toward saturation density. On the contrary, HICs can give direct experimental access also 
to supra-saturation densities. Thus, the consistency of the interpretations below and above saturation density 
is yet a challenge in order to test the robustness of the transport model predictions,
that depends on the physical input parameterization and on the different strategies in 
the simulation adopted by the given transport code. 
Furthermore, the sub-saturation density region between $0.4<\rho/\rho_0 <0.7$ is 
expected approximately to be in the same range as the crust-core transition density in the zone that 
separates the liquid core from the solid crust of a neutron star \cite{heb13,duc11}; 
this make relevant the terrestrial observables in HIC 
and nuclear structure at sub-saturation density for neutron stars studies, under the condition 
that the density range experimentally probed by a given observable 
is precisely determined and taken into account when constraining the symmetry energy.  
From this overview it results that most of the actual constrains to probe the symmetry 
energy below or around the saturation density 
are obtained by using HICs at intermediate energy, 
together with nuclear structure studies or exploiting collective motions, as Giant or Pygmy \cite{carb10} 
dipole resonances in neutron rich nuclei. 
The interplay and coherence between experimental observables and theoretical models in order to constrain the 
symmetry energy is essential but not completely achieved. 
In fact different conclusions could be often drawn from the same data by relying on transport 
simulations or changing the experimental observable, despite big efforts to test the robustness of transport model predictions within the transport model evaluation project (TMEP) \cite{Xu:2016lue,Zhang:2017esm,Ono:2019ndq,Colonna:2021xuh,TMEP:2022xjg}. Indeed, new ``ingredients'' 
as the momentum dependence of the mean-field in the isovector 
channel or the cluster correlations in fragment formation \cite{ono19} have been introduced 
in theoretical models. 
For example, when the density dependence of the symmetry energy is constrained
by means of the pre-equilibrium neutrons and protons ratio observable \cite{morf19}, also the momentum dependence in the symmetry mean field potential has to be taken into account, since it affects the density dependent neutron and proton effective mass splitting.
From the experimental point of view, 
a higher sensitivity to symmetry energy and the reduction of the uncertainty in the constraints 
is expected by the increase of the $N/Z$ of the entrance channel 
by using radioactive neutron poor (rich) beams (RIBs), by looking at multiple observables in the same experiment, so that many observables are explained simultaneously by the same analysis, 
and in parallel improving the detection capabilities:
simultaneously detection of neutron and charged particles, improved isotopic resolution, higher energy and angular resolution in new devices.



\section{Experimental investigation of the equation-of-state of isospin-asymmetric matter at high-density}
\label{sec:22}
In this section, studies constraining the high-density dependence of the symmetry energy, i.e. the isospin-asymmetric contribution to the EoS at densities above saturation density, will be reviewed, in analogy with what has been developed in sect. \ref{sec:1} for the isospin-symmetric contribution to the EoS. Here also, observables based on particle flow and pion have been used. The main difference is the use of relative observables, such as ratios or differences of neutron-proton and $\pi^{+}-\pi^{-}$ observables, in order to try to exhibit the symmetry energy effects. Kaon related studies will not be discussed since, as mentioned in the introduction, the $K^{+}/K^{0}$ yield ratio proposed by Ferini \textit{et al.} \cite{Ferini:2006je} resulted so far to be not very sensitive to the symmetry energy when used in realistic HIC \cite{FOPI:2007gvb}, even if future improvements in transport theories and in the statistical accuracy of experimental data maybe foreseen, that could change this conclusion.\\
The opportunities of comparing HIC constraints for the symmetry energy at high-density, to the ones recently offered by the multi-messenger astronomy, or using together the two classes of results in more advanced analysis scheme, will be discussed in sect. \ref{sec:2324}.\\
Given the still few results for the symmetry energy at high density, and the related uncertainties, at the end of this section, we will present also some perspectives for future studies (sect. \ref{sec:233}), aiming at improving our knowledge of the density dependence of the symmetry energy.\\

\subsection{Results of studies based on pions}
\label{sec:221}

The ratio of charged pion multiplicities has been proposed as a viable candidate for constraining
the density dependence of symmetry energy by Bao-An Li and collaborators almost two decades
ago \cite{Li:2002qx,Li:2004cq}. Soon after, experimental data for this observable, measured by the FOPI Collaboration at GSI for a few systems, in particular Au+Au, at several impact energies \cite{FOPI:2006ifg}, have become available and have been used to study the symmetry energy \cite{Xiao:2009zza,Feng:2009am,Xie:2013np,Hong:2013yva}. Extracted values for the slope parameter $L$ ranging from very soft to very stiff have revealed a rather poor understanding of pion production mechanisms in heavy-ion reactions close to threshold and possible significant model dependence between the various types of transport models employed in these studies. Additionally, a simultaneous description of pion and nucleon observables of interest for studying the nuclear matter EoS was found not to be possible for most models. In particular, using the pBUU transport model it was shown that the description of existing experimental data for elliptic flow of protons and pion multiplicities requires different momentum dependent interactions \cite{Hong:2013yva}, a feature commonly quantified in terms of the isoscalar effective mass of the nucleon. 

Efforts to understand these discrepancies have been focused on studying the impact of in-medium  modifications of the pion-nucleon interaction, the kinetic part of symmetry energy, size of the neutron skin and in-medium modification of the pion production threshold, to name just a few. The last one leads to the enhancement or suppression of particle production as result of more attractive or respectively repulsive potentials in the final state as compared to vacuum~\cite{Ferrini:2005jw,Ferini:2006je}. It has been shown to have a non-negligible impact on pion multiplicities and their ratio close to threshold \cite{Song:2015hua,Cozma:2014yna}. The inclusion of these effects is also a prerequisite for a consistent description of thermodynamic equilibrium of nuclear matter using transport models \cite{Zhang:2017nck}. For a quantitative comparison to experimental data additional effects, such as in-medium modification of resonance production cross-sections and pion optical potentials, have to be taken into account \cite{Zhang:2017mps,Zhang:2018ool,Cozma:2016qej}.

An informal collaboration of transport model practitioners currently pursuing the study of symmetry energy has been established with the goal of understanding the origin of differences between existing transport models. This effort has taken the form of transport model simulations under controlled conditions, devised to test each transport model ingredient separately. Over the span of the last 8 years several issues have been addressed. A first study has aimed at evidencing the model dependence induced by the initial state, Pauli blocking algorithm and the collision term on several nucleonic observables in HICs at impact energies of 100 and 400 MeV/nucleon \cite{Xu:2016lue}.
The next two where devoted to the study of the collision integral in a box of constant density and given temperature for the case of pure nuclear matter \cite{Zhang:2017esm} and admixture of nucleons, $\Delta$(1232) isobars and pions \cite{Ono:2019ndq}. 
Recently, an investigation of the mean-field response has been finalized and systematic differences between BUU and QMD type of transport models have been better understood~\cite{Colonna:2021xuh}. A summary of these efforts together with a brief description of each transport model used in these works can be found in Ref.~\cite{TMEP:2022xjg}. Further studies addressing
model dependence of pion production in HICs and the impact of momentum dependent interactions and the induced threshold effects on particle production are in progress.


A dedicated experiment for the purpose of studying the symmetry energy using pion observables has been proposed by the S$\pi$IRIT Collaboration. A new TPC detector, capable of accurately measuring pions down to threshold kinetic energies, has been developed \cite{Shane:2014tsa,SpRIT:2016aqk,Barney:2020mxk} and employed to gather experimental data for HICs of four combinations of Sn isotopes, in particular $^{132}$Sn+$^{124}$Sn and $^{108}$Sn+$^{112}$Sn, at an incident energy of 270 MeV/nucleon at RIBF (Japan). First results, restricted to integrated pion single- and double-ratios \cite{SpRIT:2020blg} and their transverse momentum spectra \cite{SRIT:2021gcy}, have been recently published. 

\begin{figure}[t]
\centerline{\includegraphics[width=0.85\textwidth]{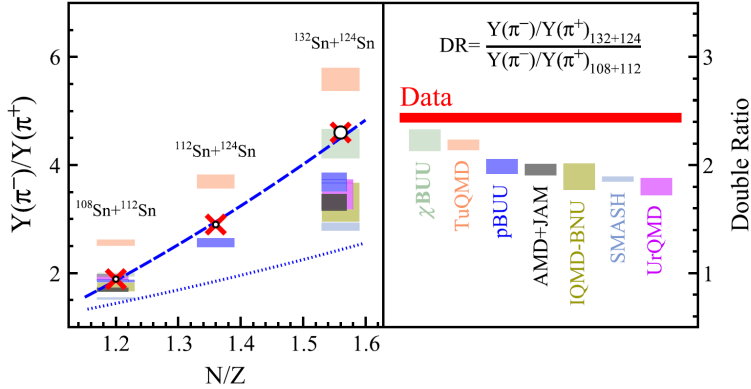}}
\caption{(reprinted from Ref.~\cite{SpRIT:2020blg} under permission) Left panel: the charged pion multiplicity ratio, as a function of N/Z, measured by the S$\pi$RIT Collaboration (crosses) is compared with predictions of seven transport models. Right panel: The experimental double ratio (horizontal bar) is compared with predictions of the same models.}
\label{fig:spiritpions}
\end{figure}

In Fig.\ref{fig:spiritpions} the experimental charged pion single and double ratios are compared with theoretical predictions, performed without any prior knowledge of experimental values, from seven transport models (3 BUU and 4 QMD codes) using their standard parameter settings. The $\chi$BUU model predictions reproduces experimental data the closest, in part due to prior adjustment of in-medium inelastic cross-sections to FOPI experimental data for pion production for Au+Au collisions at 400 MeV/nucleon. The TuQMD model reproduces the double ratio equally well but single ratios are significantly over-predicted. This can be alleviated by adjusting the strength of the poorly known isovector $\Delta$(1232) potential \cite{Cozma:2014yna}, as a result of increased sensitivity to this quantity induced by imposing energy conservation. All other models under-predict both the single ratio for neutron rich systems and the double ratio by significant margins. 

Differences among these transport models are more evident for pion multiplicities, with predictions varying by factors up to three \cite{SpRIT:2020blg}. Most models systematically over-predict or under-predict experimental values of multiplicities for all measured systems. Qualitatively, this is the result of using different nucleon optical potentials and the inclusion/omission of threshold effects. Model calculations show that low kinetic energy pions are significantly affected by the Coulomb interaction, pion optical potentials and other dynamical effects ($\it e.g.$ threshold effects). In line with that, two more recent publications show that integrated S$\pi$RIT pion multiplicities can be described by adjusting other model ingredients, such as the magnitude of short-range correlation~\cite{Yong:2021nwn} or momentum dependence of the symmetry potential~\cite{Wei:2021arw}. 

The model ingredients mentioned above were shown to have a limited impact on the high energy tail of pion spectra, allowing the study of the density dependence of the symmetry energy as well as the momentum dependence of isovector mean-field potentials \cite{Cozma:2021tfu}. Constraints for the slope parameter of symmetry energy at saturation have been extracted by comparing theoretical predictions of dcQMD model (a newer version of TuQMD) and the high transverse momentum tail ($p_T\ge$ 200 MeV/c) of pion single ratio spectra in neutron-rich $^{132}$Sn+$^{124}$Sn and neutron-deficient $^{108}$Sn+$^{112}$Sn systems \cite{SRIT:2021gcy}. To this end, parameters related to isoscalar quantities (nucleon isoscalar effective mass $m^*/m$=0.7, compressibility modulus of symmetric nuclear matter $K_0$=245 MeV and in-medium modification factor of elastic nucleon-nucleon cross-sections) have been adjusted to qualitatively reproduce experimental data for stopping, transverse and elliptic flow of protons and light fragments in HICs of impact energies spanned by the experimental data set gathered by the FOPI Collaboration \cite{FOPI:2010xrt,FOPI:2011aa}. The symmetry energy has been fixed at a sub-saturation point to S($\rho$=0.1 fm$^{-3}$)=25.5 MeV in agreement with precise constraints from nuclear structure \cite{Brown:2013mga,Zhang:2013wna}. The slope $L$ and curvature $K_{sym}$ parameters of symmetry energy have been correlated via $K_{sym}$=-488+6.728$\times L$ (MeV). Theoretical simulations with the slope $L$ and neutron-proton effective mass splitting $\Delta m^*_{np}/\delta$ varied in the ranges [15,151] MeV and [-0.33,0.33] respectively have been performed. The best description of experimental data was achieved for values of the slope $L$=79.9$\pm$37.6 MeV, independent of $\Delta m^*_{np}$. This constraint corresponds to a value of the symmetry energy at saturation $S_0$=35.3$\pm$2.8 MeV.

\subsubsection{Results from pions as measured by FOPI}
\label{sec:2211}

As previously mentioned, thermodynamic consistency of particle production in nuclear matter requires the inclusion of threshold effects~\cite{Ferrini:2005jw,Zhang:2017nck}. Only a few transport models include these important terms: RBUU (Catania)~\cite{Ferini:2006je}, RVUU/ $\chi$BUU (Texas A$\&$M)~\cite{Song:2015hua}, TuQMD/dcQMD (Bucharest)~\cite{Cozma:2014yna} and CBUU (Giessen) \cite{Teis:1996kx}. The first three been used to study the symmetry energy by comparing model prediction to FOPI experimental data: subthreshold kaon production using RBUU~\cite{FOPI:2007gvb} and pion production slightly above threshold~\cite{FOPI:2010xrt} using RVUU and TuQMD. The sensitivity of the kaon ratio has proven insufficient to lead to an unambiguous result~\cite{FOPI:2007gvb}, though confirmations of the result using other transport models is still missing. In contrast, a constraint, albeit a rather qualitative one, could be extracted using the pion ratio. These results are briefly reviewed in the following. We will restrict the presentation only to results obtained using the RVUU and TuQMD/dcQMD transport models. Results obtained by other groups~\cite{Xiao:2009zza,Feng:2009am,Xie:2013np,Hong:2013yva}, while crucial in increasing the interest in pion production as a probe for the symmetry energy and the appearance of the TMEP collaborative effort, are nowadays understood as unrealistic due to their lack of thermodynamical consistency.

\begin{figure}[t]
 \begin{minipage}[t]{0.5\linewidth}
 \vspace*{-45mm}
\begin{center}
\includegraphics[width=1\textwidth]{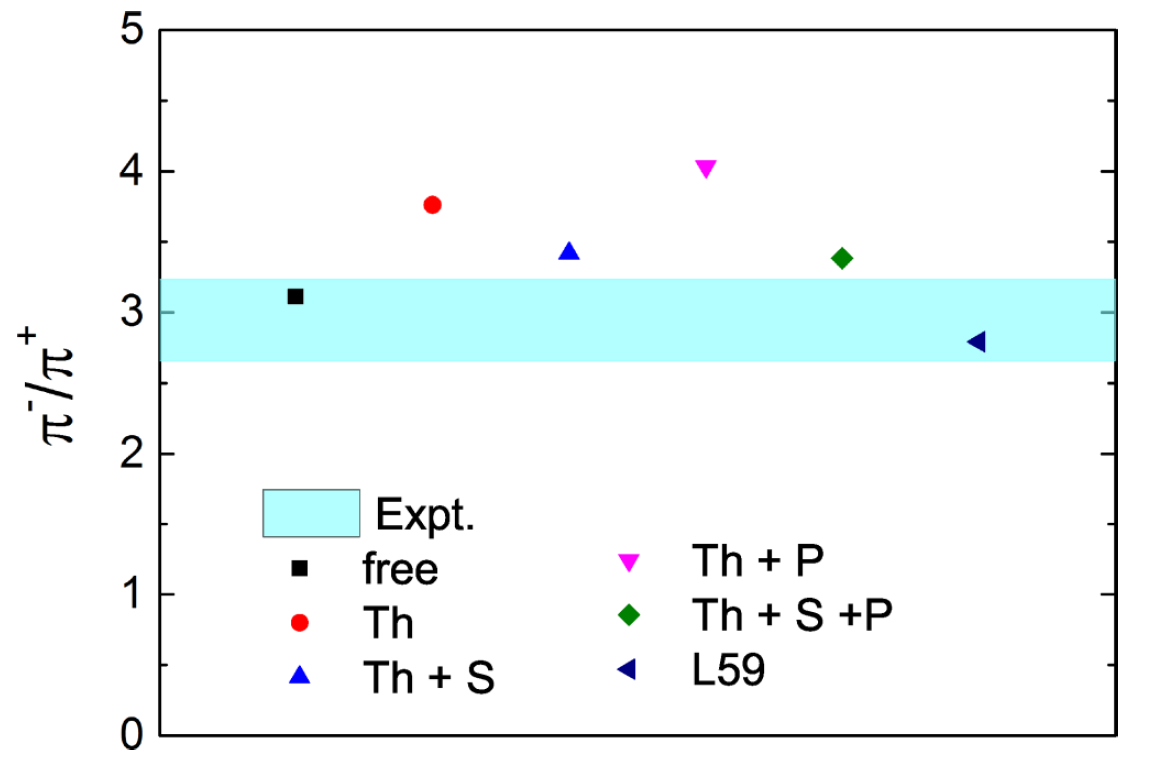}
 \end{center}
 \end{minipage}
 \hfill
 \begin{minipage}[t]{0.5\linewidth}

 \begin{center}
\includegraphics[width=1\textwidth]{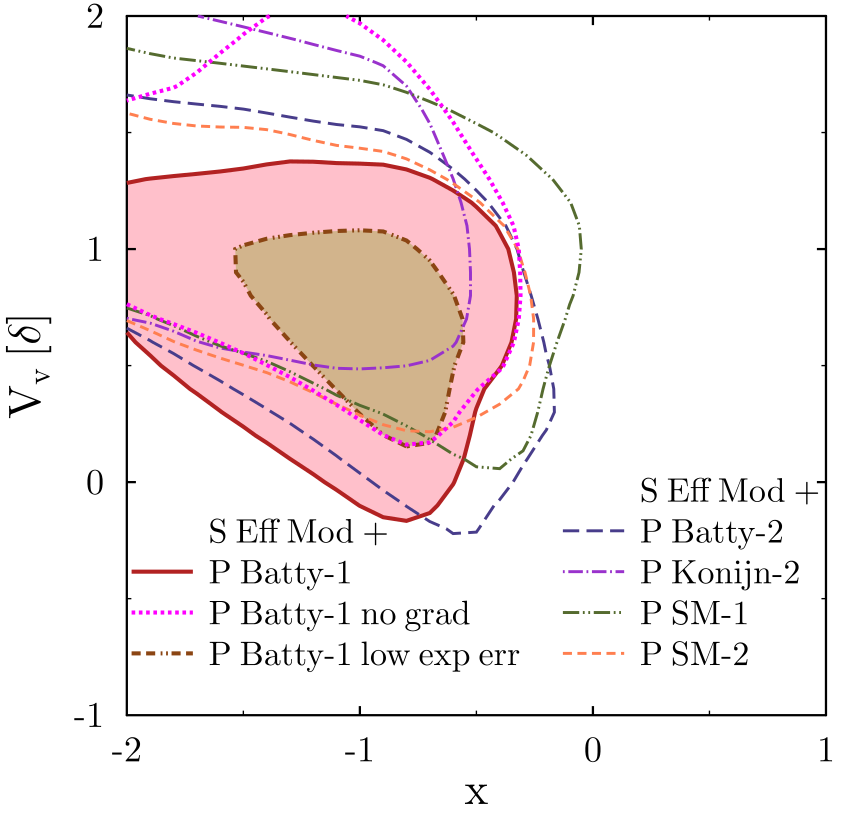} 
 \end{center}
 \end{minipage}

\caption{(reprinted from Ref.~\cite{Cozma:2016qej} under permission) Left panel: The $\pi^-/\pi^+$ multiplicity ratio in central Au+Au collisions at an impact energy of 400 MeV/nucleon using the NL$\rho$ interaction and the RVUU transport model for several choices of the pion potential (none, S, P, S+P waves), with/without threshold effects (Th) and two different values for the slope parameter $L$, 59 and 84 MeV. Reprinted from Ref.~\cite{Zhang:2017mps}, with permission. Right panel: Constraints for the slope parameter $L$ and strength of the isovector $\Delta$(1232) potential $V_v$ using the pion multiplicity and average transverse momentum ratios for central Au+Au collisions at 400 MeV/nucleon incident energy and the TuQMD model. The value of the slope parameter is given by $L$=60.5-45.0$\cdot$x [MeV]. Each contour curve represents the extracted 68$\%$ CL region for different choices of the S and P wave pion optical potentials. }

\label{fig:piratio_fopi_auau0400}
\end{figure}

The relativisticaly covariant RVUU transport model makes use of the non-linear relativistic NL$\rho$ and NL$\rho\delta$ interactions to describe the propagation and collision of nucleon, $\Delta$(1232) and pion degrees of freedom. Using the mean-field expressions for baryonic self-energies the modification of the threshold condition has been deduced and its impact on pion production has been studied~\cite{Song:2015hua}. It was found that both total multiplicity and single ratio of pions are increased substantially. Furthermore, a stiffer symmetry energy leads to a higher pion ratio, contrary to the case when in-medium threshold shifts are neglected. In-medium inelastic cross-section have been multiplied by a density dependent factor whose strength is adjusted in order to reproduce experimental total yields. Theoretical pion ratios using either the NL$\rho$ or NL$\rho\delta$ interactions are in agreement with the experimental values for Au+Au collisions.

A quantitative comparison to experimental data requires the inclusion of medium effects on $\Delta$(1232) and pion properties. For the former, this is often described as a shift of the pole mass accompanied by a modification of the decay width. For pions it amounts to the inclusion of the pion optical potential in the mean-field propagation term and of the in-medium dispersion relation when determining the pion production/absorption in elementary reactions. These effects had been included in RVUU in a subsequent publication~\cite{Zhang:2017mps} by using chiral perturbation theory results for the S-wave pion interaction and a $\Delta$-hole model for the P-wave component. It was shown that the S and P wave potentials lead to an enhancement and reduction of the single pion ratio respectively. By tuning the strength of the $\rho$ meson coupling to nucleons it was found that a symmetry energy with a slope of $L$=59 MeV reproduces experimental data the best (see the left panel of Fig.~\ref{fig:piratio_fopi_auau0400}). Similar results are obtained using the non-relativistic $\chi$BUU model which employs an interaction that reproduces the EoS of nuclear matter derived from chiral perturbation theory, including the nucleon effective potential~\cite{Zhang:2018ool}.

The TuQMD transport model has been upgraded to describe pion production realistically starting from the observation that within the traditional approach the collision term leads to a violation of energy conservation as a result of different potentials in the initial and final states. Imposing that final state kinematics is determined by obeying energy conservation naturally leads to the appearance of shifts of particle production thresholds in nuclear matter~\cite{Cozma:2014yna}. It was found that these effects have a sizable impact of pion yields and single ratios, in qualitative agreement with the RVUU model results described above. Additionally it was shown that the strength of the $\Delta$(1232) potential, in particular its isovector component, have a sizable impact on pion ratios. Consequently, constraints for the density dependence of symmetry energy cannot be unambiguously extracted from multiplicity ratio alone, given that no information about the isovector $\Delta$ potential currently exists.

This conundrum was addressed in a later publication~\cite{Cozma:2016qej}. It was shown that both the slope $L$ of symmetry energy and strength $V_v$ of the isovector $\Delta$ potential can be simultaneously determined by comparing model predictions with experimental data for both yield and transverse momentum ratios of pions. The effect of S and P wave pion optical potentials have been included in the model in order to describe the second observable reliably. The well known Ericson-Ericson parametrization is used to describe the density and isospin asymmetry of these quantities. Their free parameters have been fixed by using a combination of empirical information obtained from the study of pionic atoms and pion-nucleus scattering, results from chiral perturbation theory for the S-wave pion potentials and a Delta-hole theoretical model for the P-wave component. The inclusion of the pion potential, in particular the S-wave component, is crucial for a simultaneously accurate description of the two observables. The constraints extracted for $L$ are on average close to 90 MeV, but carry large uncertainties and are rather strongly dependent on the choice of the S-wave pion optical potential (see right panel of Fig.~\ref{fig:piratio_fopi_auau0400}). Nevertheless, a soft symmetry energy, $L <$ 50 MeV, could be excluded from the comparison model-experiment irrespective of the chosen pion potential.


\subsection{Results of studies based on elliptic flow}
\label{sec:23}

\begin{figure}
\centering
\includegraphics[width=0.6\textwidth]{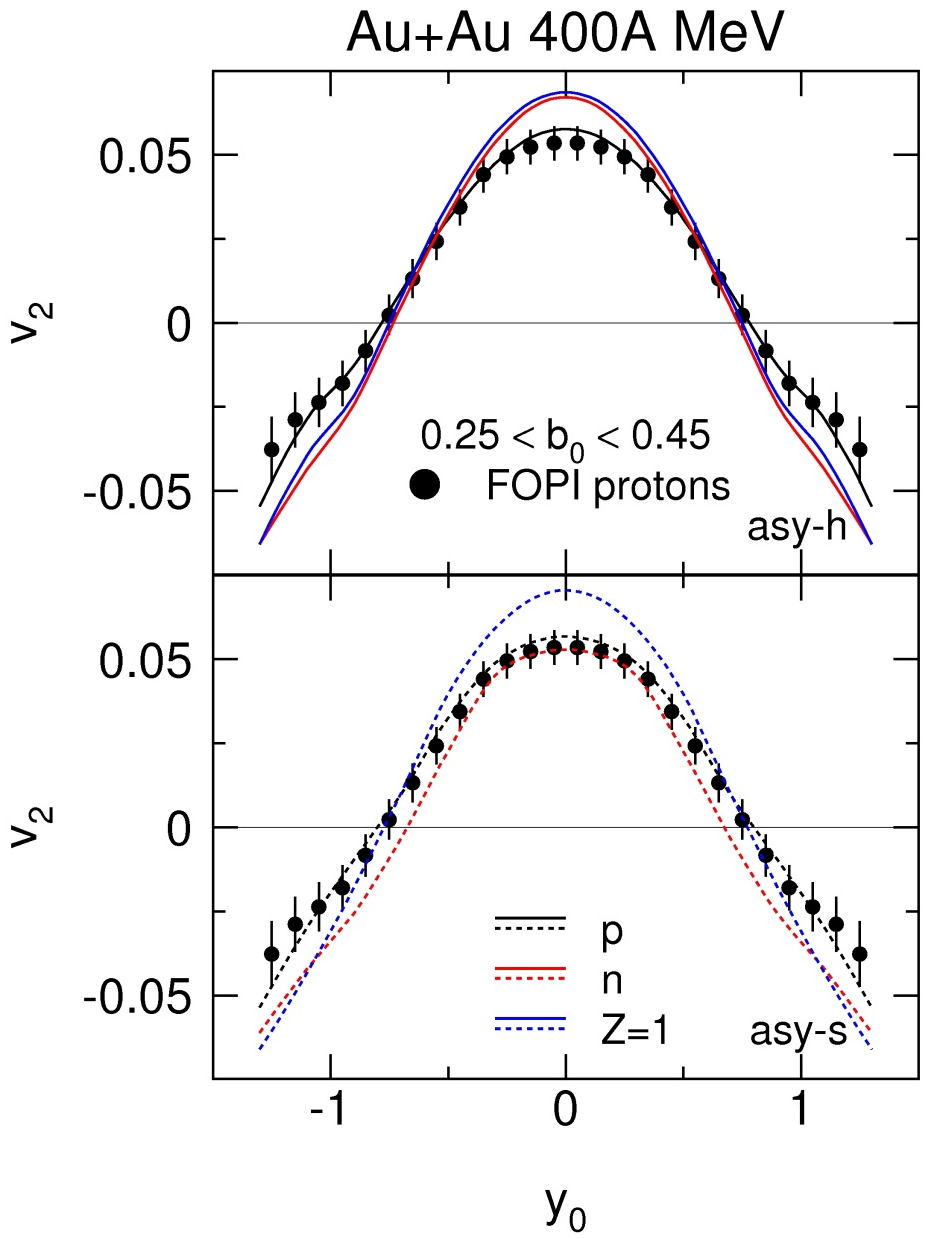}
\caption{(adapted from Ref. \cite{Russotto:2011hq}) Predictions for v\textsubscript{2} of neutrons, protons and hydrogen, as a function of scaled rapidity y\textsubscript{0} for Au+Au collisions at 400 MeV/nucleon and reduced impact parameter range $0.25<b_{0}<0.45$, as obtained by using the UrQMD transport code and stiff (top panel) or soft (bottom panel) parameterizations for the density dependence of the symmetry energy. The v\textsubscript{2} of protons as measured by the FOPI collaboration is reported on both panels of the figure (black closed circles). See the text for more details.}  
\label{fig:v2-urqmd}
\end{figure}

The elliptic flow observable has proven to be a powerful tool for studying the high-density behaviour of the symmetric part of the EoS.
But also the relative elliptic flow of neutrons and protons was expected to play an important role in probing the asymmetric part of the EoS. The first evidence for that was found by studying Au+Au collisions at 400 MeV/nucleon simulated by using the version developed by Li et al. of the Ultra-relativistic Quantum Molecular Dynamic (UrQMD) \cite{Li:2005zza,Li:2005gfa,Li:2006ez}. In this transport code, the symmetry energy is accounted by using a power-law as a function of the reduced density, that is

\begin{equation}
E_{sym}(\rho)=12(\rho/\rho_0)^\frac{2}{3}+22(\rho/\rho_0)^\gamma
\label{urqmd-par}
\end{equation}

where the exponent $\gamma$ is used to study different stiffnesses of the potential part of $E_{sym}$. Calculations were performed by using values of $\gamma$ equal to 1.5 and 0.5 to mimic a stiff and a soft symmetry energy parametrization.  
Obtained results are presented in fig. \ref{fig:v2-urqmd}, showing v\textsubscript{2} of neutrons, protons and hydrogen, as a function of scaled rapidity y\textsubscript{0}, for semi-central collisions with reduced impact parameter $0.25<b_{0}=b/b_{max}<0.45$ and for the stiff (top solid lines) and soft (bottom dashed lines) cases. For comparison the v\textsubscript{2} of protons, as measured by the FOPI collaboration \cite{FOPI:2004bfz} is reported in both panels of the figure, showing that the model reproduces the experimental data quite satisfactorily.\\
What emerges in fig. \ref{fig:v2-urqmd} is that, with a stiff $E_{sym}$, the absolute magnitude of the neutron v\textsubscript{2} at mid-rapidity is greater than for proton, and vice-versa in the soft case. Thus, a characteristic inversion of neutron/proton v\textsubscript{2} at mid-rapidity is evidenced. This effect can be qualitatively explained by the following: particles emitted at mid-rapidity and at a direction perpendicular to the reaction plane originate mainly from the expansion of the hot and dense region formed in the fireball. For neutron rich matter above saturation density, a stiffer symmetry energy generates a stronger repulsion/attraction for neutron/proton, respectively, than the softer one. Thus the neutron elliptic flow is larger in the stiffer case, reflecting the repulsion experienced by the particles in the dense region. In the case of protons the difference between stiff and soft case is smaller, compared to the neutrons, since the Coulomb and symmetry forces tend to cancel each other.\\ 
When using this finding to constrain the symmetry energy, a powerful choice is to use the ratio of neutron over proton elliptic flows rather than the the neutron elliptic flow alone. In fact, the neutron and proton elliptic flows depend on additional, generally not well fixed, ingredients of the transport models, such as the stiffness of the isoscalar EoS, the parameterization of in medium NN cross sections, the width of nucleonic wave packet, and even the initialization phase of colliding nuclei. These ingredients, since being not explicitly dependent on the nucleon type, tend to affect equally neutron and proton dynamics and can be, to some extent, cancelled when dealing with the ratio of neutron/proton v\textsubscript{2}. The impact of some of these ingredients on the ratio and the difference of neutron-proton elliptic flows were extensively studied by M.D. Cozma et al. by using the TuQMD transport code as reported in refs. \cite{Cozma:2011nr,Cozma:2013sja}. The use of the ratio of neutron/proton v\textsubscript{2} is advantageous also when dealing with experimental data, removing effects impacting measured flows independently of the particle species, such as the the correction to the underestimation of v\textsubscript{2} due to the reaction plane dispersion.
In fig. \ref{fig:v2-urqmd} it can be seen that the dependence of the Z=1 particles on the stiffness of the symmetry energy is quite similar to that of protons, indicating that also the relative behaviour of v\textsubscript{2} of neutrons and Z=1 particles can be used as a probe.\\
These findings, obtained using the UrQMD code, were confirmed using additional QMD transport models such as IQMD and TuQMD, and also by codes based on BUU approach, like the Stochastic Mean Field (SMF) of the Catania group \cite{Giordano:2010pv}, even if in that case the main attention was devoted to effects induced by the in-medium neutron and proton effective mass splitting.\\

\subsubsection{Results of FOPI-LAND experiment}
\label{sec:231}

Au+Au collisions at 400, 600 and 800 MeV/nucleon were measured at GSI in 1991 \cite{FOPI:1993wdf,FOPI:1994xpn} by coupling the Forward Wall of the FOPI apparatus \cite{GOBBI1993156} to the LAND detector \cite{LAND:1991ffr}. 
\begin{figure}
\includegraphics[width=0.75\textwidth]{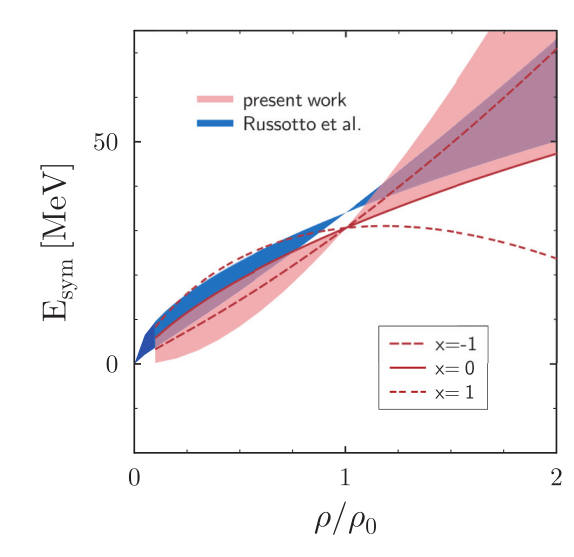}
\caption{(reprinted from Ref. \cite{Cozma:2013sja} under permission) Constraints on the density dependence of the symmetry energy obtained by comparing the FOPI-LAND experimental results with TuQMD  (Ref. \cite{Cozma:2013sja}, labeled as present work) and UrQMD (Russotto et al. \cite{Russotto:2011hq}) transport codes, together with the Gogny-inspired symmetry energy parameterization for three values of the stiffness parameter: x = -1 (stiff), x = 0, and x = 1 (soft). }
\label{fig:fopi-land-ur-tuqmd}       
\end{figure}
With the aim to study the symmetry energy at supra-saturation density, the Au+Au data taken at 400 MeV/nucleon were re-analyzed, extracting the direct and elliptic flows of neutrons and Z=1 particles for collisions with impact parameter b$<$7.5 fm. In spite of the low isotopic resolution for Z=1 particles, also the proton flows were extracted. The obtained flows were compared with UrQMD calculations, performed using the version of refs. \cite{Li:2005zza,Li:2005gfa,Li:2006ez}, for stiff and soft $E_{sym}$, as defined in the previous section, and for two forms of the elastic in medium nucleon-nucleon cross section labelled as FP1 and FP2; more details on this are available in Ref. \cite{Li:2006ez}. A good agreement between data and model predictions was found. In order to get an estimation for the $\gamma$ parameter, the transverse momentum ($p_{t}/A$) dependence  of the ratio of elliptic flows of neutrons and Z=1 particles ($v_{2}^{n}/v_{2}^{h}$) for 0.25$\leq y/y_{proj}\leq$ 0.75 was exploited. By using an interpolation between the stiff and soft prediction, the $\gamma$ parameter value was extracted for FP1 and FP2 cases. The $\gamma$ value was extracted also for the mid-peripheral interval of impact-parameters 5.5$\leq b \leq$7.5 fm and using also the ratio of neutron to proton elliptic flow, $v_{2}^{n}/v_{2}^{p}$. A serious drawback was related to the poor statistics of the available data set. A final estimation of $\gamma=0.9\pm0.4$ was obtained, including both statistical and systematic uncertainties, resulting in a value of $L=83\pm26$ MeV. This set of experimental data was also interpreted by using a new version of the UrQMD code, where Skyrme potential-energy-density functional was adopted for the mean-field part \cite{Wang2014-PhysRevC.89.044603}. This new version, with respect to the former one used in \cite{Russotto:2011hq}  where $E_{sym}$ was described by using a power-law form (eq. (\ref{urqmd-par}), adopted Skyrme interactions with similar values of the symmetric matter compressibility but different strengths of $E_{sym}$. The interpretation of FOPI-LAND experimental results with this new UrQMD version suggested values of $L$ = 89$\pm$23 MeV, very similar to the ones obtained using the former UrQMD version. \\
In a separate analysis, the experimental data were carefully compared with model predictions obtained by using the TuQMD transport code. Differing from UrQMD, in this case a Gogny-inspired momentum-dependent parameterizations of the symmetry energy was taken into account. In Refs.~\cite{Cozma:2011nr,Cozma:2013sja}, the dependence on various ingredients in the transport code of the neutron-proton elliptic flow observables (ratio and difference) was taken into account, including the compressibility modulus of isoscalar EoS, the implementation of the optical potential, the form of symmetry energy (power-law vs Gogny-inpired), the width of wave-packet in the QMD-like model. By comparing to the experimental data of the FOPI-LAND experiment, the result finally obtained was $L=122\pm57$ MeV. The comparison between this result and that obtained by using the UrQMD model is presented in fig. \ref{fig:fopi-land-ur-tuqmd}. It shows a satisfactory agreement. This is a relevant result given also the different approaches used in the two models to account for the symmetry energy, indicating the robustness and, to some extent, a signal of model independence of the proposed observable with respect to different theoretical approaches.

\subsubsection{Results of the ASY-EOS experiment}
\label{sec:232}
After the positive results obtained in the analysis of FOPI-LAND data, a new experiment was proposed at GSI, aiming to re-measure Au+Au collisions at 400 MeV/nucleon in order to improve the statistics with respect to the existing FOPI-LAND data set of \cite{FOPI:1993wdf,Russotto:2011hq}. This experiment, named  ASY-EOS experiment (S394),  was carried out at GSI on May 2011.\\  

\begin{figure}[ht!]
 \begin{minipage}[t]{0.5\linewidth}
 \begin{center}
\includegraphics[width=1\textwidth]{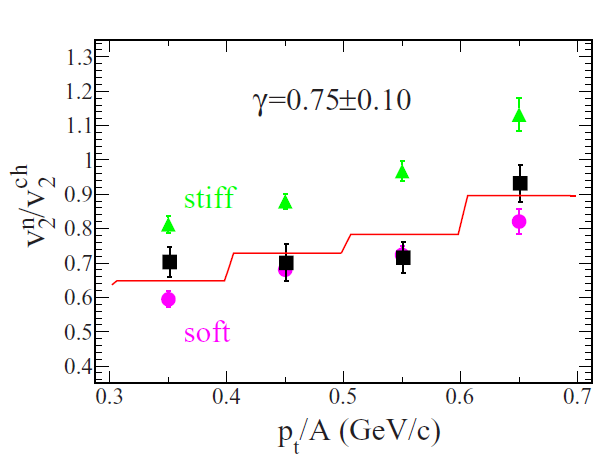}
 \end{center}
 \end{minipage}
 \hfill
 \begin{minipage}[t]{0.5\linewidth}
 \begin{center}
\includegraphics[width=1\textwidth]{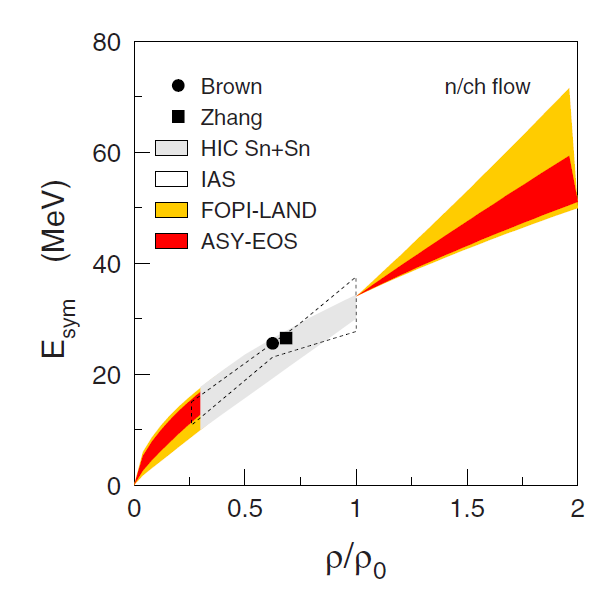} 
 \end{center}
 \end{minipage}

\caption{(reprinted from Ref. \cite{Russotto:2016ucm} under permission) Left panel: Elliptic flow ratio of neutrons over charged particles measured in the same acceptance range for semi-central ($b < 7.5$~fm) Au+Au collisions at 400~MeV/nucleon as a function of transverse momentum, $p_t/A$. The black squares represent the ASY-EOS experimental data. The green triangles and purple circles represent the UrQMD results employing a stiff ($\gamma=1.5$ in eq. (\ref{urqmd-par})  and soft ($\gamma=0.5$) density dependence of $E_{sym}$. The red solid line is the result of a linear interpolation between the predictions leading to the indicated $\gamma=0.75\pm0.10$ in Eq. \ref{urqmd-par}. Right panel: constraints deduced for the density dependence of the symmetry energy from the ASY-EOS (orange band) and FOPI-LAND (yellow band) experiments, compared also to some low-density results (see text). }
\label{fig:fig3}
\end{figure}

 The main results of the analysis of the ASY-EOS experiment is given in the left panel of fig.~\ref{fig:fig3}, showing as black squares the ratio of elliptic flows of neutrons to charged particles, $v_{2}^{n}/v_{2}^{ch}$, as a function of the transverse momentum per nucleon, for semi-central collisions of impact parameter b$\leq$7.5 fm. About this result, it is worth to say that a malfunctioning of the used ``new'' version of LAND front-end electronics produced relevant uncertainties on measured values of time of flight of each detected particle, that were solved only at probabilistic level; the result presented in fig.~\ref{fig:fig3} was obtained for a given value of the free parameter used to correct the timing problem of the LAND data (see \cite{Russotto:2016ucm} for more details). In addition, only a poor measurement of deposited energy was possible, leading to an insufficient discrimination of particles based on their charge Z. This means that only a separation between neutrons and charged particles was possible. This was a major drawback, since the hydrogen flow could not be accessed and, hence, methods to estimate the proton flow could not be implemented.

In the following we discuss the interpretation of the ASY-EOS data, using three different transport models.

\paragraph{UrQMD results}
\label{sec:2321}

In order to extract constraints on the density dependence of the symmetry energy from the ASY-EOS experiment, the UrQMD transport model was again used to describe the experimentally measured $v_{2}^{n}/v_{2}^{ch}$. Although the new UrQMD version of Ref. \cite{Wang2014-PhysRevC.89.044603} was made available at the time of the ASY-EOS data interpretations, in the main analysis of  Ref. \cite{Russotto:2016ucm} the older version of Refs. \cite{Li:2005zza,Li:2005gfa,Li:2006ez}, previously employed in the analysis of the FOPI-LAND data \cite{Russotto:2011hq}, was chosen in order to produce a result comparable to the former one. Nevertheless, as said before in Sect.~\ref{sec:231}, the interpretation of the FOPI-LAND data using the newest UrQMD version of Ref.~\cite{Wang2014-PhysRevC.89.044603} suggested a very similar value for $L$. Also in the case of the  ASY-EOS data analysis, calculations were performed for a stiff and soft choice of the potential part of $E_{sym}$, adopting the same values (1.5 and 0.5, respectively, for the $\gamma$ parameter in Eq.~\ref{urqmd-par}) used in the analyses presented in the previous sections. 
Simulations were filtered using a software replica of the experimental apparatus and compared to the experimental results. The results for the stiff and soft choice of $E_{sym}$ are plotted, against experimental values, in the left panel of Fig.~\ref{fig:fig3}. There, the exponent {$\gamma = 0.75 \pm 0.10$} was determined by fitting the measured flow ratios with a linear interpolation between the soft and stiff model predictions. Since that result was obtained for a given value of the free parameter chosen to correct the LAND timing problem, the $\gamma$ extrapolation was repeated for all the realistic values of that parameter, resulting in a $\gamma$ ranging in 0.75$\pm$0.15. The range of realistic values of that parameter was chosen by comparing the experimentally measured direct and elliptic flows with ones measured by the FOPI experiment. In addition $v_{2}^{n}/v_{2}^{ch}$ integrated over a time-of-flight range was also extracted from the  experimental data and compared to the UrQMD calculations, resulting in a $\gamma$ in the interval 0.77$\pm$0.17. The time-of-flight range was chosen to be large enough to overcome the LAND timing problems, and the $\gamma$ interval was obtained by taking into account also the variation of the upper limit of the integration window. In order to check the impact of using the flow of all the charged particle, instead of using only the Z=1 particle as done in the analysis of FOPI-LAND data, the FOPI-LAND data analysis was repeated with $v_{2}^{n}/v_{2}^{ch}$. The obtained results were, in spite of the low statistical accuracy, in agreement with the ones given by $v_{2}^{n}/v_{2}^{h}$, validating in such a way the 'forced' use of $v_{2}^{n}/v_{2}^{ch}$.\\
Before obtaining a final result, also the correction needed to take into account the effects of charge-changing processes, nuclear or instrumental, was  evaluated. This leads to misidentifications of neutrons as charged particles because of neutron-induced reactions in veto scintillators and of charged particles as neutrons, because of a missing veto signal.\\After taking into account all the corrections and systematic uncertainties, the value finally obtained was $\gamma = 0.72 \pm 0.19$, corresponding to a slope parameter $L = 72 \pm 13$ MeV. The corresponding density dependence of the symmetry energy is shown as a red area in the right panel of Fig.~\ref{fig:fig3}. It confirmed the FOPI-LAND (yellow area) result and represented an improvement of the accuracy by a factor of two. For comparison also some low density results are shown there, relative to the study of isospin diffusion by M.B. Tsang et al. \cite{PhysRevLett.102.122701}, Isobaric Analogue States of P. Danielewicz et al. \cite{Danielewicz:2013upa}, doubly magic nuclei of A. Brown \cite{Brown:2000pd} and isotope binding energy difference and neutron skin thickness of Z. Zhang et al. \cite{Zhang:2013wna}.\\ The obtained slope parameter, corresponding to a symmetry pressure $p_{0} = 3.8 \pm 0.7$ MeV fm$^{-3}$, was used to estimate the pressure in neutron star matter at saturation density \cite{Russotto:2016ucm}. Assuming a proton fraction of about 5$\%$ and adding the contribution of the degenerate electrons, the obtained value was $3.4 \pm 0.6$ MeV fm$^{-3}$.\\It has to be emphasized, however, that the UrQMD analysis of the ASY-EOS flow ratios relies on two assumptions. The expression for $E_{sym}$ (Eq. \ref{equ02}) assumes $E_{sym}(\rho_{0}) = 34$ MeV, leading to the sharp cross over of the error bands visible in Fig.~\ref{fig:fig3}. It does not reflect the present uncertainty of approximately 3 MeV of our knowledge of the symmetry energy at saturation \cite{Li:2013ola,Oertel:2016bki}. Assuming $E_{sym}(\rho_{0}) = 31$ MeV in the analysis, the value of the slope parameter is lowered to $L = 63 \pm 11$ MeV as reported in Ref. \cite{Russotto:2016ucm}. The second assumption is that of the functional form of a power law (Eq. \ref{equ02}) that, with the present results, is equivalent to assuming -70 MeV to -40 MeV for $K_{sym}$, an interval that result to be a very tight limitation with respect to the still poor knowledge of the curvature term. Thus, the final result of UrQMD interpretation of the ASY-EOS data was not totally free of model dependency biases. \\

\paragraph{TuQMD results}
\label{sec:2322}

The range of densities probed by elliptic flow ratio type of observables has been studied using the TuQMD transport model~\cite{Russotto:2016ucm}. Each of the three observables measured by the FOPI-LAND and ASY-EOS collaborations are sensitive to density values from very low up to twice saturation density. Nevertheless, it has been shown that for the neutron-to-proton elliptic flow ratio (npEFR) and the neutron-to-hydrogen (nhEFR), or similarly neutron-to charged particles (nchEFR), the maximum sensitivity lies close to 1.4-1.5$\rho_0$ and 1.0-1.1$\rho_0$ respectively. It would thus be possible to extract constraints for both the slope $L$ and curvature $K_{sym}$ parameters of the density dependence of symmetry energy by comparing transport model predictions to experimental data for npEFR and nhEFR, or alternatively npEFR and nchEFR. The results of such a study have been reported in Ref.~\cite{MDCozEPJA18}. A brief summary of the relevant model ingredients and conclusions of that investigation are presented in the following.

The TuQMD model uses a parameterization of the EoS inspired by the Gogny momentum dependent effective interaction. To allow independent variations of $L$ and $K_{sym}$ the original parameterization~\cite{Das:2002fr} has been slightly modified by the introduction of an additional density dependent but momentum independent term, proportional to the coupling parameter $D$ (dubbed the MDI2 interaction). The potential part of the EoS is given by
\begin{eqnarray}
\label{eq:eosmdi2}
\frac{E}{N}(\rho,\beta)&=&A_u(x,y)\frac{\rho(1-\beta^2)}{4\rho_0}+A_l(x,y)\frac{\rho(1+\beta^2)}{4\rho_0} \\
&&+\frac{B}{\sigma+1}\frac{\rho^{\sigma}}{\rho_0^\sigma}\,(1-x\beta^2)+
\frac{D}{3}\frac{\rho^2}{\rho_0^2}\,(1-y\beta^2) \nonumber\\
&&+\frac{1}{\rho\rho_0}\sum_{\tau,\tau'} C_{\tau \tau'}\!\!\int\!\!\int d^{\!\:3} \vec{p}\,d^{\!\:3} \vec{p}\!\;'
\frac{f_\tau(\vec{r},\vec{p}) f_{\tau'}(\vec{r},\vec{p}\!\;')}{1+(\vec{p}-\vec{p}\!\;')^2/\Lambda^2}. \nonumber
\end{eqnarray}
In the above expression $\rho$, $\beta$, $p$ and $\tau$ denote density, isospin asymmetry, momentum and isospin projection respectively. The single particle distribution function $f_\tau$ reduces to $(2/h^3)\Theta(p_F^\tau-p)$ for cold nuclear matter. The values of the 11 free parameters in Eq.~(\ref{eq:eosmdi2}) are related to properties of the EoS such as its density dependence of both symmetric and asymmetric nuclear matter, effective masses and others, see Ref.~\cite{MDCozEPJA18} for the full list. In particular, the magnitude of the symmetry energy is fixed at a sub-saturation value for density, where its most accurately known value has been extracted from experimental data for static properties of nuclei~\cite{Brown:2013mga,Zhang:2013wna}, $S(\rho=0.10\,\rm{fm}^{-3})$=25.5 MeV.

The chosen transport model considers only proton and neutron degrees of freedom. To account for cluster degrees of freedom, the final state spectra of HICs are determined using a minimum spanning tree (MST) coalescence algorithm. The $r$- and $p$-space coalescence parameters have been adjusted to optimize the description of $n$, $p$ and light cluster multiplicities in central Au+Au collisions for several impact energies for which FOPI experimental data were available~\cite{FOPI:2010xrt}, while constraining their values to intervals compatible with the finite range of the nucleon-nucleon interaction. As a result $n$ and $p$ multiplicities are overestimated, while for $^3$H, $^3$He and $^4$He the opposite holds true. Consequently, theoretical predictions for nhEFR and nchEFR are biased. Corrected values for elliptic flows of hydrogen and charged particles can be defined by replacing theoretical multiplicities by experimental ones in their expressions,
\begin{eqnarray}
\label{eq:v2corr}
 \tilde v_2^{ch}&=&\frac{M_p^{exp}\,v_2^p+\sum_{Z_i \ge 1,N_i \ge 1} M_{Z_i,N_i}^{exp}\,v_2^{Z_i,N_i}}
 {M_p^{exp}+\sum_{Z_i \ge 1,N_i \ge 1} M_{Z_i,N_i}^{exp}}\,,
\end{eqnarray}
and similarly for $\tilde v_2^{H}$. Corrections factors for elliptic flow of hydrogen and charged particles can be determined then as $f_{corr}^H=\tilde v_2^H/ v_2^H$ and $f_{corr}^{ch}=\tilde v_2^{ch}/ v_2^{ch}$ and their values can be estimated by using Eq.~\ref{eq:v2corr} and the observation that elliptic flows of individual particle species ($p$,$d$,$^3$H, etc.) are well described by the model. One arrives at the following conservative ranges for them: $f_{corr}^H$=1.075$\pm$0.05 and $f_{corr}^{ch}$=1.10$\pm$0.05. More robust approaches would make use of coalescence invariant elliptic flows~\cite{Famiano:2006rb} or employ transport models that account explicitly for light cluster degrees of freedom~\cite{Danielewicz:1991dh,Ikeno:2016xpr}.

\begin{figure}[t]
\begin{center}
\includegraphics[width=0.65\textwidth]{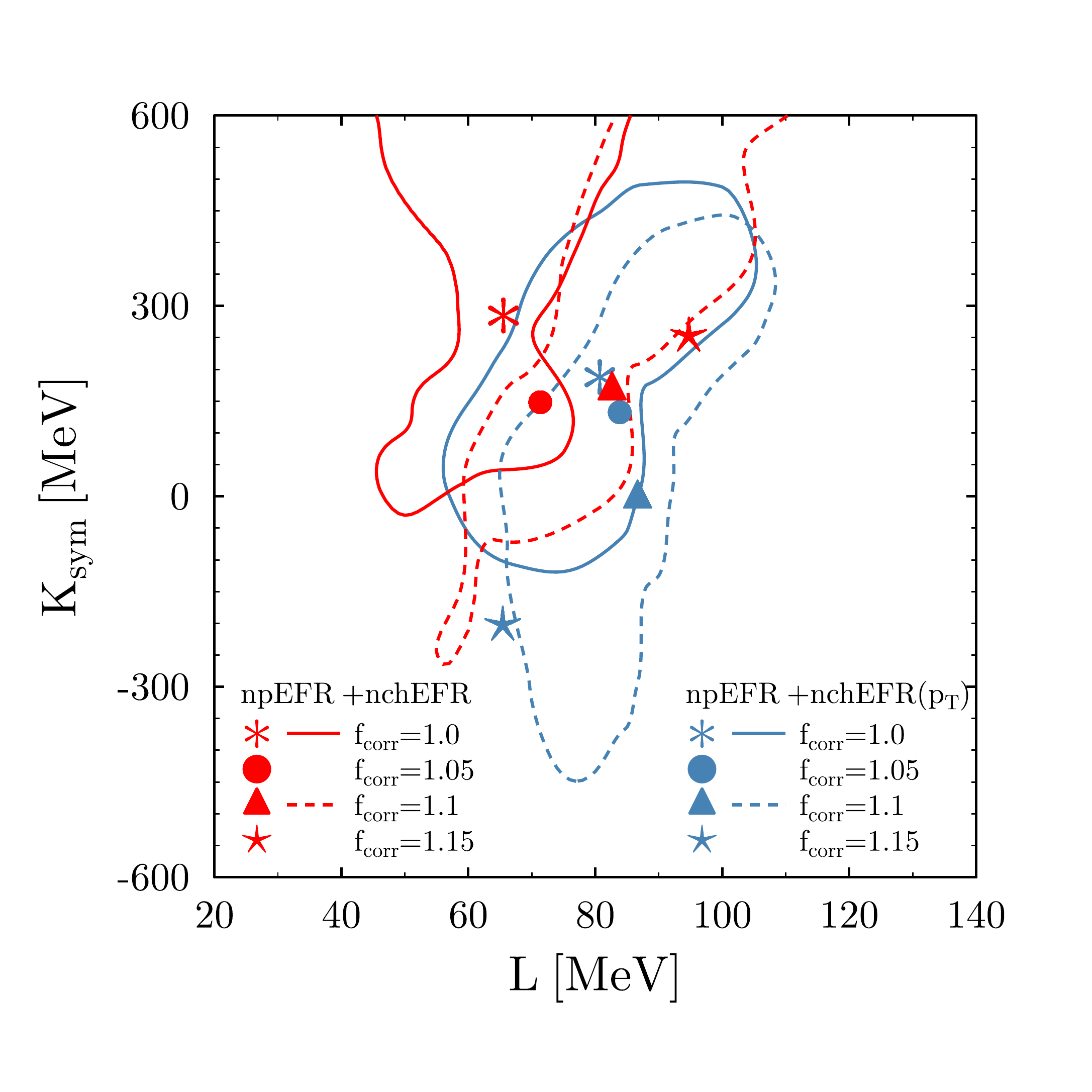}
\caption{(reprinted from Ref. \cite{MDCozEPJA18} under permission) Dependence of the extracted values for the $(L, K_{sym})$ pair on the combination of elliptic flow ratios (EFR) of different species (n=neutron, p=protons, ch=charged particles), npEFR+nchEFR and npEFR+nchEFR($p_T$). For each case the impact of the correction parameter $f_{corr}^{ch}$, used to adjust elliptic flow of charged particles, is shown. Contour curves correspond to 1 sigma confidence levels. }
\label{fig:constraint_tuqmd}
\end{center}
\end{figure}

The sensitivity of npEFR and nhEFR (similarly for nchEFR) to variations of $K_{sym}$ at constant $L$ exhibits slopes of opposite signs for the two observables (see Fig. 10 in Ref.~\cite{MDCozEPJA18}), which was shown to reflect different densities that are being probed on average. A simultaneous description of experimental data allows thus the extraction of both $L$ and $K_{sym}$. This is shown in Fig.~\ref{fig:constraint_tuqmd} using the FOPI-LAND data for npEFR and the ASY-EOS result for nchEFR. Two cases, making use of either integrated or $p_T$ dependent nchEFR, have been considered. The extracted constraint depends rather strongly on the value of the correction factor $f_{corr}^{ch}$ and chosen set of observables. Close values, well within the 1 $\sigma$ confidence limit, are nevertheless extracted for $L$ and $K_{sym}$ for values of the correction factor in the range $1.05 \leq f_{corr}^{ch} \leq 1.10$. The central values are  determined by averaging results for the two combinations of observables and the most probable value for the correction parameter, $f_{corr}$=1.10 and read: $L$=85 MeV and $K_{sym}$=96 MeV. The impact of the uncertainty in the value of $f_{corr}$ is included in the final result as a systematic error.

A systematic study of the dependence of the extracted constraint for $L$ and $K_{sym}$ on several uncertain ingredients of the model has also been performed, the result being quoted as a theoretical error of the extracted constraint. To that end the compressibility modulus has been varied in the range 210-285 MeV, a medium modification correction has been applied to elastic nucleon-nucleon cross-sections, the neutron-proton effective mass difference has been varied in the interval $\Delta m_{np}^*$=0.0-0.56$\,\beta$ and several Pauli blocking algorithms have been employed. Most notably, the stiffness of the symmetry energy was seen to be correlated to the compressibility modulus of symmetric matter, a softer $K_0$ leading to a stiffer $L$. The in-medium modification of elastic cross-sections leads to a softer symmetry energy. The others two model ingredients have a smaller impact.

The central result of the study, the constraint for the slope $L$ and curvature $K_{sym}$ parameters, reads
\begin{eqnarray}
\label{lkmdi2}
L&=&85\pm\phantom{3}22(\mathrm{exp})\pm\phantom{2}20(\mathrm{th})\pm\phantom{1}12(\mathrm{sys})\,\,\mathrm{MeV} \\
K_{sym}&=&96\pm315(\mathrm{exp})\pm170(\mathrm{th})\pm166(\mathrm{sys})\,\,\mathrm{MeV}\,. \nonumber 
\end{eqnarray}
The indicated uncertainties are of experimental, theoretical (model dependence) and systematical (underprediction of cluster-to-proton multiplicity ratios) origin. The theoretical uncertainty has been determined by adding in quadrature all model dependencies. The quoted value for the systematical error for $K_{sym}$ has been estimated as the half difference between the maximum and minimum average values for $K_{sym}$ when the $f_{corr}$ parameter is varied in the range [1.05,1.15]. In a similar fashion, a constraint solely for $L$ can be extracted from FOPI-LAND npEFR by omitting the term proportional to the coupling $D$ (cMDI2 interaction) in Eq.~\ref{eq:eosmdi2},
\begin{eqnarray}
\label{lkcmdi2_v2}
L&=& 84\pm30(\mathrm{exp})\pm 19(\mathrm{theor})\,\,\mathrm{MeV}\,
\end{eqnarray}
which is in perfect agreement with the full result in Eq.~\ref{lkmdi2} and has the advantage of not being affected by uncertainties induced by the underestimation of light cluster multiplicities.

Up to 1.5$\rho_0$ the density dependence of the SE is dominated by the slope term. Consequently the constraints extracted using cMDI2 and MDI2 potentials are of comparable accuracy in this region. However, the allowed range for $K_{sym}$ extracted from existing elliptic flow data is clearly not precise enough for the purpose of extrapolating the SE at densities above 1.5$\rho_0$. The largest contribution to the determined uncertainty originates from experimental data, particularly the FOPI-LAND npEFR. An improvement of the experimental relative accuracy for this observable, comparable to that achieved by the ASY-EOS collaboration for nchEFR, would lead to a decrease of the experimental uncertainty of $K_{sym}$ to an estimated value of 200 MeV. Further improvements may be possible if HICs are studied experimentally at bombarding energies around 250 MeV/nucleon, see Ref.~\cite{MDCozEPJA18} for details.

\paragraph{Latest results from QMD models}
\label{sec:2323}
\begin{figure}
\centering
\includegraphics[width=0.7\textwidth]{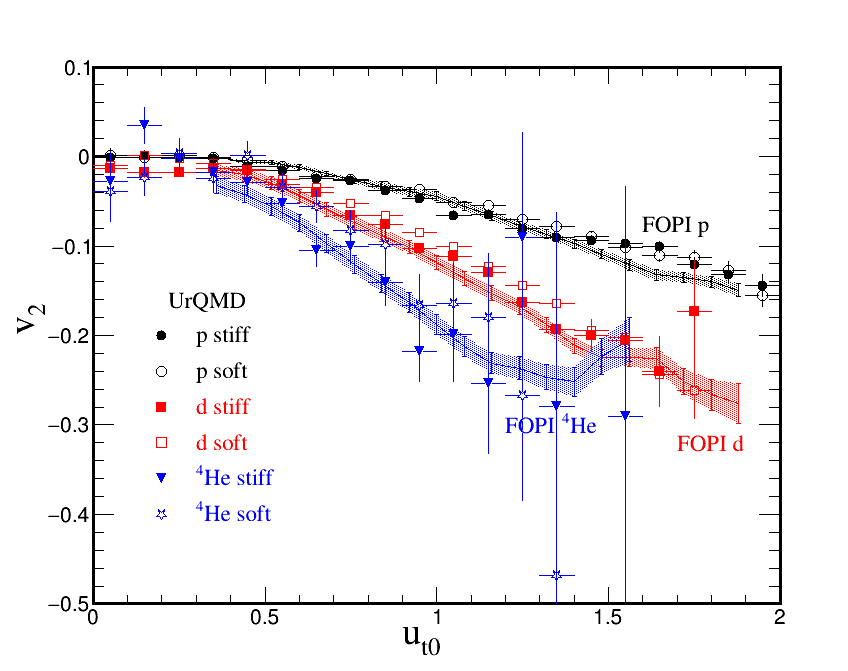}
\caption{Elliptic flow, $v_{2}$, of protons, deuterons and alpha particles, as a function of the reduced transverse momentum, $u_{t0}$, as given by the new version of UrQMD transport code for stiff (full symbols) and soft (empty symbols) $E_{sym}$ in the case of $E_{sym,0}=32 MeV$, compared to the experimental values (shaded areas) as measured by FOPI for semi-central collisions of $0.25<b_{0}=b/b_{max}<0.45$ and published in Ref. \cite{FOPI:2011aa}}
\label{fig-new-urqmd-v2}
\end{figure}
 
Here we present some new results obtained by analysing the ASY-EOS experimental data with a new version of the UrQMD model. This new version features several differences, with respect to the older ones used in \cite{Russotto:2011hq,Wang2014-PhysRevC.89.044603,Russotto:2016ucm}, regarding initialization of nuclei, description of mean-field potential, implementation of the Pauli blocking, in-medium cross section for both elastic and inelastic channels; more details can be found in \cite{Liu:2020jbg}. Each of these ingredients, taken individually, may only have a small impact on observables of interest, which however may sum up to a non-negligible effect. Simulations for the Au+Au collisions at 400 MeV/nucleon have been performed with this new version, using stiff and soft parameterizations of $E_{sym}$ as previously done, and taking into account also different values for the $E_{sym,0}$ term of eq. \ref{equ02}. A minimum-spanning tree algorithm, similar to the one used in previous works, has been used to build clusters from the final phase-space distribution of nucleons, and get, also, free neutrons and protons. The parameters used in the clusterization algorithm have been tuned by comparing the theoretical charge distribution to the one measured and published by FOPI collaboration, as done in previous works (see fig. 2. of Ref. \cite{Russotto:2011hq}). The best agreement is reached by assuming two nucleons are bound into a cluster when the relative distance, $dr$, is less than 3.8 fm and the relative momentum, $dp$, is less than 0.275 GeV/c. This $dr$ is larger, with respect to the 3 fm used in calculations with previous UrQMD versions of \cite{Russotto:2011hq,Russotto:2016ucm}, and reflects the differences in in-medium cross section, producing different distributions in the transverse momentum space. The results given by this new UrQMD version are presented in fig. \ref{fig-new-urqmd-v2}, showing the elliptic flows of proton, deuteron and $^{4}$He, as a function of reduced transverse momentum, $u_{t0}=u_{t}/u_{p}$ with $u_{t}=\beta_{t}\gamma$ and $u_{p}=\beta_{p}\gamma_{p}$, the index p
referring to the incident projectile in the c.m., as given by the simulations, for both stiff and soft cases and for $E_{sym,0}=32 MeV$, for semi-central collisions of $0.25<b_{0}<0.45$. The simulation results are compared to the experimental values measured by FOPI , as published in fig. 26 of Ref. \cite{FOPI:2011aa}. We can see there the capability of the model to reproduce the trend of the experimental data.\\What we present in the following has been obtained by using a different, with respect to the what done in \cite{Russotto:2016ucm}, way of extracting the elliptic flow of charged particles, $v_{2}^{ch}$ from the simulated data. In fact, since the calculations do not reproduce exactly the multiplicities of the different isotopes, the yield of each isotope has been renormalized in order to agree with experimental multiplicities measured by FOPI for central collisions of $b_{0}<0.15$, given in appendix A of \cite{FOPI:2010xrt}, by using the method of eq. \ref{eq:v2corr} . The assumptions that the difference between FOPI experimental data and calculation observed for central collision holds for the whole impact parameter range of interest in the flow analysis (b$<$7.5 fm), and that the model well reproduces the neutron flows and multiplicity, has been made. In \cite{Russotto:2016ucm} the renormalization was applied only to the yield of light clusters of Z$\geq$2 (see sect. V of that paper for more details). The procedure of yield renormalization for all charged particles, as here done, tends to shrink the difference between $v2_{n}/v2_{ch}$ predictions of the stiff and soft cases, leading to large relative errors on extrapolation of $\gamma$ of eq. \ref{urqmd-par}. 
\begin{figure}[t]
\centering
\includegraphics[width=0.6\textwidth]{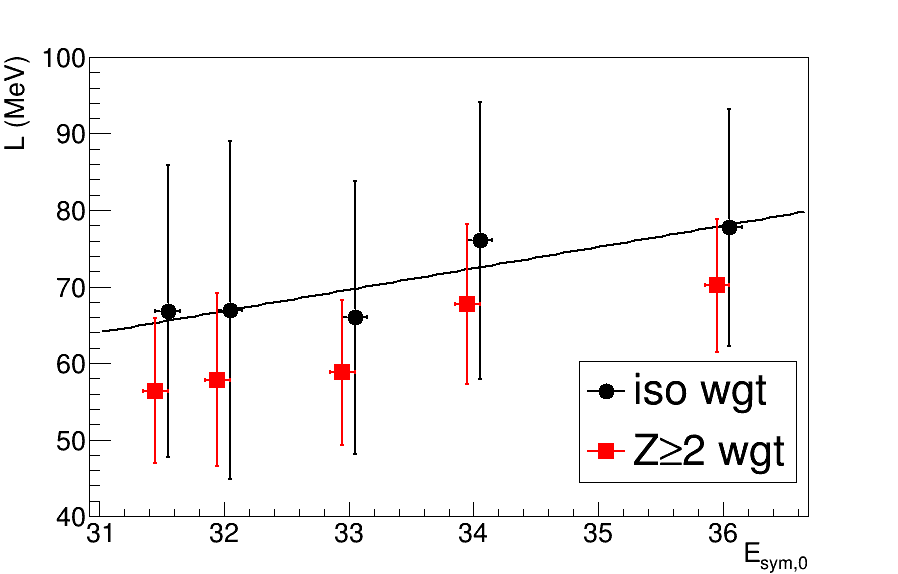}
\caption{Slope of the $E_{sym}$ around $\rho_{0}$, $L$, as a function of the value of the $E_{sym}$ at $\rho_{0}$ obtained by comparing the ASY-EOS experimental results for the $v2_{n}/v2_{ch}$ to the predictions of the new UrQMD version of \cite{Liu:2020jbg}. Black circles give results obtained by renormalizing the yield of each isotope according to the experimental multiplicities measured by FOPI in central collisions and given in Ref. \cite{FOPI:2010xrt}; the black line gives the result of a linear interpolation. Red squares have been obtained by re-normalizing only the yield of the light clusters of Z$\geq$2.}
\label{fig-new-urqmd-LS0}
\end{figure}
The results for $L$, as obtained by using the above described extrapolation method, are presented as black circles in fig. \ref{fig-new-urqmd-LS0}, and as a function of different values of $E_{sym,0}$. The black line represents the result of a linear fit, that is $L$ = -22.56+2.79$\times E_{sym,0}$~[MeV]. For comparison, results obtained by applying the yield re-normalization only to particles heavier than hydrogen, similar to what done previously in \cite{Russotto:2016ucm}, are also shown as red squares. The point at $E_{sym,0}$ = 34 MeV denoted by a red symbol differs slightly, by about 4 MeV, from the result obtained in the previous analysis of \cite{Russotto:2016ucm}, being a good proof of stability of the obtained results against both, changes in the transport code (differences between UrQMD version used in \cite{Russotto:2011hq,Russotto:2016ucm} and the new one of \cite{Liu:2020jbg}), and different analysis method (isotope-by-isotope yield renormalization). New results obtained using the isotope-by-isotope renormalization method show larger errors and stiffer results for the $E_{sym}$, being however in good agreement with the ones obtained re-normalizing only the $Z\geq2$ particles and, in general, with what obtained in the former interpretations of the ASY-EOS data described in the previous sections. For $E_{sym,0}$ = 34 MeV, the renormalization method leads to $L=76\pm18$, to be compared to $L = 72\pm13$ MeV of \cite{Russotto:2016ucm}.\\ 

\begin{figure}[t]
\centering
\includegraphics[width=0.6\textwidth]{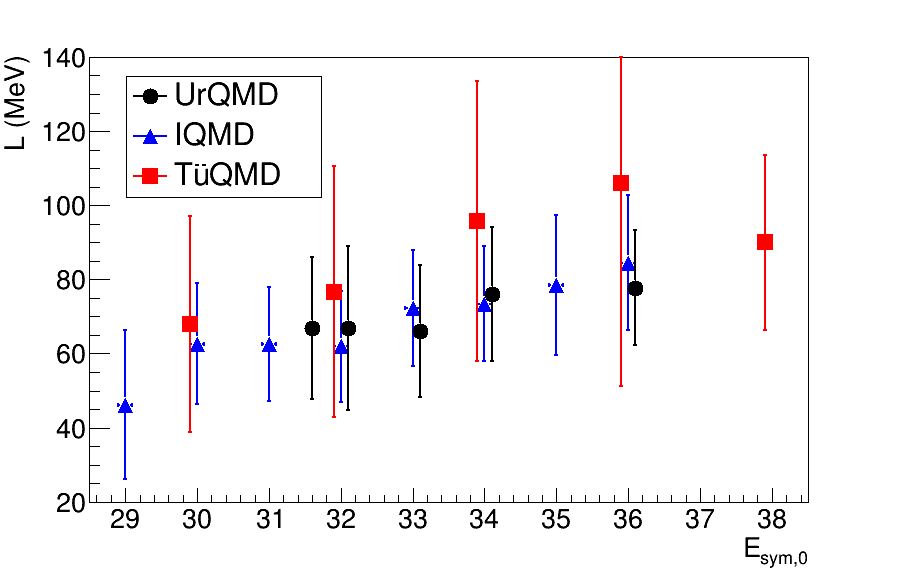}
\caption{Slope of the $E_{sym}$ around $\rho_{0}$, $L$, as a function of the value of the $E_{sym}$ at $\rho_{0}$ obtained by comparing the ASY-EOS experimental results for the $v2_{n}/v2_{ch}$ to the predictions of the new UrQMD version of \cite{Liu:2020jbg} and to the IQMD and TuQMD ones. This has been obtained, as in fig. \ref{fig-new-urqmd-LS0}, by renormalizing the yield of each isotope according to the experimental multiplicities measured by FOPI in central collisions. }
\label{fig-new-urqmd-iqmd-tuqmd-LS0}
\end{figure}
As a proof of further stability and robustness of these new results presented here, fig. \ref{fig-new-urqmd-iqmd-tuqmd-LS0} compare the UrQMD results of fig. \ref{fig-new-urqmd-LS0} to the ones obtained with the IQMD and TuQMD transport codes. In both cases (IQMD and TuQMD) the above mentioned "renormalization" has been applied. We can see evidences of agreement between results obtained with the three different transport codes. This is a further signal of the convergence of results, as previously seen in sect. \ref{sec:231}, when using the elliptic flow ratio at 400 MeV/nucleon and the prediction of different transport codes, featuring differences in both the basic ingredients (in-medium cross section for both elastic and inelastic channels, description of mean-field potential, procedures used to recognized the free nucleons and build clusters, etc.) and the technical implementation of physical features (e.g. implementation of the Pauli blocking, etc.). 
Further detailed comparisons will be pursued in the future, aiming to face up the model dependence of the proposed observable along the whole relevant range of incident beam energy for this kind of studies.

\subsection{Connection of ASY-EOS results with multi-messenger astrophysics results}
\label{sec:2324}

The last couple of years have witnessed the emergence of symmetry energy
studies at high density making use of novel results obtained through gravitational waves (GW) and neutron stars (NS)  X-rays satellite based observations. In, fact the knowledge of the dense nuclear matter EoS is particularly relevant for astrophysics \cite{ozel16a,dege18,latt21}, where the EoS completes the set of structure equations that allow to derive the mass-radius (M-R) relations for NS and allows modelization of the supernova
explosions. 

Since there is a one-to-one correspondence between the EoS and the M-R relation \cite{lind92}, astrophysics offers a complementary and competitive way of determining the EoS through  simultaneous measurements of masses and radii of NS. Provided a sufficient precision is reached, a measured M-R relation might allow to derive or pick the right EoS through the ``reverse engineering'' approach. The idea and current status is presented in Fig. \ref{figmr} which will be discussed in more detail later on.

A characteristic steep dependence of the mass on radius over a broad range of
masses for nucleonic EoSs 
implies that a precise measurement of the radius alone, assuming some typical or
canonical mass around 1.44 M$_{\odot}$ (a Chandrasekhar limit), can also be
sufficient to select the best matching EoS. Some EoSs are also ruled out by the limit imposed by the most massive NS measured so far.

Simultaneous measurement of the mass and radius is a very challenging process
involving many assumptions and uncertainties. The first candidates for a
simultaneous measurement of the mass and radius were the low-mass X-ray binary (LMXB) 
systems in globular clusters where X-ray emissions are powered by the
accretion process and the distance is relatively well known (see e.g.
\cite{LAT07,Gal08}).

The new generation of instruments such as NICER allows also for a radius measurement through X-ray pulse profile modeling for some millisecond pulsars (MSP). 
These highly spun-up (``recycled'') pulsars exhibit thermal emissions powered by rotation. In some cases they allow for independent accurate mass measurements.

Generally, measuring or inferring the radius of a NS consists in measuring the
thermal flux of photons (usually X-rays) emitted from the NS surface. From
that, assuming some emission pattern (e.g. a black body), atmospheric
composition, interstellar absorption, knowing the distance and knowing or
assuming the mass for the gravitational redshift correction, one can extract the radius.

Another source of the NS radii are the binary neutron star merger events
detected through the gravitational waves (GW) generated at the latest stage of the in-spiral (see e.g. \cite{bali21} for a recent review). The GW170817 event reported by the LIGO and Virgo Collaborations provided the first detection of GW from the coalescence of a neutron star binary system, constituting a new opportunity for probing the properties of neutron rich nuclear matter at extreme conditions found in the interior of those compact stars. The analysis of detected GW allowed to determine the tidal deformability, $\Lambda$, which is related to the strength of the quadrupole mass deformation of a star due to the stress caused by its companion's gravity and thus depends on the EoS.
The $\Lambda$ is highly sensitive to the neutron star radius, being proportional to the fifth power of the areal radius of the star. Additional information on the post merger evolution of the remnant can also be gained from the possible electromagnetic emissions (or their lack) at various wavelengths and from neutrino signals. These multi-messenger  observations open a new era in exploration of the high density EoS.

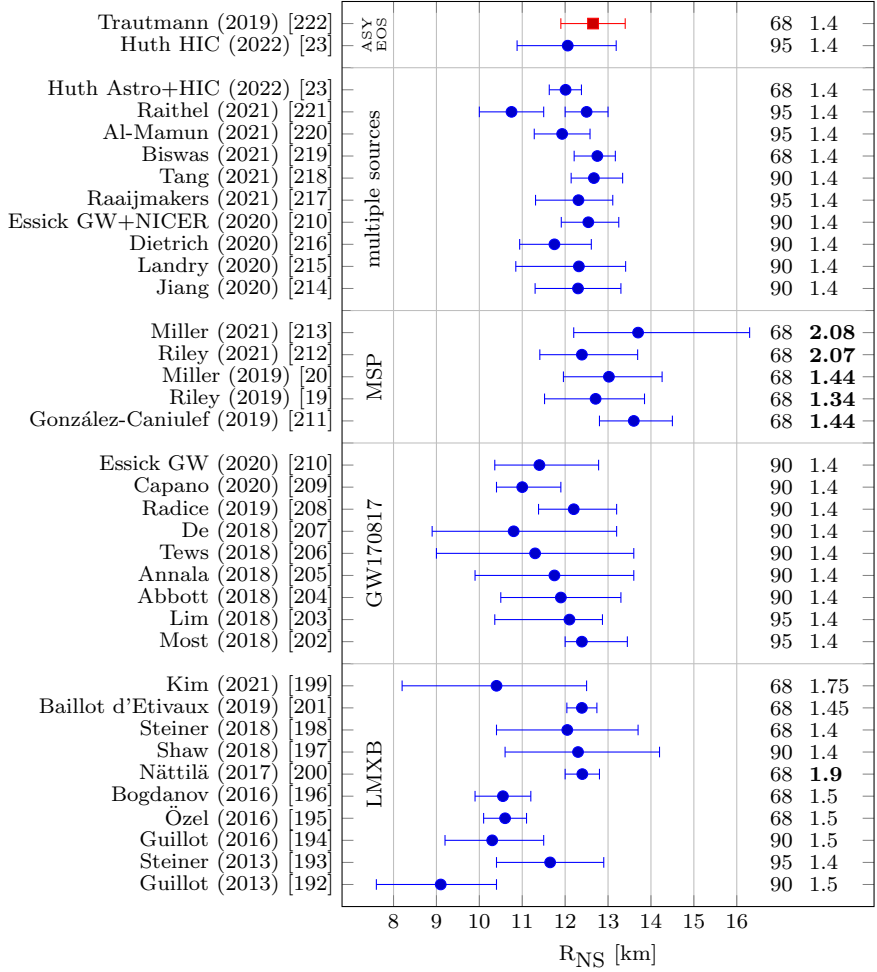
\begin{figure}[ht!]
  \centering
  \begin{tikzpicture}[scale=1],
    \centering
    \begin{axis}[
	height=12cm,
	width=7cm,  
	xmin=6.8,
	xmax=19.2,
	ymin=0,
	ymax=41,
	scale only axis,
	xmajorgrids,
	xlabel near ticks,
	xlabel={R$_{\mbox{NS}}$ [km]},
	yticklabels={
	Guillot (2013)\cite{gui13},
	Steiner (2013)\cite{Steiner:2012xt},
	Guillot (2016)\cite{gui16},
	\"{O}zel (2016)\cite{ozel16},
	Bogdanov (2016)\cite{bog16},
	N\"{a}ttil\"{a} (2017)\cite{nat17},
	Shaw (2018)\cite{shaw18},
	Steiner (2018)\cite{ste18},
	Baillot d'Etivaux (2019)\cite{bai19},
	Kim (2021)\cite{kim21},
	,
	Most (2018)\cite{mos18},
	Lim (2018)\cite{lim18},
	Abbott (2018)\cite{LIGOScientific:2018cki},
	Annala (2018)\cite{ann18},
	Tews (2018)\cite{tews18},
	De (2018)\cite{de18},
	Radice (2019)\cite{rad19},
	Capano (2020)\cite{cap20},
	Essick GW (2020)\cite{ess20},
	,
	Gonz\'{a}lez-Caniulef (2019)\cite{gon19},
	Riley (2019)\cite{Riley:2019yda},
	Miller (2019)\cite{Miller:2019cac},
	Riley (2021)\cite{ril21},
	Miller (2021)\cite{mil21},
	,
	Jiang (2020)\cite{jiang20},
	Landry (2020)\cite{lan20},
	Dietrich (2020)\cite{die20},
	Essick GW+NICER (2020)\cite{ess20},
	Raaijmakers (2021)\cite{raa21},
	Tang (2021)\cite{tan21},
	Biswas (2021)\cite{bis21},
	Al-Mamun (2021)\cite{alm21},
	Raithel (2021)\cite{rai21},
	Huth Astro+HIC (2022)\cite{Huth:2021bsp},
	,
	Huth HIC (2022)\cite{Huth:2021bsp},
	Trautmann (2019)\cite{Trautmann:2019gmh}
	},
	xtick={8,9,...,16},
	ytick={1,...,41},
	y tick label style={rotate=0,anchor=east}]
	\addplot+[blue,only marks][error bars/.cd,x dir=both, x explicit]
	  coordinates {
	    (9.1,1) -= (1.5,0) += (1.3,0) 
	    (11.65,2) +- (1.25,0) 
	    (10.3,3) -= (1.1,0) += (1.2,0)	  
	    (10.6,4) +- (0.5,0)	  
	    (10.55,5) +- (0.65,0)     
	    (12.4,6) +- (0.4,0)	  
	    (12.3,7) -= (1.7,0) += (1.9,0) 
	    (12.05,8) +- (1.65,0)	  
	    (12.39,9) +- (0.35,0)     
	    (10.4,10) -= (2.2,0) += (2.1,0)	  
	    (12.39,12)  -= (0.39,0) += (1.06,0)     
	    (12.10,13) -= (1.74,0) += (0.77,0)     
	    (11.9 ,14)  -= (1.4,0) += (1.4,0)	 
	    (11.75,15) -= (1.85,0) += (1.85,0)   
	    (11.3 ,16)  -= (2.3,0) += (2.3,0)	
	    (10.8 ,17)  -= (1.9,0) += (2.4,0)	 
	    (12.2 ,18)  -= (0.82,0) += (1.0,0)   
	    (11.0 ,19)  -= (0.6,0) += (0.9,0)   
	    (11.4 ,20)  -= (1.04,0) += (1.38,0) 

	    (13.6,22)  -= (0.8,0) += (0.9,0)  
	    (12.71,23) -= (1.19,0) += (1.14,0)     
	    (13.02,24) -= (1.06,0) += (1.24,0)     
	    (12.39,25) -= (0.98,0) += (1.30,0)     
	    (13.7,26) -= (1.5,0) += (2.6,0)     
	    (12.3 ,28) -= (1,0) += (1,0)       
	    (12.32,29) -= (1.47,0) += (1.09,0)    
	    (11.75,30) -= (0.81,0) += (0.86,0)    
	    (12.54,31) -= (0.63,0) += (0.71,0)    
	    (12.31,32) -= (1,0) += (0.8,0)    
	    (12.67,33) -= (0.53,0) += (0.67,0)    
	    (12.75,34) -= (0.54,0) += (0.42,0)    
	    (11.93,35) -= (0.65,0) += (0.65,0)   
	    (10.75,36) -= (0.75,0) += (0.75,0) 
	    (12.5 ,36) -= (0.5,0) += (0.5,0)	 
	    (12.01,37) -= (0.38 ,0) += (0.37,0)  
	    (12.06,39) -= (1.18,0) += (1.13,0)	 
	  };
	\addplot+[red,only marks][error bars/.cd,x dir=both, x explicit]
	  coordinates {
	    (12.65,40) -= (0.75,0) += (0.75,0)	 
	  };
	  
\draw [lightgray] (axis cs:7.0,11) -- (axis cs:19.0,11);
\draw [lightgray] (axis cs:7.0,21) -- (axis cs:19.0,21);
\draw [lightgray] (axis cs:7.0,27) -- (axis cs:19.0,27);
\draw [lightgray] (axis cs:7.0,38) -- (axis cs:19.0,38);

\node [rotate=90, anchor=center] at (axis cs:7.55,6   ) {LMXB}; 
\node [rotate=90, anchor=center] at (axis cs:7.55,16  ) {GW170817}; 
\node [rotate=90, anchor=center] at (axis cs:7.55,24.) {MSP}; 
\node [rotate=90, anchor=center] at (axis cs:7.55,32.5) {multiple sources}; 
\node [rotate=90, anchor=center,font=\tiny] at (axis cs:7.38,39.5) {ASY}; 
\node [rotate=90, anchor=center,font=\tiny] at (axis cs:7.72,39.5) {EOS}; 

        \node [rotate=0, anchor=west] at (axis cs:17.5, 1) {1.5}; 
        \node [rotate=0, anchor=east] at (axis cs:17.5, 1) {90}; 
        \node [rotate=0, anchor=west] at (axis cs:17.5, 2) {1.4};
        \node [rotate=0, anchor=east] at (axis cs:17.5, 2) {95};
        \node [rotate=0, anchor=west] at (axis cs:17.5, 3) {1.5};
        \node [rotate=0, anchor=east] at (axis cs:17.5, 3) {90};
        \node [rotate=0, anchor=west] at (axis cs:17.5, 4) {1.5};
        \node [rotate=0, anchor=east] at (axis cs:17.5, 4) {68};
        \node [rotate=0, anchor=west] at (axis cs:17.5, 5) {1.5};
        \node [rotate=0, anchor=east] at (axis cs:17.5, 5) {68};
        \node [rotate=0, anchor=west] at (axis cs:17.5, 6) {\bf{1.9}}; 
        \node [rotate=0, anchor=east] at (axis cs:17.5, 6) {68}; 
        \node [rotate=0, anchor=west] at (axis cs:17.5, 7) {1.4}; 
        \node [rotate=0, anchor=east] at (axis cs:17.5, 7) {90}; 
        \node [rotate=0, anchor=west] at (axis cs:17.5, 8) {1.4}; 
        \node [rotate=0, anchor=east] at (axis cs:17.5, 8) {68}; 
        \node [rotate=0, anchor=west] at (axis cs:17.5, 9) {1.45}; 
        \node [rotate=0, anchor=east] at (axis cs:17.5, 9) {68}; 
        \node [rotate=0, anchor=west] at (axis cs:17.5,10) {1.75}; 
        \node [rotate=0, anchor=east] at (axis cs:17.5,10) {68}; 
		
        \node [rotate=0, anchor=west] at (axis cs:17.5,12) {1.4}; 
        \node [rotate=0, anchor=east] at (axis cs:17.5,12) {95}; 
        \node [rotate=0, anchor=west] at (axis cs:17.5,13) {1.4}; 
        \node [rotate=0, anchor=east] at (axis cs:17.5,13) {95}; 

        \node [rotate=0, anchor=west] at (axis cs:17.5,14) {1.4}; 
        \node [rotate=0, anchor=east] at (axis cs:17.5,14) {90}; 
        \node [rotate=0, anchor=west] at (axis cs:17.5,15) {1.4}; 
        \node [rotate=0, anchor=east] at (axis cs:17.5,15) {90}; 
        \node [rotate=0, anchor=west] at (axis cs:17.5,16) {1.4}; 
        \node [rotate=0, anchor=east] at (axis cs:17.5,16) {90}; 
        \node [rotate=0, anchor=west] at (axis cs:17.5,17) {1.4}; 
        \node [rotate=0, anchor=east] at (axis cs:17.5,17) {90}; 
        \node [rotate=0, anchor=west] at (axis cs:17.5,18) {1.4}; 
        \node [rotate=0, anchor=east] at (axis cs:17.5,18) {90}; 
        \node [rotate=0, anchor=west] at (axis cs:17.5,19) {1.4}; 
        \node [rotate=0, anchor=east] at (axis cs:17.5,19) {90}; 
        \node [rotate=0, anchor=west] at (axis cs:17.5,20) {1.4}; 
        \node [rotate=0, anchor=east] at (axis cs:17.5,20) {90}; 

        \node [rotate=0, anchor=west] at (axis cs:17.5,22) {\bf{1.44}}; 
        \node [rotate=0, anchor=east] at (axis cs:17.5,22) {68}; 
        \node [rotate=0, anchor=west] at (axis cs:17.5,23) {\bf{1.34}};
        \node [rotate=0, anchor=east] at (axis cs:17.5,23) {68};
        \node [rotate=0, anchor=west] at (axis cs:17.5,24) {\bf{1.44}};
        \node [rotate=0, anchor=east] at (axis cs:17.5,24) {68};
        \node [rotate=0, anchor=west] at (axis cs:17.5,25) {\bf{2.07}};
        \node [rotate=0, anchor=east] at (axis cs:17.5,25) {68};
        \node [rotate=0, anchor=west] at (axis cs:17.5,26) {\bf{2.08}}; 
        \node [rotate=0, anchor=east] at (axis cs:17.5,26) {68}; 

        \node [rotate=0, anchor=west] at (axis cs:17.5,28) {1.4}; 
        \node [rotate=0, anchor=east] at (axis cs:17.5,28) {90}; 
        \node [rotate=0, anchor=west] at (axis cs:17.5,29) {1.4}; 
        \node [rotate=0, anchor=east] at (axis cs:17.5,29) {90}; 
        \node [rotate=0, anchor=west] at (axis cs:17.5,30) {1.4}; 
        \node [rotate=0, anchor=east] at (axis cs:17.5,30) {90}; 
        \node [rotate=0, anchor=west] at (axis cs:17.5,31) {1.4}; 
        \node [rotate=0, anchor=east] at (axis cs:17.5,31) {90}; 
        \node [rotate=0, anchor=west] at (axis cs:17.5,32) {1.4}; 
        \node [rotate=0, anchor=east] at (axis cs:17.5,32) {95}; 
        \node [rotate=0, anchor=west] at (axis cs:17.5,33) {1.4}; 
        \node [rotate=0, anchor=east] at (axis cs:17.5,33) {90}; 
        \node [rotate=0, anchor=west] at (axis cs:17.5,34) {1.4}; 
        \node [rotate=0, anchor=east] at (axis cs:17.5,34) {68}; 
        \node [rotate=0, anchor=west] at (axis cs:17.5,35) {1.4}; 
        \node [rotate=0, anchor=east] at (axis cs:17.5,35) {95}; 
        \node [rotate=0, anchor=west] at (axis cs:17.5,36) {1.4}; 
        \node [rotate=0, anchor=east] at (axis cs:17.5,36) {95}; 
        \node [rotate=0, anchor=west] at (axis cs:17.5,37) {1.4}; 
        \node [rotate=0, anchor=east] at (axis cs:17.5,37) {68}; 

        \node [rotate=0, anchor=west] at (axis cs:17.5,40) {1.4}; 
        \node [rotate=0, anchor=east] at (axis cs:17.5,40) {68}; 

        \node [rotate=0, anchor=west] at (axis cs:17.5,39) {1.4}; 
        \node [rotate=0, anchor=east] at (axis cs:17.5,39) {95}; 
	\end{axis}

\end{tikzpicture}%
      
\caption{Neutron star radii ($R_{NS}$) from different analyses of LMXB, from
GW170817 event, from millisecond pulsars, from mixed sources of information and
from the interpretation of the results of the heavy ion ASY-EOS experiment, from
bottom to top, giving the year of the publication and the reference. The numbers to the right of the symbols denote the confidence
levels (in \%) for the measured/inferred radii and the masses of the NS in solar
mass units. The masses emphasized in boldface are the inferred/measured ones.
The remaining masses are the assumed or averaged ones.}
\label{fig_rns} 
\end{figure}

\begin{figure}[ht]
\includegraphics[width=0.99\textwidth]{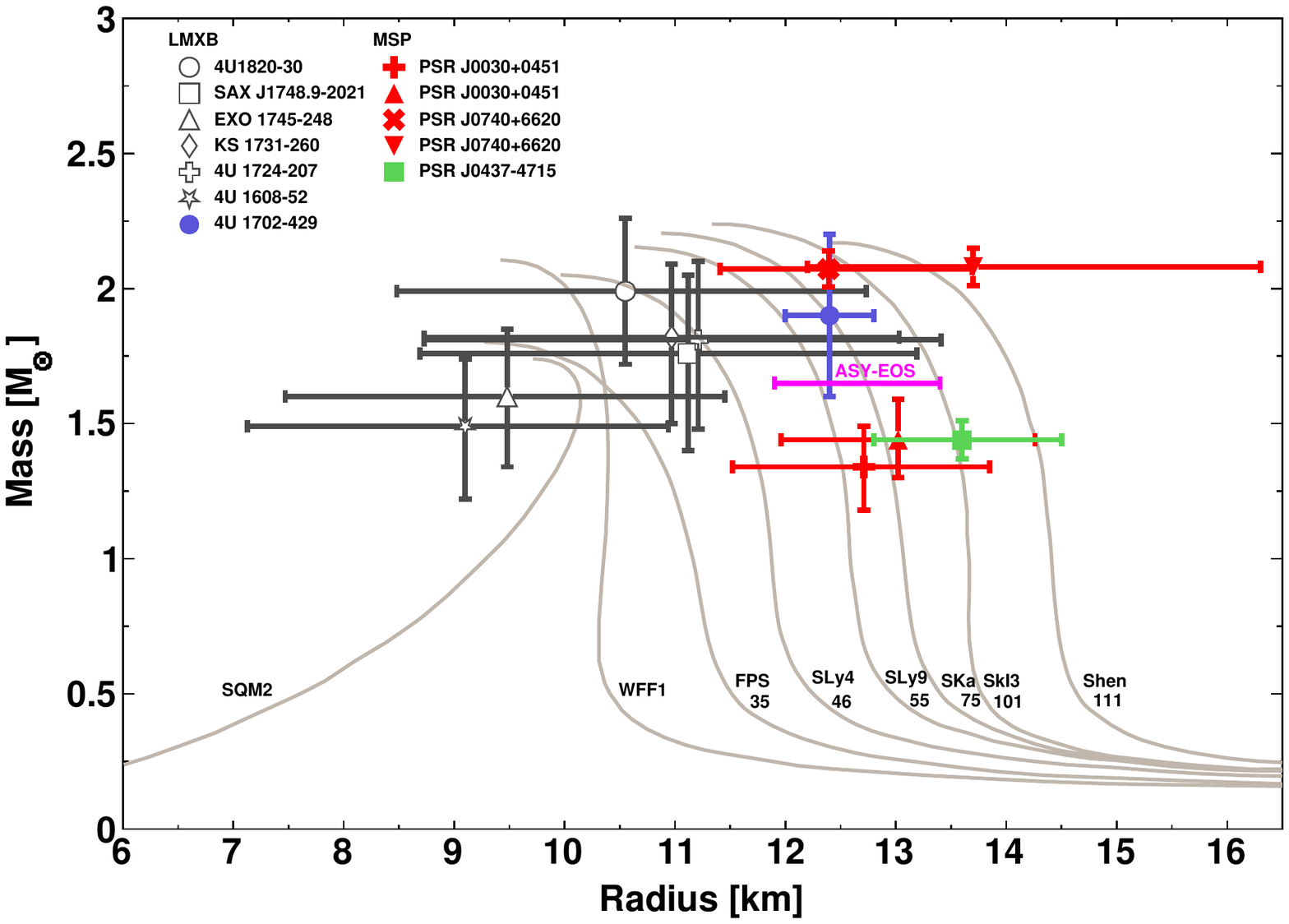}

\caption{Masses and radii for selected individual NS. The data for LMXB were
taken from \cite{kim21} (open symbols) and from \cite{nat17} (filled blue
circle). The data for millisecond pulsars measured by NICER (red symbols) were
taken from \cite{Riley:2019yda} and \cite{ril21} for J0030 and J0740 (filled crosses)
and from \cite{Miller:2019cac} and \cite{mil21} (filled triangles) for the same pulsars,
respectively). The data point for J0437 MSP measured by HST+ROSAT (green filled
square) was taken from \cite{gon19}. The radius range labeled as {\bf \textcolor{magenta}{ASY-EOS}} was
obtained by translating the symmetry energy slope parameter range obtained from
the ASY-EOS heavy-ion experiment \cite{Russotto:2016ucm} to the NS radius range
\cite{Trautmann:2019gmh}. It corresponds to a NS mass of 1.4 M$_{\odot}$ but has
been displaced arbitrarily in vertical direction for clarity. The labeled gray
lines represent the M-R relations for selected EoSs. The lines for SQM2 and
WFF1 were taken from \cite{dege18}. The line for FPS was taken from \cite{sot22}
and the remaining lines were taken from \cite{sot22d} (see these references for
description of the lines). The numbers below the acronyms of the EoSs (where
available) denote the corresponding values of the symmetry energy $L$ parameter in
MeV.}

\label{figmr}
\end{figure}

Figure \ref{fig_rns} shows an overview and recent progress in NS radius
measurements/inferences from the above sources of information. It updates the
systematics presented in \cite{luk18}.

Starting the survey from the bottom of Fig. \ref{fig_rns}, the first two points
represent the analyses of the same data from five quiescent LMXB. They yielded
slightly inconsistent results for the NS radius: $R_{NS} = 9.1^{+1.3}_{-1.5}$ km
(0.86 - 2.42 M$_{\odot}$) \cite{gui13} and $10.4 <R_{NS} < 12.9$ km (1.4
M$_{\odot}$) \cite{Steiner:2012xt}. The main differences in the analyses concerned the
assumptions about the composition of the atmosphere, about the constancy of the
radius for all NS, the distance uncertainty and different statistical inference
methods. The discrepancy between these two classes of results seems to persist up to now despite the subsequent
reanalyses including more qLMXB sources (given by the next entries from the bottom of Fig. \ref{fig_rns}): $R_{NS} = 10.3^{+1.2}_{-1.1}$ km
\cite{gui16} or adding also thermonuclear bursters: $10.1 <R_{NS} < 11.1$ km
(1.5 M$_{\odot}$) \cite{ozel16} and $9.9 <R_{NS} < 11.2$ km (1.5 M$_{\odot}$)
\cite{bog16} on lower radius side, to $12^{+1.9}_{-1.7}$ km (1.4 M$_{\odot}$)
\cite{shaw18} and $10.4 < R_{NS} < 13.7$ km (1.4 M$_{\odot}$) \cite{ste18} on
the other side. In particular, Ref. \cite{kim21} shows a recent reanalysis of
the sources used in \cite{ozel16}, but takes in addition into account the
uncertainties in chemical composition of the photosphere and in the touchdown
radius. More specifically, the authors allow in their model that the photosphere not 
necessarily falls back on the NS surface after expansion due to a thermonuclear burst. 
The radius of the photosphere after fall back (at touchdown) is allowed to be larger or equal
to the radius of the NS. In addition the authors vary the hydrogen mass fraction
in the H-He plasma of the photosphere in their analysis.
The presented point ($R_{NS} = 10.4^{+2.1}_{-2.2}$ km,
M$=1.75^{+0.27}_{-0.30}$ M$_{\odot}$) represents a mean value of the results
from Bayesian analysis for the six analysed sources. The inferred radius is close to the result of \cite{ozel16} but
has a much larger uncertainty resulting from the additional degrees of freedom
taken into account in the analysis. The large error bar also almost covers the range of radii
spanned by the pioneering results of \cite{gui13} and \cite{Steiner:2012xt}.

The point \cite{nat17} ($R_{NS}$ = 12.4 $\pm$ 0.4 km, M = 1.9 $\pm$ 0.3
M$_{\odot}$) obtained for an individual X-ray bursting NS in binary system (4U
1702-429) demonstrates the achievable precision up to now for this kind of
sources. It shows quite a good agreement with the NICER result of \cite{ril21}
(see also Fig. \ref{figmr})
.

Another very precise result for quiescent LMXB was obtained in \cite{bai19}. The
analysis included also the sources taken into account in
\cite{gui13,Steiner:2012xt,gui14,shaw18,ste18}. The averaged value for two sets of
distances yields the $R_{NS}$ = 12.39 $\pm$ 0.35 km for M = 1.45 M$_{\odot}$ and
thus joins the larger radius class of results. The analysis performed in
\cite{bai19} allowed also to extract the most likely value of the $L$ parameter,
which amounts to   37.2$_{-8.9}^{+9.2}$ MeV and points to a relatively soft
symmetry energy.

The results obtained for quiescent and bursting LMXB systems indicate the
importance of the model dependence in the data analysis and a strong
contribution of systematic uncertainties. Generally, looking at the bottom part
of Fig. \ref{fig_rns}, one might conclude that LMXBs provide radii in the range
from about 8 to 14 km with a tendency to converge towards a narrower range
between 11 and 14 km (2$\sigma$). The results of \cite{nat17} and \cite{bai19}
demonstrate that a 3\% precision (1$\sigma$) is achievable, however overall, the
LMXB results suffer from a large systematic scatter.

The results for the NS radius inferred from the gravitational wave signal
GW170817 \cite{LIGOScientific:2017vwq} and its electromagnetic counterpart
\cite{gwem17} are presented in the next to bottom part of Fig. \ref{fig_rns} and
refer to the canonical NS mass of 1.4 M$_{\odot}$. The main constraint from the
GW170817 event comes from the tidal deformability, $\Lambda$
found to be less than 800 in the original paper \cite{LIGOScientific:2017vwq}.
This constraint together with the requirement that the EoS should support the 2 M$_{\odot}$ neutron stars yielded the radius range of $9.9 < R_{NS} < 13.6$ km
in \cite{ann18}. 

Using a numerous family of the EoSs and the Bayesian inference with the
information on the lower and upper bounds on $\Lambda$ as well as on the maximum
mass of the NS, a most probable value of $R_{NS} = 12.39^{+1.06}_{-0.39}$ km has
been obtained in \cite{mos18}. Similar analysis in \cite{lim18} gave $R_{NS} =
12.10^{+0.77}_{-1.74}$ km. Refs. \cite{LIGOScientific:2018cki} and \cite{de18}
attempted at deriving more tight constraints on $\Lambda$ than in
\cite{LIGOScientific:2017vwq} by applying an additional condition on the maximum
mass of the NS \cite{Dem10} and on the distribution of the measured masses. They
arrived at the values of $R_{NS} = 11.9^{+1.4}_{-1.4}$ km and $R_{NS} =
10.8^{+2.4}_{-1.9}$ km, respectively. 

Ref. \cite{tews18} uses the chiral EFT interactions and QMC methods to derive
the dense matter EoS, reduce the $\Lambda$ range and predict the corresponding
NS radius to fall between 9.0 km $< R_{NS} <$ 13.6 km.

By combining the gravitational wave and electromagnetic data and performing a
Bayesian parameter estimation of the GW170817 the authors of \cite{rad19}
arrived at the value of 12.2$_{-0.8}^{+1.0} \pm 0.2$ km for the $R_{NS}$.

Finally, two more recent analyses yield the values of 11.0$^{+0.9}_{-0.6}$ km
\cite{cap20} and 11.40$_{-1.04}^{+1.38}$ km \cite{ess20} for the $R_{NS}$,
within frameworks similar to that of \cite{tews18}.

Summarizing this non-exhaustive set of results based on GW170817 and its post
merger emissions, one can conclude that they span the range of radii from about 9 to 13.5 km (1.6-2$\sigma$), consistent with the results from LMXBs. The result from \cite{cap20} indicates that the NS merger events can provide the NS radius estimates with about 6-7\% precision (1$\sigma$). It is also worth to notice that in \cite{Krastev:2018nwr,Zhang:2018vbw} the authors stated that while the tidal polarizability $\Lambda$ depends strongly on the details of the symmetry energy, different trends of $E_{sym}(\rho)$ may lead to very similar values of $\Lambda$. Thus, measuring $\Lambda$ alone may not uniquely determine the density dependence of the symmetry energy; both nuclear laboratory experiments and astrophysical observations are therefore necessary to break this degeneracy. At similar conclusion was reached in \cite{Forbes-PhysRevD.100.083010} where it was shown how observations of gravitational waves from binary neutron star mergers can be combined with insights from nuclear physics to obtain useful constraints on EoS of dense matter between one and two times the nuclear saturation density. 

The middle part of  Fig. \ref{fig_rns} shows the recent results obtained for
three millisecond pulsars (PSR J0437-4715 \cite{gon19}, PSR J0030+0451
\cite{Riley:2019yda,Miller:2019cac} and PSR J0740+6620 \cite{ril21,mil21}) using
the thermal emission pulse profile modeling. The result of \cite{gon19}, $R_{NS}
= 13.6^{+0.9}_{-0.8}$ km, (M=1.44 $\pm$ 0.07 M$_{\odot}$) has been obtained from
the HST and the ROSAT data, while the radii and masses of the other two pulsars
were obtained using the NICER data (\cite{Riley:2019yda}: $R_{NS} =
12.71^{+1.14}_{-1.19}$ km, M=1.34$^{+0.15}_{-0.16}$ M$_{\odot}$,
\cite{Miller:2019cac}: $R_{NS} = 13.02_{-1.06}^{+1.24}$ km,
M=1.44$_{-0.14}^{+0.15}$ M$_{\odot}$, \cite{ril21}: $R_{NS} =
12.39^{+1.30}_{-0.98}$ km, M=2.072$_{-0.066}^{+0.067}$ M$_{\odot}$ and
\cite{mil21}: $R_{NS} = 13.7_{-1.5}^{+2.6}$ km, M=2.08 $\pm$ 0.07 M$_{\odot}$). 
Including additional constraint in the latter result the authors of \cite{mil21}
arrive at the value of $R_{NS} = 12.35 \pm 0.75$ km. It is more compatible with
the result of \cite{ril21}, nevertheless the discrepancy between the raw values
indicates the importance of the systematic uncertainties in MSP analyses, too.
The results are compatible with the larger radius class of LMXBs and with the
upper side of the GW170817 results.  Overall, the pulse profile modeling for
MSPs can provide the radii with the precision of about 6-15\% (1$\sigma$).

Multiple sources of information related to the symmetry energy from the low
energy nuclear experiments and data, from the high energy experiments as well as
the information from LMXBs, gravitational wave events or MSPs allow to perform
Bayesian type of analyses that use different combinations of these and other
constraints to infer the properties of the EoS and the NS radii. Some recent
results of such analyses are presented in the next to top part of Fig.
\ref{fig_rns} . 

Generally, refs.
\cite{jiang20,lan20,die20,ess20,raa21,tan21,bis21,alm21,rai21,Huth:2021bsp} give
consistent results, however as might be expected, the results get easily biased
depending on the constraint used. This can be noticed for the two points from
\cite{ess20}, where adding the NICER constraint in addition to the GW one shifts
the radius towards higher values. Similar effect is visible from the results of
\cite{rai21}, where the lower radius value is constrained by the LMXB data and
the higher value point is constrained by the NICER data. We also emphasize the
role of the heavy-ion collision (HIC) constraint by presenting the two results
of \cite{Huth:2021bsp}, the upper point constrained by HIC alone and the lower one
constrained in addition by astrophysical data. These results will be discussed in more detail later on.
The most constrained result, using the astrophysical and HIC information
presents the sharpest radius value so far: $R_{NS} = 12.01_{-0.38}^{+0.37}$ km,
with about 3\% precision.

Overall, the survey of astrophysical constraints on the $R_{NS}$ shows a
progress in the precision of the data, the quality of the models and in the data
analysis techniques. The uncertainty of inference of the $R_{NS}$ has been
reduced from about 25\% down to 6-3\% in the latest results. This denotes the
reduction of the uncertainty of $L$ parameter, when derived from R, from about 100\% to 24-12\%. The factor
of 4 in conversion of the relative uncertainties comes from the approximate conversion 
formula (see \cite{Lattimer:2015nhk} eq. (3), \cite{tews17} eq. (52) or \cite{rai21} eq. (6)), 
where $L \propto R^{4}$ and thus $\Delta L/L = 4 \Delta R/R$. Thus, the accuracy of derivation of $L$ from
the astrophysical measurements of $R_{NS}$ becomes comparable to the accuracy of $L$ derived from
the heavy ion collision data, as shown in the next paragraph.

The point at the top of Fig. \ref{fig_rns} has been obtained in
\cite{Trautmann:2019gmh} by converting the symmetry energy slope parameter range from the ASY-EOS heavy-ion experiment \cite{Russotto:2016ucm} to the NS radius range, using the known tight correlation between the symmetry pressure $p_{0}$ and the radius $R_{1.4}$ of a canonical neutron star of 1.4 \(M_\odot\) \cite{Lattimer:2015nhk}. The corresponding $R_{NS}$ value amounts to $12.65 \pm 0.75$ km
. Its 6\% precision has been obtained from the $L$ value measured with about 20\% accuracy. While it may be difficult to improve the current 3\% precision in the $R_{NS}$
extraction from the astrophysical data, improving the terrestrial results by
more than a factor of two in the $L$ extraction is within a reach with new detectors, experimental techniques and models.  It is, however, important to stress that also this result relies on the two assumptions made in the first UrQMD analysis of ASY-EOS data and discussed in sect. \ref{sec:2321}.


As a complement to Fig. \ref{fig_rns}, Fig. \ref{figmr} shows the masses and
radii for a few selected individual NS from LMXB and MSP, including the latest
NICER results. In addition, the ASY-EOS radius range (see description of Fig.
\ref{fig_rns}) obtained on the basis of the heavy ion experiment is presented.
The points are put on top of some ``typical'' M-R relations for selected EoSs
(see \cite{dege18,sot22,sot22d} for a detailed description of the lines). The
M-R relations taken from \cite{sot22,sot22d} are of particular interest since
they provide a direct link to the symmetry energy slope parameter $L$ values (see
the numbers below the labels in MeV).

As stated above, the precision of the astrophysical results improves, however,
definitely more precise data points are needed to better constrain the nuclear
EoS. Despite the fact that the elaborate Bayesian techniques allow to combine
different data and models in optimal ways, and provide weights for the EoSs that
permit to construct the most likely M-R relations and to effectively shrink the
measured uncertainties, the presented results call for more statistics and more
coherence in the data.

Restricting to the most recent and most precise results presented in Fig.
\ref{figmr}, one can state that the range of the symmetry energy slope
parameters constrained by the data is still broad, from about 40 to about 110
MeV. Thus, the uncertainty is still large, of the order of 40-50\%.

What follows below is a more detailed discussion of some specific results.\\

\begin{figure}[ht]
\includegraphics[width=1.0\textwidth]{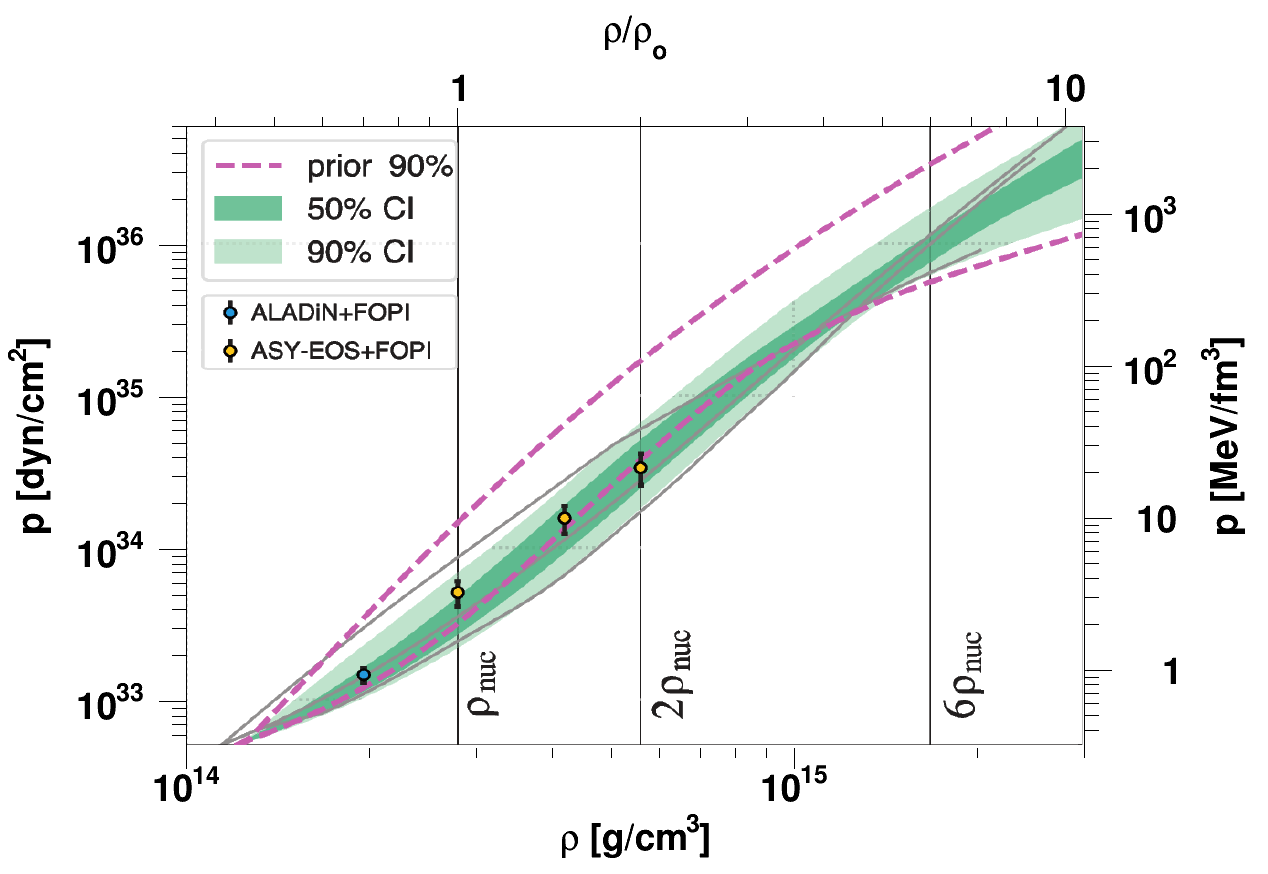}
\caption{(adapted from Ref. \cite{LIGOScientific:2018cki}) Marginalized posterior (green bands) and prior (purple dashed lines) for the pressure p as a function of the rest-mass density $\rho$ of the NS interior as obtained in \cite{LIGOScientific:2018cki}, with superimposed (orange and green points) constraints obtained from HICs discussed in this paper as deduced assuming a 3$\%$ proton fraction. All the points take into account FOPI constraints for symmetric nuclear matter from both IQMD and UrQMD analyses, while the green one at 0.7$\rho_{0}$ takes into account also the ALADiN constraint for $E_{sym}$ at sub-saturation densities and the orange ones in the region between 1 and 2 $\rho_{0}$ also the ASY-EOS constraint for $E_{sym}$. The dark (light) shaded green regions corresponds to the 50$\%$ (90$\%$) posterior credible level and the purple dashed lines show the 90$\%$ prior credible interval. Vertical lines correspond to once, twice, and six times the nuclear saturation density. Plotted in addition in gray are some representative EoS models of Wiringa \textit{et al.} \cite{Wiringa:1988tp}, Akmal \textit{et al.} \cite{Akmal:1998cf} and Lackey \textit{et al} \cite{Lackey:2005tk};  see fig. 2 of Ref. \cite{LIGOScientific:2018cki} for more details.   
}
\label{fig:fig-Prho}
\end{figure}

Going back to the information obtained from by GW detection in binary NS merger events, one of the main result is reported here in fig. \ref{fig:fig-Prho}, adapted from fig. 2 of \cite{LIGOScientific:2018cki}, showing the results for the pressure p as a function of rest mass density of the NS interior as obtained in that work. That resulted in an estimation of 
pressure of 
$21.8^{+16.9}_{-10.6}$~MeV/fm$^{3}$ at 2 $\rho_{0}$. A second event of  binary neutron star merger GW190425 has been more recently reported in \cite{LIGOScientific:2020aai}; there a pressure of 19-80 MeV/fm$^{3}$ was estimated for neutron star matter at $2\rho_{0}$. From that figure  we can see that the 90$\%$ confidence interval for the pressure of neutron star matter at $\rho_{0}$ is about between 1.2 and 3.7 MeV/fm$^{3}$. It follows that the pressure estimation of 3.4$\pm$0.6 MeV/fm$^{3}$, obtained in \cite{Russotto:2016ucm} from the UrQMD analysis of the ASY-EOS experimental results, is consistent with the GW170817, falling in the higher part of 90$\%$ confidence interval; a similar result was obtained by comparing to the results from neutron-star observations of Ref. \cite{Steiner:2012xt}, where the ASY-EOS result was located within the upper half of the 95$\%$ confidence interval obtained in that study.\\ 
In addition, another comparison of results obtained from GW170817 with the one given by HICs is presented in fig \ref{fig:fig-Prho}. In fact, green and orange points of that figure present constraints obtained from HIC collisions, fully discussed in the previous sections of this paper, as deduced assuming a 3$\%$ proton fraction. All points take into account FOPI constraints for the symmetric nuclear matter contribution deduced from IQMD and UrQMD analysis of the experimental data, as it has been validated between 0.7 up to $\approx 3 \rho_0$. The symmetry energy contribution comes from two different experimental HIC results, according to densities where they give the highest precision. The green point at 0.7$\rho_{0}$ and the orange ones between 1 and 2 $\rho_{0}$ use respectively the ALADiN (as recently obtained by A. Le Fèvre et al. in \cite{LeFevre22}) and ASY-EOS constraints for $E_{sym}$. Apart from the 1 $\rho_{0}$ point, which lies at the high pressure boundary but within the 90$\%$ confidence interval, the other 3 points show a remarkable agreement. And also the errors appear to be very competitive, providing a proof of the effectiveness of HIC data.\\
As a further examples of comparison of the ASY-EOS results with  recent results triggered by multi-messenger astronomy, in \cite{Zhang:2018bwq} Zhang and Li produced a restricted EoS parameter space using observational constraints on the radius, maximum mass, tidal polarizability and causality condition of neutron stars, resulting in an estimation of the $E_{sym}$ at $2\rho_{0}$ of $46.9\pm10.1$~MeV. It is interesting to note that in a subsequent paper \cite{Zhang:2019fog}, the authors show that the observation of a 2.17 \(M_\odot\) neutron star reduces the error to $\pm 9$ MeV and point out that it is unlikely that even heavier neutron stars will be observed because the value 2.17 \(M_\odot\) is already close to the theoretical maximum according to several studies. In Ref. \cite{Xie:2019sqb} a Bayesian analysis of GW170817 and quiescent low-mass X-ray binaries radii suggested the value of $E_{sym}$ at $2\rho_{0}$ to lie in the interval $31-51$ MeV. The ASY-EOS constraint of \cite{Russotto:2016ucm} extracted using the UrQMD transport model suggests 51-60 MeV interval for the $E_{sym}$ at $2\rho_{0}$, and this is in reasonable agreement with above cited values of the Refs. \cite{Zhang:2018bwq,Zhang:2019fog,Xie:2019sqb}.\\


As written above, several studies, combining the constraints on the symmetry energy from GW, astrophysical observations and the ones obtained in terrestrial laboratory, have been recently published, showing the importance and effectiveness of this  "multiple sources" approach. As a relevant example, we will discuss here the results obtained in a very recent work of S. Huth et al.  \cite{Huth:2021bsp}. 

\begin{figure}[ht]
\includegraphics[width=1\textwidth]{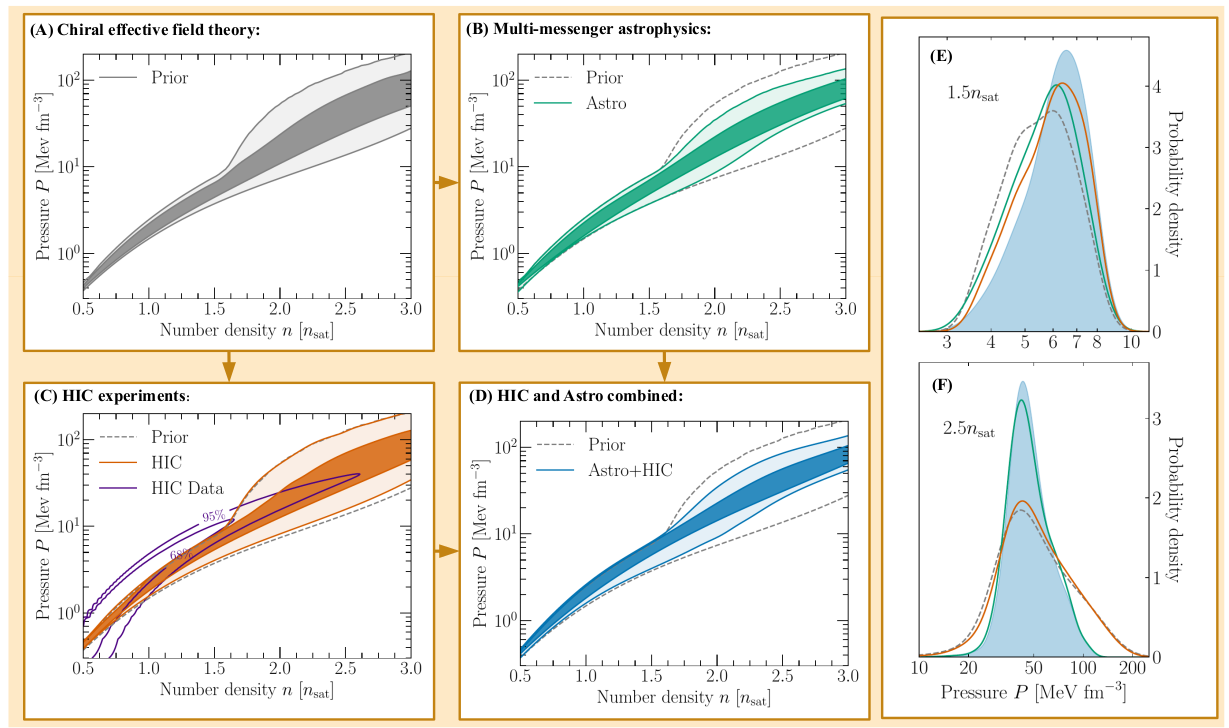}
\caption{(reprinted from Ref. \cite{Huth:2021bsp} under permission) Evolution of the pressure as a function of baryon number density for the EoS prior (A, gray), when including only data from multi-messenger neutron-star observations (B, green), when including only HIC data (C, orange), and when combining both (D, blue). The shading corresponds to the 95\% and 68\% credible intervals (lightest to darkest). The impact of the HIC experimental constraint (HIC Data, purple lines at 95\% and 68\%) on the EoS is shown in panel C. In panels (B) through (D),  the 95\% prior bound is also shown for comparison (gray dashed lines). Panels (E) and (F) show the distributions for the pressure at, respectively, 1.5 and 2.5 $\rho_0$ at the different stages (lines colour corresponds to the one used in the other four panels) of the analysis, with the combined multi-messenger+HIC region shaded in light-blue. See Ref. \cite{Huth:2021bsp} for more details.}
\label{fig:fig-huth21-a}
\end{figure}

In detail, Ref. \cite{Huth:2021bsp} uses Bayesian inference to combine data from astrophysical multi-messenger observations of neutron stars and HICs, as measured in FOPI and ASY-EOS experiments, with microscopic nuclear theory calculations, to improve the constraints of dense matter. The authors start from a "prior" EoS, as determined by using local $\chi EFT$ interactions, as shown in panel A of Fig. \ref{fig:fig-huth21-a}. The large uncertainties for the obtained EoS at higher densities are then partially reduced by adding constraints from astrophysics observations and HIC data. 
To that end, mass measurements of massive neutron stars PSRJ0348+4042 and PSR J1614-2230, information from X-ray pulse-profile modelling of PSR J0030+0451 and PSR J0740+6620 as obtained by NICER and X-ray Multi-Mirror Mission (XMM-Newton) and information from the two neutron-star mergers GW170817 and GW190425, observed by Advanced Ligo and Advanced Virgo, are used to get multi-messenger astrophysics constraint on the high-density neutron star EoS and reduce its uncertainties, as shown in panel B of Fig. \ref{fig:fig-huth21-a}. In addition, HIC data from FOPI and ASY-EOS experiments are used to impose a HIC constraint on the prior EoS, as shown in panel C of Fig. \ref{fig:fig-huth21-a}. 
Finally, both astrophysical and HIC constraints are combined to get a final EoS constraint, as shown in panel D of Fig. \ref{fig:fig-huth21-a}. The final result for the radius of a 1.4 solar mass neutron stars is $12.01_{-0.38}^{+0.37}$ km at 68\% uncertainty. In panel (E) it is clearly seen how the inclusion of HIC data leads to an increase in the pressure in the 1.5 $\rho_{0}$ region, where the HIC experimental sensitivity is higher, shifting the neutron-star radii towards larger values, consistent with recent NICER observations \cite{mil21,raa21}. Instead, pressure in the 2.5 $\rho_{0}$ region, shown in pane (F), is essentially determined by the astrophysical observations, since the HIC sensitivity at those densities is very small. However, this finding shows that constraints from HIC experiments show a remarkable consistency with multi-messenger observations and provide complementary information on nuclear matter at intermediate densities. The work shows how joint analyses can shed light on the properties of neutron-rich supra-nuclear matter over the density range probed in neutron stars. The work also suggests that a model independent analysis of HIC data and a reduction of experimental uncertainty have a great potential to provide EoS information complementary to astrophysics; the larger uncertainties in the EoS above 1.5 $\rho_0$, as clearly seen in Fig. \ref{fig:fig-huth21-a}, indicates the importance of precise measurements of the symmetry energy in that high-dense regime. This region of densities is also the one of most importance for the modelling of neutron-star and the one where model uncertainties are larger. In addition, constraints from  HICs, as shown in this paper, are scarce in this density region. This highlights the importance of measurements of the symmetry energy around 1.5-2.5 $\rho_0$; plans in that direction will be discussed in the next section.

 \begin{figure}[ht!]
 \includegraphics[width=0.7\textwidth]{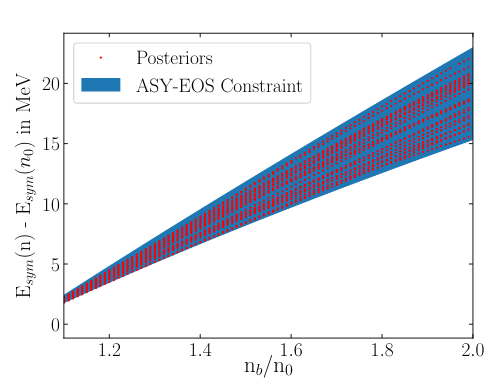}
 \caption{(reprinted from Ref. \cite{Ghosh:2021bvw} under permission) Symmetry energy, minus its value at saturation densities, as a function of reduced density (n\textsubscript{b} is the baryon density and n\textsubscript{0} indicates the nuclear saturation density),  of the prior EoSs (dashed red curves), used by Ghosh \textit{et al.} in a Bayesian scheme taking into account constraints from $\chi EFT$, observations of mass and  radius  of  neutron-stars and GW observations of binary neutron star merger, lying inside the band allowed by the ASY-EOS results in the range of 1.1-2.0 $\rho_{0}$. See Ref. \cite{Ghosh:2021bvw} for more details.}
 \label{fig:fig-Gho21}
 \end{figure}

In another recent paper by  Ghosh \textit{et al.} \cite{Ghosh:2021bvw}, ASY-EOS data have been again used to investigate the use of multi-physics constraints for studying the neutron-matter EoS. Similarly to the work of Ref. \cite{Huth:2021bsp}, Ghosh \textit{et al.} used a Bayesian scheme to investigate the EoS, with a Relativistic Mean Field approach to determine the prior. The constraints taken into account in the main analysis, in order to determine the posterior, were chiral effective field theory results for low densities and multi-messenger astrophysical observations for high densities. The HIC results, from KaoS, FOPI and ASY-EOS experiments, were taken into accounts as constraints for intermediate densities, but they were treated differently from $\chi EFT$ and astrophysical ones, because of the problem of their model-dependence. As an example, Fig. \ref{fig:fig-Gho21} shows the set of selected EoSs (dashed red curves) when requiring agreement to the ASY-EOS results in the range of 1.1-2.0 $\rho_{0}$ (blue band), amounting to about $40\%$ of the prior EoSs. Posteriors with and without HIC results were extracted and compared, finding few differences between the two cases, but considering implications of HIC results on final EoSs with caution. However, also this work emphasizes that precise and model-independent HIC data can play a relevant role in determining the neutron-star EoS, especially in the region of 1-2 $\rho_0$. Comparing the works of Huth \textit{et al.} and Ghosh \textit{et al.}, the former uses a softer EoS from $\chi EFT$ up to 1.5 $\rho_{0}$, getting stiffer when including HIC constraints, with the final results being mainly ruled by $E_{sym}$; instead the RMF model of Ghosh \textit{et al.} favors isospin-symmetric matter at high densities, and the final results is less ruled by the $E_{sym}$ contribution. \\
 

The results on multi-physics analyses discussed here are surely not exhaustive for what concerns recent advances in neutron-stars EoS determination. As we have seen this is a very hot topic and new results are appearing at short intervals. However, the works discussed emphasize that HIC data, such as those made available by the FOPI and ASY-EOS experiment, can not only give laboratory constraints on EoS asymmetric matter, to be compared with the astrophysical ones, but can be used together with the latter to provide better constraints on EoS. The development of model independent analyses of the HIC data and new and more precise experiments able to investigate regions of high-densities, around 2 $\rho_0$, thus appears to be mandatory.

\subsection{Perspectives}
\label{sec:233}
In this section we will outline some possible future developments that represent natural follow-ups of studies presented in this paper, with the caveat of not representing an exhaustive report on future possibilities. 
\subsubsection{Pion studies}\label{sec:2331}
The up-to-date studies of the EoS of nuclear matter using pion production close to threshold have relied on transport models that describe particle emission exclusively via nucleon resonance excitation. This mechanism induces a strong dependence of multiplicities and single ratios on the poorly known in-medium resonance potentials, reducing the sensitivity to the symmetry energy. The impact of the assumption of resonance production  grows weaker for energetic pions and, consequently, the study of the symmetry energy was possible using the high transverse momentum region of pion spectra measured by the S$\pi$RIT Collaboration \cite{SRIT:2021gcy}. It has however been shown, using microscopical models, that close to threshold pions are predominantly produced in non-resonant nucleon-nucleon collisions~\cite{Engel:1996ic,Shyam:1996id}. Including such a production mechanism in transport models, while challenging from a technical perspective, would reduce the sensitivity to resonance potentials considerably. Consequently, the lower transverse momentum parts of pion spectra could be used for more accurate constraints for the symmetry energy above saturation.

Experimental data for pion production in Au+Au at an impact energy of 1.23 GeV/nucleon have been published recently by the HADES Collaboration~\cite{HADES:2020ver}. Comparisons with five transport models for rapidity and transverse mass spectra were also provided, revealing important discrepancies among theoretical models but also deviations from experimental values. The reasons for these difficulties may originate in different momentum dependent interactions and resonance potentials considered in each model. Furthermore, the inclusion of all relevant decay channels of heavier resonances, in particular the two-pion ones, may be of relevance for a proper description of rapidity spectra. 

A recent study, employing the dcQMD transport model, has shown that the high transverse momentum part of the pion single ratio is sensitive to the symmetry energy even at impact energies far above threshold~\cite{Cozma:2021tfu}. The high accuracy of the HADES pion transverse mass spectra opens the possibility of studying the symmetry energy at densities above twice saturation. Preliminary estimations  using the dcQMD model \cite{dcQMD22} suggest a sensitivity of the pion single ratio to the asy-EoS stiffness of the order of 30$\%$, which is significantly larger than the experimental accuracy of single ratios of about $\pm 2\%$. It is expected that data for Ag+Ag collisions at 1.58 AGeV will be soon made available by the HADES collaboration. A comparison of the predictions of four transport models (UrQMD, PHSD, PHQMD and SMASH) concerning integrated multiplicities, rapidity distributions and transverse mass spectra for proton, pions and other strange particles was presented in Ref. \cite{Reichert:2021ljd}. Despite different treatments of the HIC dynamics among the models, consistent results were obtained for the bulk of investigated hadrons, except for the $\Xi^-$ case. However, at such impact energies the pion single ratio, such as other observables related to particle productions, is sensitive to resonance potentials over the entire transverse mass range, making such a study challenging though still feasible. For a successful attempt, a transport model that can realistically describe pion production from threshold up into the 1-2 GeV/nucleon region is required.

On the experimental side, the S$\pi$RIT Collaboration aims at enlarging the experimental database for pion production close to threshold by measuring $^{136}$Xe+$^{124}$Sn at 334 MeV/nucleon (RIKEN) and 200 MeV/nucleon (FRIB). A low energy beam scan (200-800 MeV/nucleon) including Au+Au collisions
currently planned by the HADES Collaboration, would improve the situation considerably if an accuracy comparable to one achieved for the same system at 1.23 GeV/nucleon can be reached. It can thus be foreseen that in the mid-term future a rather accurate determination of the symmetry energy at densities close to 1.5$\rho_0$ can be accomplished from studying pion production close to threshold. A proper understanding of pion production above 1.0 GeV/nucleon would add constraints in the 2.0-2.5$\rho_0$ density range. 

\subsubsection{Elliptic flow studies}\label{sec:2332}
The ASY-EOS experimental results proved the effectiveness of the
$v_{2}^n/v_2^{ch}$ ratio in constraining the high-density behaviour of the $E_{sym}$, and validated the experimental approach used there. But, even if the experiment was surely successful in reaching its main aim, two drawbacks have to be noticed: 

\begin{itemize}
\item the problems of LAND electronics, briefly discussed here, but extensively reported in \cite{Russotto:2016ucm}, did not allow to get a better statistics and, above all, did not allow neither discrimination of charged particles, nor to implement analysis to  separate hydrogen isotopes and extract the proton flow;
\item it was conceived as a first/exploratory experiment for a future extensive campaign. In fact, although the ASY-EOS determination of $E_{sym}$ is estimated to have reached the supra-saturation density region, which was already a unique achievement, there remains a strong need for constraining it further, with an emphasis on the density region around $\sim$2 $\rho_{0}$. SIS@GSI provides a unique tool to probe such densities with HICs. Unfortunately, because of the upgrade phase ongoing at GSI/FAIR in the last years, strongly reducing beam-time opportunities for users and of the recent difficulties related to the pandemic a new experimental campaign, expected to be the natural follow-up of the ASY-EOS experiment, has still not been firmly planned.     
\end{itemize}

\begin{figure}[ht]
\centering
\includegraphics[width=1.\textwidth]{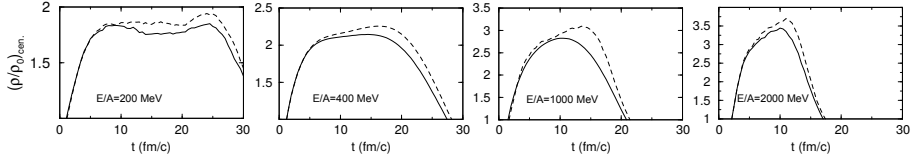} 
\caption{(adapted from Ref. \cite{Li:2002yda} under permission) Evolution of the central baryon density in $^{132}$Sn+$^{124}$Sn collisions at beam energies from 200 to 2000 MeV/nucleon for b = 1 fm, as predicted by the hadronic transport model of \cite{Li:2002yda} using a stiff (solid curves) and a super-soft (dashed curves) $E_{sym}$.}
\label{fig:dens-bali}
\end{figure}

The next desirable step is to measure the elliptic flow of neutron, proton and light clusters in Au+Au collisions over a wide range of beam energies from 250 to 1000-1500 MeV/nucleon, in order to provide tighter constraints on the slope parameter $L$ and new limits on K$_{sym}$, the currently poorly constrained symmetry energy curvature parameter. For that a new campaign \cite{Russotto:2021mpu} has been proposed representing a unique set of measurements, presently possible only at the GSI/FAIR facility, in view of the available range of beam energies and the existing instrumentation.\\
A relevant novelty of the proposed experiment is the direct measurement of n/p elliptic flow ratio, made possible by the capabilities of the NeuLAND detector. Some details about this new detector are presented below. We remind that, as presented in Fig. 19 of \cite{Russotto:2016ucm}, according to the TuQMD model for the Au+Au case at 400 MeV/nucleon, the v\textsubscript{2}-ratio of neutrons with respect to charged-particles or hydrogen is mainly sensitive to a region centred slightly above saturation density, while the n/p ratio sensitivity is centred around 1.4$\rho_{0}$. Thus, the direct measurement of n/p ratio, instead of n/charged particles ratio, will enable probing of higher density region than was done before, at the same incident energy of 400 MeV/nucleon. This by itself  will be a significant achievement. In addition, the n/p elliptic-flow ratio is less affected by the procedure used to build the clusters, the so called clusterization algorithms, from the final stage of the HIC as simulated by the theoretical models, allowing thus the reduction of model dependence related to that stage of the analysis.\\ 
The second important feature of the proposed campaign is the measurement of excitation function in a relevant beam energy range, from 250 to 1000 MeV/nucleon. Microscopic transport calculations predict that for a short time ($\sim$20~fm/c) densities of up to $\sim$3 times the saturation density can be reached in the central zone of a HIC even at moderate incident energies of $\sim$1 AGeV \cite{Li:2002yda}. This is shown in Fig.~\ref{fig:dens-bali}, presenting the time evolution of the central baryon density in $^{132}$Sn+$^{124}$Sn collisions at beam energies from 200 to 2000 MeV/nucleon for b = 1 fm, as predicted by the hadronic transport model of Ref. \cite{Li:2002yda} using a stiff (solid curves) and a super-soft (dashed curves) $E_{sym}$. One can notice that the supersoft $E_{sym}$ leads to slightly higher densities of about 10–15$\%$, due to the softening of the nuclear EoS with the supersoft $E_{sym}$ relative to the stiff one. Additional arguments related to the maximum probed density are already given in the fig.~\ref{fig:FOPI_density}. It shows that the maximum densities in the innermost center of the collision may reach up to 3.5$\rho_0$ at 1.5~AGeV, while the densities probed by the mid-rapidity protons during the heavy ion collision may extend up to twice the saturation density, irrespective of the transverse momentum cut \cite{LeFevre:2015paj}. 
These IQMD calculations were used to estimate the range of densities to which elliptic flow observable is most sensitive to, that is the "force-weighted" average density shown as error bars in fig.~\ref{fig:FOPI_density}.  It shows that for Au+Au collisions at 1~AGeV, the typical densities directly influencing the flow, through the impact of the mean-field on the time evolution of the reaction, span between $\rho_0$  and 2.2$\rho_0$. Thus, according to the models, raising the beam energy allows to probe $E_{sym}$ at higher densities. Nevertheless, as demonstrated in \cite{LeFevre:2016vpp}, and already mentioned in Sect.~\ref{sec:121}, higher densities reached in the fireball play also an indirect role in the strength of the elliptic flow, because they determine how fast the later expansion of the fireball will occur and modify the elliptic flow by the interplay between expanding fireball and flowing away spectators. Therefore, at 1~AGeV, some influence of the EoS at 3$\rho_0$ on $v_2$ cannot be ruled out.  In the most compressed phase of the collision, it is expected that due to the symmetry energy, the neutron part in the fireball will expand faster than that of the protons. Thus they will interact differently with the spectators, which will result in a different elliptic flow.  However, as the incident energy is increased the fraction of nucleons excited into baryonic resonances, mainly $\Delta(1232)$ at energies of 1 GeV/nucleon and below, in the highly compressed phase of the collision reaches values in the neighborhood of $20\%$ \cite{PhysRevC.47.R2467}. This has a potential impact on the time evolution of the reaction that may leave a comparable imprint on the $E_{sym}$ around 2 $\rho_{0}$, depending on the chosen observable. Thus, one should be aware that the highest density reached during a HIC is not necessarily equivalent to the density that can effectively be probed, i.e. without the knowledge of additional systematic theoretical uncertainties, for the purpose of constraining the high density dependence of the symmetry energy.  The proposed measurement could effectively provide powerful data to explore also these delicate aspects.\\
In addition, with an unambiguous proton identification one should be able not only to constrain the slope of $E_{sym}$, $L$, but also its curvature, $K_{sym}$. The latter is the parameter least constrained experimentally and theoretically  so far. In Fig. 13 of \cite{MDCozEPJA18} it was shown that for Au+Au semi-central collisions neutron-proton elliptic flow ratio sensitivities to $L$ and $K_{sym}$ reach maxima at 600 and 250 MeV/nucleon, respectively, thus at quite different energies. Note that the potential terms that are proportional to $L$ and $K_{sym}$ have different dependencies on density and, consequently, the forces generated by these two terms attain their maximum effectiveness at different regions of density; in a simplified way, the $L$ term is proportional to the isospin asymmetry $\delta$, while the $K_{sym}$ term to $\delta\rho$. This explains why the maximum sensitivity for $L$ and $K_{sym}$ do not occur for the same incident energies. This is another strong reason for measuring excitation functions of neutron-proton elliptic flow observable.\\

It was also proposed to measure, at the same time, yields and isotopic compositions of clusters. This is of 
importance 
for advanced tuning of clusterization algorithms used in the transport models in order to get more realistic predictions.
Ideally, cluster degrees of freedom should be included explicitly in transport models as they have direct impact on predictions of neutron-to-proton yields and flows, with clusters acting as absorbers of otherwise free nucleons, as well as pionic observables~\cite{Ikeno:2016xpr}. Nevertheless, such approaches are very challenging from a technical point of view. As a consequence, only the effects of light clusters, up to at most $^4$He, have been included explicitly in the collision integral of only a few existing transport models~\cite{Danielewicz:1991dh,Ono:2013aaa}. Alternative methods, such as coalescence invariant proton or neutron spectra, which are defined as weighted sums of proton or neutron content of light clusters with $Z\leq$2, have been proposed~\cite{Famiano:2006rb} and used successfully~\cite{Coupland-PhysRevC.94.011601,morf19} in studying the EoS.
It is worth to notice that transport models also predict a sensitivity of the ratio of yields and flows of light isobar nuclei to the high density behavior of the symmetry energy \cite{Famiano:2006rb,Giordano:2010pv,Zhang:2014sva,Coupland-PhysRevC.94.011601}, that could be measured and used as an additional probe of the EoS.
It follows that precise data on the inter-related phenomena of clustering and neutron and proton emissions as well as correlations between them could represent strong constraints to nuclear transport theories.\\

\begin{figure}
\centering
\includegraphics[width=0.495\textwidth]{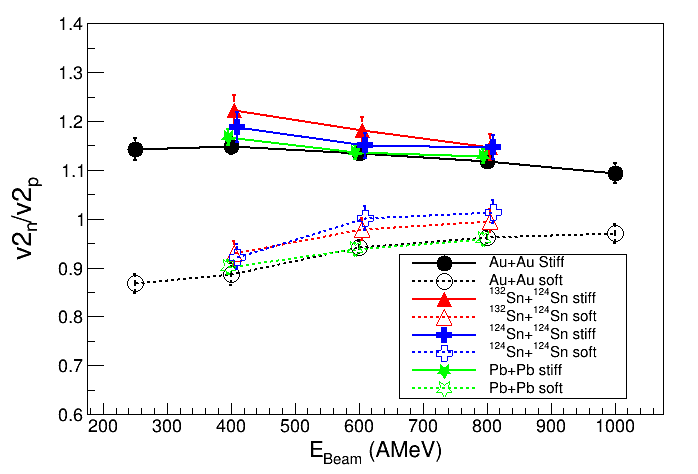}
\includegraphics[width=0.495\textwidth]{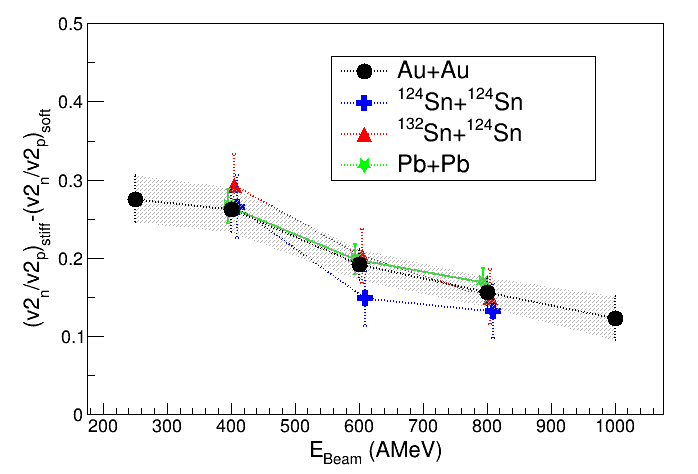}
\caption{Left panel: Excitation functions of neutron-to-proton elliptic flow
ratios, $v2_{n}/v2_{p}$, at mid-rapidity for semi-central Au+Au,
$^{132}$Sn+$^{124}$Sn, $^{124}$Sn+$^{24}$Sn and Pb+Pb collisions, as predicted by the
UrQMD model for stiff and soft $E_{sym}(\rho)$. Right panel: differences between
the stiff and soft results, providing a quantitative evaluation of the $v2_{n}/v2_{p}$ sensitivity to the symmetry energy.}

\label{fig:fig7}
\end{figure}

\begin{figure}[ht!]
\includegraphics[width=0.5\textwidth]{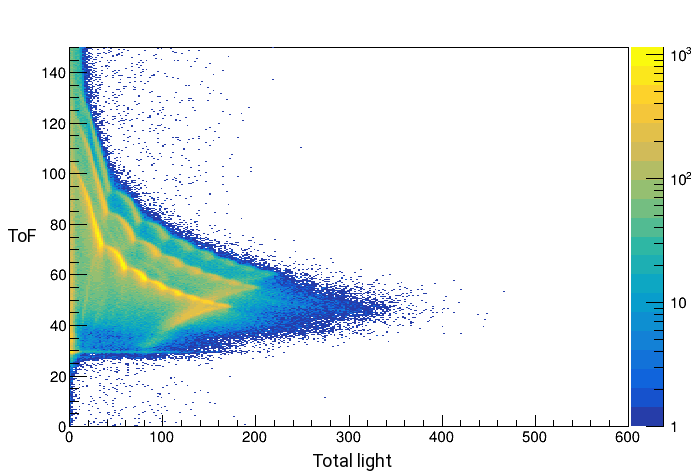}
\includegraphics[width=0.5\textwidth]{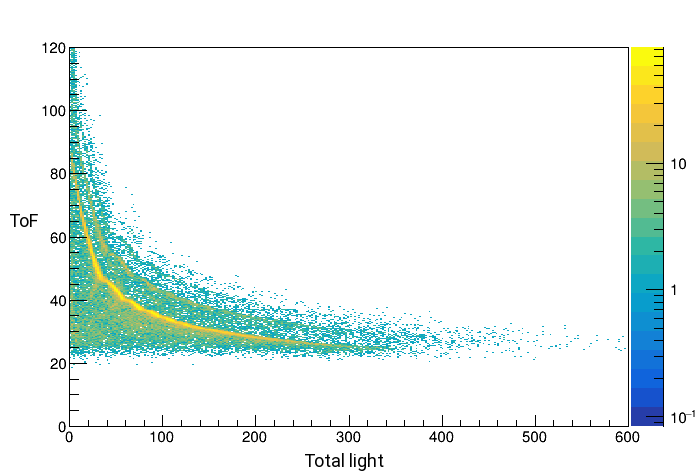}
\caption{Left panel: Time-of-flight (in ns) vs total light deposit in the
NeuLAND demonstrator as obtained from Sn+Sn @ 270 MeV/nucleon measurement at RIKEN;
Right panel: Time-of-flight vs total light collected in 12 double-planes of
NeuLAND from the simulation of Au+Au reaction at 400 MeV/nucleon. In both panels, separated ridges corresponding to proton, deuteron and triton can be clearly seen. In the simulation, also the low populated He region is recognizable (courtesy of I. Ga\v spari\'c).}
\label{fig:neuland}
\end{figure}

Simulations of semi-central Au+Au collisions at 250, 400, 600, 800 and 1000 MeV/nucleon and, for comparison, neutron rich $^{132}$Sn+$^{124}$Sn, $^{124}$Sn+$^{124}$Sn and Pb+Pb  systems at 400, 600 and 800 MeV/nucleon have been carried out by using the same version of the UrQMD transport model used in \cite{Russotto:2016ucm}.  The neutron-to-proton elliptic flow ratio, $v_{2}^{n}/v_{2}^{p}$, at mid-rapidity ($0.4<y_{lab}/y_{proj}<0.6$), with a stiff ($\gamma$=1.5) and a soft ($\gamma$=0.5) parametrizations of the potential part of the $E_{sym}$ for semi-central $(b_{0}<0.54)$ collisions is shown, as a function of the incident beam energy, in the left panel of Fig.~\ref{fig:fig7}. The difference of predictions, made by using soft and stiff choices for the symmetry energy, for this quantity provides a measure of the sensitivity of the proposed observable to the symmetry energy, and is shown in the right panel of the same figure.
The obtained sensitivity is at the level of $\sim$30\% at 250 and 400 MeV/nucleon decreasing at higher beam energies. This reflects the fact that the mean-field contribution decreases at higher energies and the two-body collisions start to dominate. There, multiple rescattering of the nucleons in the more dense interaction region and the stronger contribution of $\Delta$ resonance excitation, with related opening of the pionic degree of freedom, start to play the major role in determining the asymptotic pattern of the neutron ad proton distributions, reducing the strength of the symmetry energy effect.  Nevertheless, up to 1 GeV/nucleon the sensitivity  of the proposed observable is $\sim$15\%, while a measurement can easily reach a $\sim$5\% accuracy, allowing a discrimination between stiff and soft choices. A similar conclusion can be drawn from  a recent paper \cite{WANG2020135249} where UrQMD simulations with 11 selected Skyrme forces were performed. In Fig. 3 of that paper, the ratio between the elliptic flow parameter of free neutrons and protons was plotted as a function of the slope parameter $L$ for Au+Au collision from 400 to 1000 MeV/nucleon. The highest sensitivity was obtained at 400 MeV/nucleon, while the sensitivity at 1000 MeV/nucleon was reduced by a factor 2. According to the results shown in Fig.~\ref{fig:fig7},  the sensitivity of the Au+Au systems is very similar to the one of the other neutron rich systems, as in the case of $^{132}$Sn radioactive ion beams. Heavier systems, such as Pb and Au, allow for smaller statistical uncertainties given the highest neutron and proton multiplicities per event and an equal number of simulated events for each system. This suggest that Au+Au choice still represents one of the best choice for future experiments, given also the detailed studies available in literature, as the ones performed by using the FOPI detector, to be used as reference data for check and comparison. It is also important to stress the differences in trends (slopes) observed in left panel of Fig.~\ref{fig:fig7}. For the soft $E_{sym}$, the ratios increase with the energy while, for the stiff EoS the trend is opposite. This proves the needs for measuring the excitation functions of these observable.\\ 
Similar sensitivity studies have been performed using two additional transport models, IQMD~\cite{Hartnack:1997ez} and TuQMD ~\cite{Cozma:2011nr}, providing similarly strong arguments for extending measurement of flow observable toward higher energies. In a preliminary study, shown in sect. \ref{sec:2323}, the constraints extracted for the slope parameter $L$ by comparing predictions of each of the three transport models to ASY-EOS data, were observed to agree with one another at a $\sim$5-15$\%$ level. Larger discrepancies between model predictions have been observed at higher impact energies, requiring further investigations. It is clear that the divergence of results among the different transport codes constitutes a relevant issue to be investigated and solved, before unambiguous and accurate information regarding the EoS at high-densities can be extracted from experimental data. 
A collaborative effort devoted to understanding differences between model predictions for neutron/proton elliptic flow like observables, in a similar fashion to the one initiated for pionic observables \cite{Xu:2016lue,Zhang:2017esm,Ono:2019ndq,Colonna:2021xuh}, would be extremely beneficial/relevant for a positive outcome of the ASY-EOS experimental program.\\
A strong argument in favor of a new experimental measurement is provided by the possibility of relevant improvements in the experimental setup. In fact, as a relevant novelty, the R3B collaboration at GSI has finalized in these years the building of a new next-generation neutron detector, NeuLAND \cite{R:2021lxa}, being an improved device with respect to the old LAND detector. In fact, the NeuLAND detector features a higher detection efficiency, a higher resolution, and a larger multi-neutron-hit resolving power. This is achieved by a highly granular design of plastic scintillators, avoiding insensitive converter material. The final implementation of the  detector will consist of 3000 individual sub-modules with a size of 5$\times$5$\times$250 cm$^{3}$, arranged in 30 double planes with 100 sub-modules, with alternate horizontal and vertical positioning, providing an active face size of 250$\times$250 cm$^{2}$ and a total depth of 3 m. Given the reached performance, the NeuLAND detector will give a unique opportunity to measure the neutron and LCPs in the same angular regions. The outstanding calorimetric properties of NeuLAND will allow  protons and other hydrogen isotopes to be relatively well separated, and will give access to the neutron vs proton observables. The NeuLAND demonstrator was a part of the S$\pi$IRIT experiment \cite{SpRIT:2020blg} carried out at RIKEN in 2016 and the capability of resolving both protons and neutrons was clearly demonstrated there. The identification plot of hydrogen isotopes in the demonstrator (4 double planes, 40 cm total thickness) is presented in the left panel of Fig. \ref{fig:neuland}.
The p, d, t lines are clearly resolved up to the punch-through energy (about 260 MeV for protons) above which the characteristic back-bendings occur. The 13 double planes of NeuLAND now available, resulting in a total depth of 130 cm, will assure stopping of protons up to about 500 MeV. A simulated identification plot for the Au+Au collisions at 400 MeV/nucleon is presented in the right panel of Fig. \ref{fig:neuland}. Indeed, no punch-through segments are observed at this energy and the p, d, t lines clearly stick out of the secondary reaction and multi-hit background. The simulations include tracking, the secondary reaction losses, multiple Coulomb scattering, light propagation in plastic scintillators and quenching effects. The estimated efficiency for proton identification amounts to about 64\% at 200 MeV and 36\% at 400 MeV. Taking into account the thickness of the NeuLAND calorimeter and the secondary reaction and scattering probability, the estimated efficiencies are still impressive.
The one neutron interaction probability is about 70\% at 400MeV. Taking into account also the reconstruction efficiency a five-neutron event is recognized with correct neutron multiplicity with a probability of about 20 to 30\% (200 to 1000 MeV). \\

\begin{figure}[ht!]
 \begin{minipage}[t]{0.50\linewidth}
 \begin{center}
\includegraphics[width=0.89\textwidth]{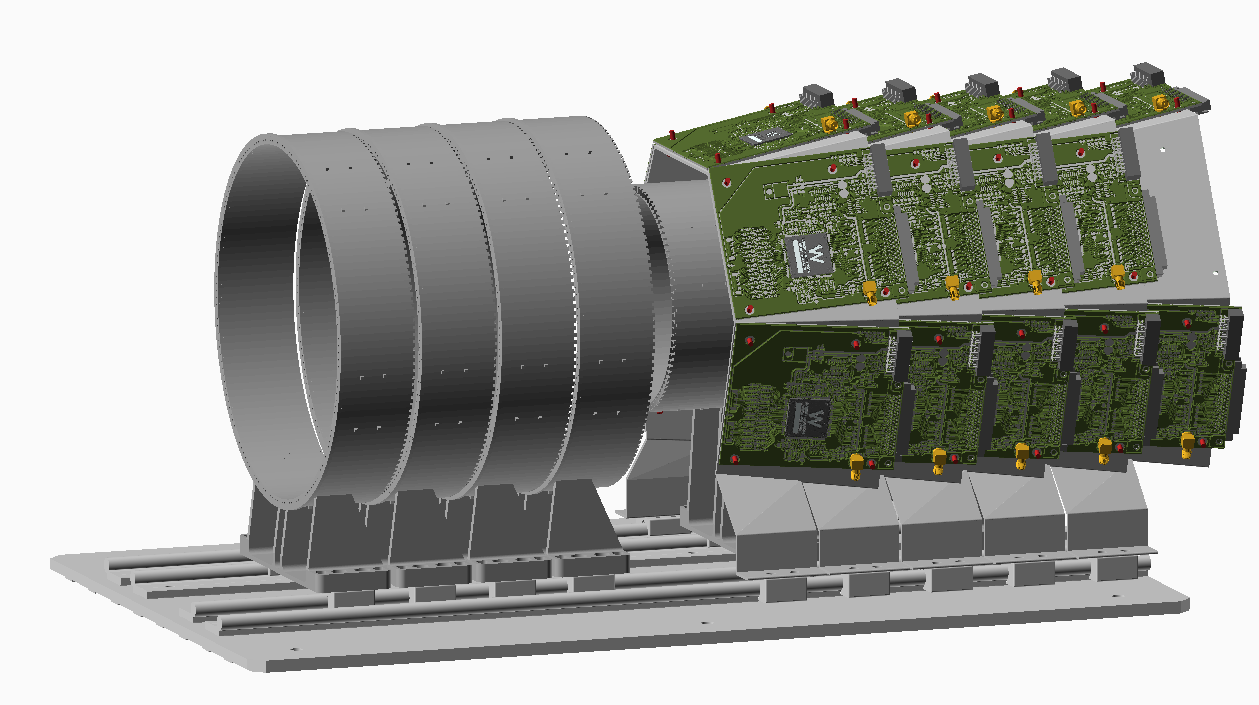}
 \end{center}
 \end{minipage}
 \hfill
 \begin{minipage}[t]{0.45\linewidth}
    \vspace*{-30.5mm}
 \begin{center}
\includegraphics[width=0.95\textwidth]{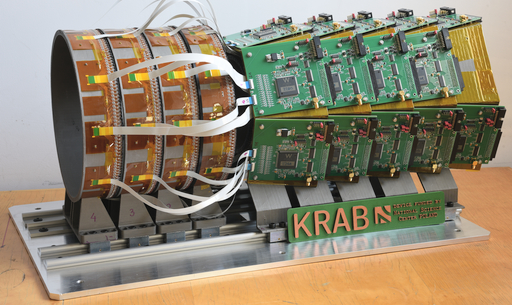}
 \end{center}
 \end{minipage}

\caption{Left panel: The design of the KRAB detector including the CITIROC
boards; Right panel: The actual view of the device as of the end of March 2022.}

\label{fig:krabbing}
\end{figure}

Another advantage of a new experiment will be obtained by using a new device, the KRAkow Barrel (KRAB), explicitly developed  in view of future flow measurements, Fig. \ref{fig:krabbing}. The KRAB, covering polar angles from $30^{\circ}$ to $165^{\circ}$ with $\sim$87\% geometrical efficiency and with $\sim$5\% multi-hit probability, will be used similarly to the MicroBall detector of the ASY-EOS experiment, but will be also able to provide a fast trigger signal based on the multiplicity threshold as well as, thanks to its high segmentation, a very precise azimuthal distributions for charged particles in the c.m. backward hemisphere, indispensable for high resolution estimates of the reaction plane. The main features of the KRAB detector are: 5 rings of 4$\times$4 mm$^{2}$ fast scintillating fibers read out by the SiPMs. This new device will be sufficiently large for radioactive beams and sufficiently small and lightweight in order not to disturb neutrons, having the minimum and maximum internal radii of 6.9 and 11.5 cm and a length of $\sim$50 cm. It will consist of 4$\times$160 segments in forward rings and 96 segments in the backward ring with a total of 736 channels. 
It will then be able to produce a fast trigger based on total multiplicity in an angular region of $\theta>30^{\circ}$, where a strong correlation between the multiplicity and the magnitude of the impact parameter is expected from model predictions. It is expected that KRAB, as NeuLAND,  should greatly improve the quality of the data with respect to the previous ASY-EOS experiment.

Figure \ref{fig:krabbing} presents the design and the current status of construction of the KRAB detector.

In addition to these new devices, the set-up will be completed by the use of CHIMERA CsI(Tl) rings and of a wall of plastic scintillators, like the R$^3$B New Time-of-Flight Wall (TOFD) detector of the R3B collaboration, at small angles, providing information needed to characterize the reaction centrality and reaction plane orientation, and of KraTTA hodoscope at mid-rapidity, placed symmetrically to the NeuLAND detector, but on the opposite side, to measure yields and flows of proton and light clusters.
Using the above described devices, a set-up similar to the ones employed by the ASY-EOS experiment can be configured, but with overall improved performances, being, as said before, a strong reason motivating future experimental efforts.

\section{Conclusions}
\label{sec:24}
 
The study of the nuclear matter EoS is an important topic in modern nuclear physics. It has triggered a considerable amount of works in the last decades on, both, theoretical and experimental sides. Among the latter, studies performed by means of HICs play a relevant role, being able to provide very accurate information. 

In this review, we have focused on that subset of results concerning the high-density part of the EoS, carried out at the beam energies around the 1 GeV/nucleon as typical of the SIS18 accelerator of the GSI laboratory. This field of research has also seen in the last years the arrival of new and very interesting data coming from astrophysical observations of neutron stars by means of gravitational waves and satellite-based X-rays observations. 

The isospin-symmetric contribution of the EoS has been extensively studied by means of HICs. Studies on kaon production showed high-sensitivity and stability against model ingredients, and allowed to get clear indications in favour of a soft EoS. In contrast, pion studies were not effective in constraining the isospin-symmetric contribution of the EoS, due to the difficulties of the transport models to exhaustively reproduce different aspects of the experimental data. The direced flow measurements carried out at LBL, AGS and BNL and at energies between 1 and 10 GeV/nucleon gave further indications for a soft EoS at densities around 3-4 $\rho_0$. The softness of the EoS at densities around 2$\rho_0$ was precisely confirmed by the FOPI experiment, by means of extensive studies on the elliptic flows of protons and light clusters for Au+Au collisions at several beam energies between 400 and 1500 MeV/nucleon. Values of $K_{0}$ compressibility modulus of in the range 190-220 MeV with uncertainties of the order of $\sim20\%$ were obtained and can be considered as a well established result of HIC studies.

In contrast, a full knowledge of the isospin dependent component of the EoS with a comparable accuracy requires additional effort, both on the experimental and theoretical sides. Most of the studies devoted to investigate the symmetry energy at densities below the saturation one show a significant agreement for a $E_{sym}$ of 23-26 MeV around 0.7 $\rho_0$ and values of the $E_{sym}$ slope around $\rho_0$, $L$, of $\sim 70\pm 20$ MeV. The recent work of Lynch and Tsang \cite{lyn21}, evidencing the importance of using each $E_{sym}$ constraints at the specific sensitive density, has given a further strong indication in favour of symmetry energy being well constrained from nuclear structure and reactions studies below saturation density. Latest PREX-II results suggests a stiffer slope of the symmetry energy at $\rho_0$, thus, in tension with the current understanding of the EoS. And this is currently a matter of debate.
However, further works will be advantageous to get an, even better, coherent and exhaustive description of the symmetry energy at sub-saturation densities. Additional work on models is required in order to get a complete and coherent description of the several reaction observables and to reach a better agreement among constraints obtained by using different observables and/or models. Instead, only few works explicitly constraining symmetry energy above the saturation densities by using HIC data have been performed so far. 

First studies based on integrated pion multiplicity measured by the FOPI experiment close to the threshold reached contradictory results when using different models, which could be often traced back to different treatments of in-medium pion production. Later, works of the S$\pi$RIT collaboration and the code-comparison project have allowed to get a better understanding of the requirements for realistic pion production modelling and to select observables, related to the high-kinetic energy part of the pion spectra, able to provide reliable constraint of the symmetry energy at high densities. Recent results, interpreting the high transverse momentum tail of the pion single ratio spectra by using the dcQMD trasport model, indicated $L=79.9\pm37.6$ MeV and a sensitivity around 1.5 $\rho_0$. A confirmation of this result using additional transport models is of great importance and subject of current efforts of the code comparison collaboration.

Elliptic flow studies, based on the ratio of neutron-to-proton and neutron-to-charged particles, proved to be very effective for investigating the high-density behaviour of symmetry energy. Transport model simulation with QMD models indicated the robustness of the elliptic flow ratio observable with respect to many ingredients, such as compressibility of symmetric matter EoS, width of wave-packet, implementation of the optical potential, of the transport models. Extensive Studies using BUU or SMF type of models compared to the existing set of data  have not yet been published, but are necessary for a thorough study of possible model dependence of obtained results.
First studies based on Au+Au data at 400 MeV/nucleon of FOPI-LAND experiment, employing both neutron-to-proton and neutron-to-hydrogen flow ratios,  suggested $L=83\pm26$ MeV when using the UrQMD transport model, and $L=122\pm57$ MeV and $K_{sym}=229\pm363$ MeV when using the TuQMD model. The two results, despite the different approaches used in the two models, have a substantial overlap. A new data set for the same system and energy, with a better statistics, was made available with the ASY-EOS experiment. A first study based on the neutron-to-charged particles elliptic flow ratio and the UrQMD model indicated $L=72\pm13$ MeV. The effectiveness in testing the symmetry energy at densities above the saturation one was also demonstrated. The analysis based on the TuQMD model suggested $L$=85$~\pm$22(exp) $\pm$ 20(th) $\pm$ 12(sys) MeV and $K_{sym}$=96~$\pm$315(exp) $\pm$ 170(th) $\pm$ 166(sys) MeV. In this paper we have also presented  a comparison of results obtained by using IQMD, TuQMD and a new UrQMD version, using a common analysis scheme, investigating the dependence of the slope of the symmetry energy on the $E_{sym,0}$ value. Obtained values for $L$ from the 3 models increase as a function of $E_{sym,0}$ and show a remarkable agreement. For $E_{sym,0}=32$ MeV the resulting $L$ range is about $70_{-25}^{+35}$ MeV. 

ASY-EOS and FOPI data have allowed a HIC based estimation of pressure-density relation for neutron star matter and  radius of a canonical neutron star. An overview of recent progress in neutron star radius measurement has been given where the HIC results have been compared to results obtained from low-mass X-ray binary system, X-ray modelling for millisecond pulsars, gravitational waves and multiple source analyses. The noticeable progress made in the last years has been highlighted with the uncertainty being pushed down to the level of $\sim 5\%$. HIC based results present a remarkable agreement with astrophysical ones and underline the potential of new and even more accurate future studies. In addition, a recent study of the EoS of neutron matter, based on a joint Bayesian analysis of microscopic chiral effective field calculations, astrophysical constraints and FOPI and ASY-EOS constraints has been discussed in some detail. This study has given the most accurate radius value among the reviewed results, $R_{NS} = 12.01_{-0.38}^{+0.37}$ km, with about 3\% precision. This study has shown also that HIC studies play a relevant role in determining the neutron matter EoS in the multi-messenger astronomy era. The need of new  heavy-ion studies has been argued, probing the region around 1-2 $\rho_0$, where theoretical model uncertainties are larger, where laboratory constraints are still scarce, and where HICs have a large impact on the neutron matter EoS. 

In view of that, the potential of new experimental measurements has been discussed. For the case of pion production, more accurate experimental measurements close to production threshold are planned at several laboratories worldwide. Regarding flow, needs of new measurements at energies different from the 400 MeV/nucleon explored until now, would allow to extend the symmetry energy studies to higher densities, constraining both slope and curvature with improved precision.

It is expected that the huge amount of work in this field of research, of which a small but relevant part has been reviewed in this paper, will allow to get even more precise constraints on the EoS in the forthcoming years. 



\begin{acknowledgements}
Work supported by the Polish National Science Centre, contract No. UMO-2017/25/B/ST2/02550, by the French-German Collaboration Agreements 03-45 and 13-70 between IN2P3 - DSM/CEA and GSI.\\
The authors wish to thank I. Ga\v spari\'c (RBI, Zagreb, Croatia) for providing NeuLAND plots in sect. \ref{sec:2332}, QingFeng Li (Institute of Modern Physics, Chinese Academy of Sciences, Lanzhou, China and School of Science, Huzhou University, Huzhou, China) and Yongjia Wang (School of Science, Huzhou University, Huzhou, China) for providing through the years UrQMD calculations, and Wolfgang Trautmann (GSI, Darmstadt, Germany) for his fundamental role in the whole ASY-EOS project. We wish also to thank the  Referee for the critical reading of the manuscript and valuable suggestions.
\end{acknowledgements}

%
%

\bibliographystyle{sn-mathphys.bst}  
\bibliography{NewRev.bib}   

\providecommand{\noopsort}[1]{}\providecommand{\singleletter}[1]{#1}%

\begin{thebibliography}{245}
\ifx \bisbn   \undefined \def \bisbn  #1{ISBN #1}\fi
\ifx \binits  \undefined \def \binits#1{#1}\fi
\ifx \bauthor  \undefined \def \bauthor#1{#1}\fi
\ifx \batitle  \undefined \def \batitle#1{#1}\fi
\ifx \bjtitle  \undefined \def \bjtitle#1{#1}\fi
\ifx \bvolume  \undefined \def \bvolume#1{\textbf{#1}}\fi
\ifx \byear  \undefined \def \byear#1{#1}\fi
\ifx \bissue  \undefined \def \bissue#1{#1}\fi
\ifx \bfpage  \undefined \def \bfpage#1{#1}\fi
\ifx \blpage  \undefined \def \blpage #1{#1}\fi
\ifx \burl  \undefined \def \burl#1{\textsf{#1}}\fi
\ifx \doiurl  \undefined \def \doiurl#1{\url{https://doi.org/#1}}\fi
\ifx \betal  \undefined \def \betal{\textit{et al.}}\fi
\ifx \binstitute  \undefined \def \binstitute#1{#1}\fi
\ifx \binstitutionaled  \undefined \def \binstitutionaled#1{#1}\fi
\ifx \bctitle  \undefined \def \bctitle#1{#1}\fi
\ifx \beditor  \undefined \def \beditor#1{#1}\fi
\ifx \bpublisher  \undefined \def \bpublisher#1{#1}\fi
\ifx \bbtitle  \undefined \def \bbtitle#1{#1}\fi
\ifx \bedition  \undefined \def \bedition#1{#1}\fi
\ifx \bseriesno  \undefined \def \bseriesno#1{#1}\fi
\ifx \blocation  \undefined \def \blocation#1{#1}\fi
\ifx \bsertitle  \undefined \def \bsertitle#1{#1}\fi
\ifx \bsnm \undefined \def \bsnm#1{#1}\fi
\ifx \bsuffix \undefined \def \bsuffix#1{#1}\fi
\ifx \bparticle \undefined \def \bparticle#1{#1}\fi
\ifx \barticle \undefined \def \barticle#1{#1}\fi
\bibcommenthead
\ifx \bconfdate \undefined \def \bconfdate #1{#1}\fi
\ifx \botherref \undefined \def \botherref #1{#1}\fi
\ifx \url \undefined \def \url#1{\textsf{#1}}\fi
\ifx \bchapter \undefined \def \bchapter#1{#1}\fi
\ifx \bbook \undefined \def \bbook#1{#1}\fi
\ifx \bcomment \undefined \def \bcomment#1{#1}\fi
\ifx \oauthor \undefined \def \oauthor#1{#1}\fi
\ifx \citeauthoryear \undefined \def \citeauthoryear#1{#1}\fi
\ifx \endbibitem  \undefined \def \endbibitem {}\fi
\ifx \bconflocation  \undefined \def \bconflocation#1{#1}\fi
\ifx \arxivurl  \undefined \def \arxivurl#1{\textsf{#1}}\fi
\csname PreBibitemsHook\endcsname

\bibitem{refId0}
\begin{barticle}
\bauthor{\bsnm{{Li, Bao-An}}},
\bauthor{\bsnm{{Ramos, \`Angels}}},
\bauthor{\bsnm{{Verde, Giuseppe}}},
\bauthor{\bsnm{{Vida\~na, Isaac}}}:
\batitle{Topical issue on nuclear symmetry energy}.
\bjtitle{Eur. Phys. J. A}
\bvolume{50}(\bissue{2}),
\bfpage{9}
(\byear{2014}).
\doiurl{10.1140/epja/i2014-14009-x}
\end{barticle}
\endbibitem

\bibitem{BARAN2005335}
\begin{barticle}
\bauthor{\bsnm{Baran}, \binits{V.}},
\bauthor{\bsnm{Colonna}, \binits{M.}},
\bauthor{\bsnm{Greco}, \binits{V.}},
\bauthor{\bsnm{{Di Toro}}, \binits{M.}}:
\batitle{Reaction dynamics with exotic nuclei}.
\bjtitle{Physics Reports}
\bvolume{410}(\bissue{5}),
\bfpage{335}--\blpage{466}
(\byear{2005}).
\doiurl{10.1016/j.physrep.2004.12.004}
\end{barticle}
\endbibitem

\bibitem{Li:2008gp}
\begin{barticle}
\bauthor{\bsnm{Li}, \binits{B.-A.}},
\bauthor{\bsnm{Chen}, \binits{L.-W.}},
\bauthor{\bsnm{Ko}, \binits{C.M.}}:
\batitle{Recent progress and new challenges in isospin physics with heavy-ion
  reactions}.
\bjtitle{Physics Reports}
\bvolume{464}(\bissue{4}),
\bfpage{113}--\blpage{281}
(\byear{2008}).
\doiurl{10.1016/j.physrep.2008.04.005}
\end{barticle}
\endbibitem

\bibitem{Burgio:2020fom}
\begin{botherref}
\oauthor{\bsnm{Burgio}, \binits{G.F.}},
\oauthor{\bsnm{Vida\~na}, \binits{I.}}:
{The Equation of State of Nuclear Matter: From Finite Nuclei to Neutron Stars}.
Universe
\textbf{6}(8)
(2020).
\doiurl{10.3390/universe6080119}
\end{botherref}
\endbibitem

\bibitem{Piekarewicz:2018gtd}
\begin{barticle}
\bauthor{\bsnm{Piekarewicz}, \binits{J.}}:
\batitle{{Nuclear Astrophysics in the Multimessenger Era: A Partnership Made in
  Heaven}}.
\bjtitle{Acta Phys. Polon. B}
\bvolume{50}(\bissue{3}),
\bfpage{239}
(\byear{2019})
{\href{https://arxiv.org/abs/1812.04438}{{arXiv:1812.04438}}}
{[nucl-th]}.
\doiurl{10.5506/APhysPolB.50.239}
\end{barticle}
\endbibitem

\bibitem{BLAIZOT1980171}
\begin{barticle}
\bauthor{\bsnm{Blaizot}, \binits{J.P.}}:
\batitle{Nuclear compressibilities}.
\bjtitle{Physics Reports}
\bvolume{64}(\bissue{4}),
\bfpage{171}--\blpage{248}
(\byear{1980}).
\doiurl{10.1016/0370-1573(80)90001-0}
\end{barticle}
\endbibitem

\bibitem{Youngblood-PRL.82.691}
\begin{barticle}
\bauthor{\bsnm{Youngblood}, \binits{D.H.}},
\bauthor{\bsnm{Clark}, \binits{H.L.}},
\bauthor{\bsnm{Lui}, \binits{Y.-W.}}:
\batitle{Incompressibility of nuclear matter from the giant monopole
  resonance}.
\bjtitle{Phys. Rev. Lett.}
\bvolume{82},
\bfpage{691}--\blpage{694}
(\byear{1999}).
\doiurl{10.1103/PhysRevLett.82.691}
\end{barticle}
\endbibitem

\bibitem{Fuchs:2005yn}
\begin{barticle}
\bauthor{\bsnm{Fuchs}, \binits{C.}},
\bauthor{\bsnm{Wolter}, \binits{H.H.}}:
\batitle{{Modelization of the EOS}}.
\bjtitle{Eur. Phys. J. A}
\bvolume{30},
\bfpage{5}--\blpage{21}
(\byear{2006}).
\doiurl{10.1140/epja/i2005-10313-x}
\end{barticle}
\endbibitem

\bibitem{Dani02Esymm}
\begin{barticle}
\bauthor{\bsnm{Danielewicz}, \binits{P.}},
\bauthor{\bsnm{Lacey}, \binits{R.}},
\bauthor{\bsnm{Lynch}, \binits{W.G.}}:
\batitle{{Determination of the equation of state of dense matter}}.
\bjtitle{Science}
\bvolume{298},
\bfpage{1592}
(\byear{2002}).
\doiurl{10.1126/science.1078070}
\end{barticle}
\endbibitem

\bibitem{Brown:2000pd}
\begin{barticle}
\bauthor{\bsnm{Brown}, \binits{B.A.}}:
\batitle{{Neutron radii in nuclei and the neutron equation of state}}.
\bjtitle{Phys. Rev. Lett.}
\bvolume{85},
\bfpage{5296}--\blpage{5299}
(\byear{2000}).
\doiurl{10.1103/PhysRevLett.85.5296}
\end{barticle}
\endbibitem

\bibitem{Roca-Maza:2011qcr}
\begin{barticle}
\bauthor{\bsnm{Roca-Maza}, \binits{X.}},
\bauthor{\bsnm{Centelles}, \binits{M.}},
\bauthor{\bsnm{Vi\~nas}, \binits{X.}},
\bauthor{\bsnm{Warda}, \binits{M.}}:
\batitle{{Neutron Skin of $^{208}\mathrm{Pb}$, Nuclear Symmetry Energy, and the
  Parity Radius Experiment}}.
\bjtitle{Phys. Rev. Lett.}
\bvolume{106},
\bfpage{252501}
(\byear{2011}).
\doiurl{10.1103/PhysRevLett.106.252501}
\end{barticle}
\endbibitem

\bibitem{PhysRevLett.102.122701}
\begin{barticle}
\bauthor{\bsnm{Tsang}, \binits{M.B.}},
\bauthor{\bsnm{Zhang}, \binits{Y.}},
\bauthor{\bsnm{Danielewicz}, \binits{P.}},
\bauthor{\bsnm{Famiano}, \binits{M.}},
\bauthor{\bsnm{Li}, \binits{Z.}},
\bauthor{\bsnm{Lynch}, \binits{W.G.}},
\bauthor{\bsnm{Steiner}, \binits{A.W.}}:
\batitle{{Constraints on the Density Dependence of the Symmetry Energy}}.
\bjtitle{Phys. Rev. Lett.}
\bvolume{102},
\bfpage{122701}
(\byear{2009}).
\doiurl{10.1103/PhysRevLett.102.122701}
\end{barticle}
\endbibitem

\bibitem{STEINER2005325}
\begin{barticle}
\bauthor{\bsnm{Steiner}, \binits{A.W.}},
\bauthor{\bsnm{Prakash}, \binits{M.}},
\bauthor{\bsnm{Lattimer}, \binits{J.M.}},
\bauthor{\bsnm{Ellis}, \binits{P.J.}}:
\batitle{Isospin asymmetry in nuclei and neutron stars}.
\bjtitle{Physics Reports}
\bvolume{411}(\bissue{6}),
\bfpage{325}--\blpage{375}
(\byear{2005}).
\doiurl{10.1016/j.physrep.2005.02.004}
\end{barticle}
\endbibitem

\bibitem{Fattoyev-PRL.120.172702}
\begin{barticle}
\bauthor{\bsnm{Fattoyev}, \binits{F.J.}},
\bauthor{\bsnm{Piekarewicz}, \binits{J.}},
\bauthor{\bsnm{Horowitz}, \binits{C.J.}}:
\batitle{{Neutron Skins and Neutron Stars in the Multimessenger Era}}.
\bjtitle{Phys. Rev. Lett.}
\bvolume{120},
\bfpage{172702}
(\byear{2018}).
\doiurl{10.1103/PhysRevLett.120.172702}
\end{barticle}
\endbibitem

\bibitem{Horowitz:2014bja}
\begin{barticle}
\bauthor{\bsnm{Horowitz}, \binits{C.J.}},
\bauthor{\bsnm{Brown}, \binits{E.F.}},
\bauthor{\bsnm{Kim}, \binits{Y.}},
\bauthor{\bsnm{Lynch}, \binits{W.G.}},
\bauthor{\bsnm{Michaels}, \binits{R.}},
\bauthor{\bsnm{Ono}, \binits{A.}},
\bauthor{\bsnm{Piekarewicz}, \binits{J.}},
\bauthor{\bsnm{Tsang}, \binits{M.B.}},
\bauthor{\bsnm{Wolter}, \binits{H.H.}}:
\batitle{{A way forward in the study of the symmetry energy: experiment,
  theory, and observation}}.
\bjtitle{J. Phys. G}
\bvolume{41},
\bfpage{093001}
(\byear{2014})
{\href{https://arxiv.org/abs/1401.5839}{{arXiv:1401.5839}}}
{[nucl-th]}.
\doiurl{10.1088/0954-3899/41/9/093001}
\end{barticle}
\endbibitem

\bibitem{Drischler:2020yad}
\begin{barticle}
\bauthor{\bsnm{Drischler}, \binits{C.}},
\bauthor{\bsnm{Melendez}, \binits{J.A.}},
\bauthor{\bsnm{Furnstahl}, \binits{R.J.}},
\bauthor{\bsnm{Phillips}, \binits{D.R.}}:
\batitle{{Quantifying uncertainties and correlations in the nuclear-matter
  equation of state}}.
\bjtitle{Phys. Rev. C}
\bvolume{102}(\bissue{5}),
\bfpage{054315}
(\byear{2020}).
\doiurl{10.1103/PhysRevC.102.054315}
\end{barticle}
\endbibitem

\bibitem{LIGOScientific:2017vwq}
\begin{barticle}
\bauthor{\bsnm{Abbott}, \binits{B.P.}}, \betal:
\batitle{{GW170817: Observation of Gravitational Waves from a Binary Neutron
  Star Inspiral}}.
\bjtitle{Phys. Rev. Lett.}
\bvolume{119}(\bissue{16}),
\bfpage{161101}
(\byear{2017}).
\doiurl{10.1103/PhysRevLett.119.161101}
\end{barticle}
\endbibitem

\bibitem{LIGOScientific:2020aai}
\begin{barticle}
\bauthor{\bsnm{Abbott}, \binits{B.P.}}, \betal:
\batitle{{GW190425: Observation of a Compact Binary Coalescence with Total Mass
  $\sim 3.4 M_{\odot}$}}.
\bjtitle{Astrophys. Journal Letters}
\bvolume{892}(\bissue{1}),
\bfpage{3}
(\byear{2020}).
\doiurl{10.3847/2041-8213/ab75f5}
\end{barticle}
\endbibitem

\bibitem{Riley:2019yda}
\begin{barticle}
\bauthor{\bsnm{Riley}, \binits{T.E.}}, \betal:
\batitle{{A ${NICER}$ View of {PSR} J0030+0451: Millisecond Pulsar Parameter
  Estimation}}.
\bjtitle{The Astrophysical Journal}
\bvolume{887}(\bissue{1}),
\bfpage{21}
(\byear{2019}).
\doiurl{10.3847/2041-8213/ab481c}
\end{barticle}
\endbibitem

\bibitem{Miller:2019cac}
\begin{barticle}
\bauthor{\bsnm{Miller}, \binits{M.C.}}, \betal:
\batitle{{ PSR J0030+0451 Mass and Radius from NICER Data and Implications for
  the Properties of Neutron Star Matter}}.
\bjtitle{The Astrophysical Journal}
\bvolume{887}(\bissue{1}),
\bfpage{24}
(\byear{2019}).
\doiurl{10.3847/2041-8213/ab50c5}
\end{barticle}
\endbibitem

\bibitem{Ghosh:2021bvw}
\begin{barticle}
\bauthor{\bsnm{Ghosh}, \binits{S.}},
\bauthor{\bsnm{Chatterjee}, \binits{D.}},
\bauthor{\bsnm{Schaffner-Bielich}, \binits{J.}}:
\batitle{{Imposing multi-physics constraints at different densities on the
  neutron Star Equation of State}}.
\bjtitle{Eur. Phys. J. A}
\bvolume{58}(\bissue{3}),
\bfpage{37}
(\byear{2022}).
\doiurl{10.1140/epja/s10050-022-00679-w}
\end{barticle}
\endbibitem

\bibitem{Zhang:2018bwq}
\begin{barticle}
\bauthor{\bsnm{Zhang}, \binits{N.-B.}},
\bauthor{\bsnm{Li}, \binits{B.-A.}}:
\batitle{{{Extracting Nuclear Symmetry Energies at High Densities from
  Observations of Neutron Stars and Gravitational Waves}}}.
\bjtitle{Eur. Phys. J. A}
\bvolume{55}(\bissue{3}),
\bfpage{39}
(\byear{2019}).
\doiurl{10.1140/epja/i2019-12700-0}
\end{barticle}
\endbibitem

\bibitem{Huth:2021bsp}
\begin{barticle}
\bauthor{\bsnm{Huth}, \binits{S.}}, \betal:
\batitle{{Constraining Neutron-Star Matter with Microscopic and Macroscopic
  Collisions}}.
\bjtitle{Nature}
\bvolume{606},
\bfpage{276}--\blpage{280}
(\byear{2022})
{\href{https://arxiv.org/abs/2107.06229}{{arXiv:2107.06229}}}
{[nucl-th]}.
\doiurl{10.1038/s41586-022-04750-w}
\end{barticle}
\endbibitem

\bibitem{DeFilippo:2012qd}
\begin{barticle}
\bauthor{\bsnm{De~Filippo}, \binits{E.}},
\bauthor{\bsnm{Pagano}, \binits{A.}},
\bauthor{\bsnm{Russotto}, \binits{P.}}, \betal:
\batitle{{Correlations between emission timescale of fragments and isospin
  dynamics in ${}^{124}$Sn+${}^{64}$Ni and ${}^{112}$Sn+${}^{58}$Ni reactions
  at 35 $A$ MeV}}.
\bjtitle{Phys. Rev. C}
\bvolume{86},
\bfpage{014610}
(\byear{2012}).
\doiurl{10.1103/PhysRevC.86.014610}
\end{barticle}
\endbibitem

\bibitem{Danielewicz:2013upa}
\begin{barticle}
\bauthor{\bsnm{Danielewicz}, \binits{P.}},
\bauthor{\bsnm{Lee}, \binits{J.}}:
\batitle{{Symmetry Energy II: Isobaric Analog States}}.
\bjtitle{Nucl. Phys. A}
\bvolume{922},
\bfpage{1}--\blpage{70}
(\byear{2014}).
\doiurl{10.1016/j.nuclphysa.2013.11.005}
\end{barticle}
\endbibitem

\bibitem{Brown:2013mga}
\begin{barticle}
\bauthor{\bsnm{Brown}, \binits{B.A.}}:
\batitle{{{Constraints on the Skyrme Equations of State from Properties of
  Doubly Magic Nuclei}}}.
\bjtitle{Phys. Rev. Lett.}
\bvolume{111}(\bissue{23}),
\bfpage{232502}
(\byear{2013}).
\doiurl{10.1103/PhysRevLett.111.232502}
\end{barticle}
\endbibitem

\bibitem{Zhang:2013wna}
\begin{barticle}
\bauthor{\bsnm{Zhang}, \binits{Z.}},
\bauthor{\bsnm{Chen}, \binits{L.-W.}}:
\batitle{{Constraining the symmetry energy at subsaturation densities using
  isotope binding energy difference and neutron skin thickness}}.
\bjtitle{Phys. Lett. B}
\bvolume{726},
\bfpage{234}--\blpage{238}
(\byear{2013}).
\doiurl{10.1016/j.physletb.2013.08.002}
\end{barticle}
\endbibitem

\bibitem{Tamii-PhysRevLett.107.062502}
\begin{barticle}
\bauthor{\bsnm{Tamii}, \binits{A.}},
\bauthor{\bsnm{Poltoratska}, \binits{I.}},
\bauthor{\bparticle{von} \bsnm{Neumann-Cosel}, \binits{P.}},
\bauthor{\bsnm{Fujita}, \binits{Y.}}, \betal:
\batitle{Complete electric dipole response and the neutron skin in
  $^{208}\mathrm{Pb}$}.
\bjtitle{Phys. Rev. Lett.}
\bvolume{107},
\bfpage{062502}
(\byear{2011}).
\doiurl{10.1103/PhysRevLett.107.062502}
\end{barticle}
\endbibitem

\bibitem{Pieka-PhysRevC.85.041302}
\begin{barticle}
\bauthor{\bsnm{Piekarewicz}, \binits{J.}},
\bauthor{\bsnm{Agrawal}, \binits{B.K.}},
\bauthor{\bsnm{Col\`o}, \binits{G.}},
\bauthor{\bsnm{Nazarewicz}, \binits{W.}},
\bauthor{\bsnm{Paar}, \binits{N.}},
\bauthor{\bsnm{Reinhard}, \binits{P.-G.}},
\bauthor{\bsnm{Roca-Maza}, \binits{X.}},
\bauthor{\bsnm{Vretenar}, \binits{D.}}:
\batitle{Electric dipole polarizability and the neutron skin}.
\bjtitle{Phys. Rev. C}
\bvolume{85},
\bfpage{041302}
(\byear{2012}).
\doiurl{10.1103/PhysRevC.85.041302}
\end{barticle}
\endbibitem

\bibitem{Li-PhysRevLett.99.162503}
\begin{barticle}
\bauthor{\bsnm{Li}, \binits{T.}}, \betal:
\batitle{{Isotopic Dependence of the Giant Monopole Resonance in the Even-$A$
  $^{112-124}\mathrm{Sn}$ Isotopes and the Asymmetry Term in Nuclear
  Incompressibility}}.
\bjtitle{Phys. Rev. Lett.}
\bvolume{99},
\bfpage{162503}
(\byear{2007}).
\doiurl{10.1103/PhysRevLett.99.162503}
\end{barticle}
\endbibitem

\bibitem{Li-PhysRevC.81.034309}
\begin{barticle}
\bauthor{\bsnm{Li}, \binits{T.}}, \betal:
\batitle{{Isoscalar giant resonances in the Sn nuclei and implications for the
  asymmetry term in the nuclear-matter incompressibility}}.
\bjtitle{Phys. Rev. C}
\bvolume{81},
\bfpage{034309}
(\byear{2010}).
\doiurl{10.1103/PhysRevC.81.034309}
\end{barticle}
\endbibitem

\bibitem{Amorini:2008sm}
\begin{barticle}
\bauthor{\bsnm{Amorini}, \binits{F.}}, \betal:
\batitle{{{Isospin Dependence of Incomplete Fusion Reactions at 25
  MeV/Nucleon}}}.
\bjtitle{Phys. Rev. Lett.}
\bvolume{102},
\bfpage{112701}
(\byear{2009}).
\doiurl{10.1103/PhysRevLett.102.112701}
\end{barticle}
\endbibitem

\bibitem{lyn21}
\begin{barticle}
\bauthor{\bsnm{Lynch}, \binits{W.G.}},
\bauthor{\bsnm{Tsang}, \binits{M.B.}}:
\batitle{Decoding the density dependence of the nuclear symmetry energy}.
\bjtitle{Physics Letters B}
\bvolume{830},
\bfpage{137098}
(\byear{2022}).
\doiurl{10.1016/j.physletb.2022.137098}.
\bcomment{arXiv:2106.10119}
\end{barticle}
\endbibitem

\bibitem{adh21}
\begin{barticle}
\bauthor{\bsnm{Adhikari}, \binits{D.}}, \betal:
\batitle{{Accurate Determination of the Neutron Skin Thickness of
  $^{208}\mathrm{Pb}$ through Parity-Violation in Electron Scattering}}.
\bjtitle{Phys. Rev. Lett.}
\bvolume{126},
\bfpage{172502}
(\byear{2021}).
\doiurl{10.1103/PhysRevLett.126.172502}
\end{barticle}
\endbibitem

\bibitem{SRIT:2021gcy}
\begin{barticle}
\bauthor{\bsnm{Estee}, \binits{J.}}, \betal:
\batitle{{Probing the Symmetry Energy with the Spectral Pion Ratio}}.
\bjtitle{Phys. Rev. Lett.}
\bvolume{126}(\bissue{16}),
\bfpage{162701}
(\byear{2021}).
\doiurl{10.1103/PhysRevLett.126.162701}
\end{barticle}
\endbibitem

\bibitem{Russotto:2016ucm}
\begin{barticle}
\bauthor{\bsnm{Russotto}, \binits{P.}}, \betal:
\batitle{{Results of the ASY-EOS experiment at GSI: The symmetry energy at
  suprasaturation density}}.
\bjtitle{Phys. Rev. C}
\bvolume{94}(\bissue{3}),
\bfpage{034608}
(\byear{2016}).
\doiurl{10.1103/PhysRevC.94.034608}
\end{barticle}
\endbibitem

\bibitem{Xiao:2008vm}
\begin{barticle}
\bauthor{\bsnm{Xiao}, \binits{Z.}},
\bauthor{\bsnm{Li}, \binits{B.-A.}},
\bauthor{\bsnm{Chen}, \binits{L.-W.}},
\bauthor{\bsnm{Yong}, \binits{G.-C.}},
\bauthor{\bsnm{Zhang}, \binits{M.}}:
\batitle{{Circumstantial Evidence for a Soft Nuclear Symmetry Energy at
  Suprasaturation Densities}}.
\bjtitle{Phys. Rev. Lett.}
\bvolume{102},
\bfpage{062502}
(\byear{2009}).
\doiurl{10.1103/PhysRevLett.102.062502}
\end{barticle}
\endbibitem

\bibitem{Ferini:2006je}
\begin{barticle}
\bauthor{\bsnm{Ferini}, \binits{G.}},
\bauthor{\bsnm{Gaitanos}, \binits{T.}},
\bauthor{\bsnm{Colonna}, \binits{M.}},
\bauthor{\bsnm{Di~Toro}, \binits{M.}},
\bauthor{\bsnm{Wolter}, \binits{H.H.}}:
\batitle{{Isospin effects on sub-threshold kaon production at intermediate
  energies}}.
\bjtitle{Phys. Rev. Lett.}
\bvolume{97},
\bfpage{202301}
(\byear{2006}).
\doiurl{10.1103/PhysRevLett.97.202301}
\end{barticle}
\endbibitem

\bibitem{Russotto:2011hq}
\begin{barticle}
\bauthor{\bsnm{Russotto}, \binits{P.}}, \betal:
\batitle{{Symmetry energy from elliptic flow in $^{197}$Au $+ ^{197}$Au}}.
\bjtitle{Phys. Lett. B}
\bvolume{697},
\bfpage{471}--\blpage{476}
(\byear{2011}).
\doiurl{10.1016/j.physletb.2011.02.033}
\end{barticle}
\endbibitem

\bibitem{FOPI:2007gvb}
\begin{barticle}
\bauthor{\bsnm{Lopez}, \binits{X.}}, \betal:
\batitle{{Isospin dependence of relative yields of K$^+$ and K$^0$ mesons at
  1.528 AGeV}}.
\bjtitle{Phys. Rev. C}
\bvolume{75},
\bfpage{011901}
(\byear{2007}).
\doiurl{10.1103/PhysRevC.75.011901}
\end{barticle}
\endbibitem

\bibitem{Xiao:2009zza}
\begin{barticle}
\bauthor{\bsnm{Xiao}, \binits{Z.}},
\bauthor{\bsnm{Li}, \binits{B.-A.}}, \betal:
\batitle{{Circumstantial Evidence for a Soft Nuclear Symmetry Energy at
  Suprasaturation Densities}}.
\bjtitle{Phys.Rev.Lett.}
\bvolume{102},
\bfpage{062502}
(\byear{2009}).
\doiurl{10.1103/PhysRevLett.102.062502}
\end{barticle}
\endbibitem

\bibitem{Feng:2009am}
\begin{barticle}
\bauthor{\bsnm{Feng}, \binits{Z.-Q.}},
\bauthor{\bsnm{Jin}, \binits{G.-M.}}:
\batitle{{Probing high-density behavior of symmetry energy from pion emission
  in heavy-ion collisions}}.
\bjtitle{Phys. Lett. B}
\bvolume{683},
\bfpage{140}
(\byear{2010}).
\doiurl{10.1016/j.physletb.2009.12.006}
\end{barticle}
\endbibitem

\bibitem{Xie:2013np}
\begin{barticle}
\bauthor{\bsnm{Xie}, \binits{W.-J.}},
\bauthor{\bsnm{Su}, \binits{J.}},
\bauthor{\bsnm{Zhu}, \binits{L.}},
\bauthor{\bsnm{Zhang}, \binits{F.-S.}}:
\batitle{{Symmetry energy and pion production in the Boltzmann-Langevin
  approach}}.
\bjtitle{Phys. Lett. B}
\bvolume{718},
\bfpage{1510}
(\byear{2013}).
\doiurl{10.1016/j.physletb.2012.12.021}
\end{barticle}
\endbibitem

\bibitem{Hong:2013yva}
\begin{barticle}
\bauthor{\bsnm{Hong}, \binits{J.}},
\bauthor{\bsnm{Danielewicz}, \binits{P.}}:
\batitle{{Subthreshold pion production within a transport description of
  central Au + Au collisions}}.
\bjtitle{Phys. Rev. C}
\bvolume{90}(\bissue{2}),
\bfpage{024605}
(\byear{2014}).
\doiurl{10.1103/PhysRevC.90.024605}
\end{barticle}
\endbibitem

\bibitem{Xu:2016lue}
\begin{barticle}
\bauthor{\bsnm{Xu}, \binits{J.}}, \betal:
\batitle{{Understanding transport simulations of heavy-ion collisions at 100A
  and 400A MeV: Comparison of heavy-ion transport codes under controlled
  conditions}}.
\bjtitle{Phys. Rev. C}
\bvolume{93}(\bissue{4}),
\bfpage{044609}
(\byear{2016}).
\doiurl{10.1103/PhysRevC.93.044609}
\end{barticle}
\endbibitem

\bibitem{Zhang:2017esm}
\begin{barticle}
\bauthor{\bsnm{Zhang}, \binits{Y.-X.}}, \betal:
\batitle{{Comparison of heavy-ion transport simulations: Collision integral in
  a box}}.
\bjtitle{Phys. Rev. C}
\bvolume{97}(\bissue{3}),
\bfpage{034625}
(\byear{2018}).
\doiurl{10.1103/PhysRevC.97.034625}
\end{barticle}
\endbibitem

\bibitem{Ono:2019ndq}
\begin{barticle}
\bauthor{\bsnm{Ono}, \binits{A.}}, \betal:
\batitle{{Comparison of heavy-ion transport simulations: Collision integral
  with pions and \ensuremath{\Delta} resonances in a box}}.
\bjtitle{Phys. Rev. C}
\bvolume{100}(\bissue{4}),
\bfpage{044617}
(\byear{2019}).
\doiurl{10.1103/PhysRevC.100.044617}
\end{barticle}
\endbibitem

\bibitem{Colonna:2021xuh}
\begin{barticle}
\bauthor{\bsnm{Colonna}, \binits{M.}}, \betal:
\batitle{{Comparison of heavy-ion transport simulations: Mean-field dynamics in
  a box}}.
\bjtitle{Phys. Rev. C}
\bvolume{104}(\bissue{2}),
\bfpage{024603}
(\byear{2021}).
\doiurl{10.1103/PhysRevC.104.024603}
\end{barticle}
\endbibitem

\bibitem{TMEP:2022xjg}
\begin{barticle}
\bauthor{\bsnm{Wolter}, \binits{H.}}, \betal:
\batitle{{Transport model comparison studies of intermediate-energy heavy-ion
  collisions}}.
\bjtitle{Prog. Part. Nucl. Phys.}
\bvolume{125},
\bfpage{103962}
(\byear{2022})
{\href{https://arxiv.org/abs/2202.06672}{{arXiv:2202.06672}}}
{[nucl-th]}.
\doiurl{10.1016/j.ppnp.2022.103962}
\end{barticle}
\endbibitem

\bibitem{SpRIT:2020blg}
\begin{barticle}
\bauthor{\bsnm{Jhang}, \binits{G.}}, \betal:
\batitle{{Symmetry energy investigation with pion production from Sn+Sn
  systems}}.
\bjtitle{Phys. Lett. B}
\bvolume{813},
\bfpage{136016}
(\byear{2021}).
\doiurl{10.1016/j.physletb.2020.136016}
\end{barticle}
\endbibitem

\bibitem{Li:2002qx}
\begin{barticle}
\bauthor{\bsnm{Li}, \binits{B.-A.}}:
\batitle{{Probing the high density behavior of nuclear symmetry energy with
  high-energy heavy ion collisions}}.
\bjtitle{Phys. Rev. Lett.}
\bvolume{88},
\bfpage{192701}
(\byear{2002}).
\doiurl{10.1103/PhysRevLett.88.192701}
\end{barticle}
\endbibitem

\bibitem{Cozma:2013sja}
\begin{barticle}
\bauthor{\bsnm{Cozma}, \binits{M.D.}},
\bauthor{\bsnm{Leifels}, \binits{Y.}},
\bauthor{\bsnm{Trautmann}, \binits{W.}},
\bauthor{\bsnm{Li}, \binits{Q.}},
\bauthor{\bsnm{Russotto}, \binits{P.}}:
\batitle{{Toward a model-independent constraint of the high-density dependence
  of the symmetry energy}}.
\bjtitle{Phys. Rev. C}
\bvolume{88}(\bissue{4}),
\bfpage{044912}
(\byear{2013}).
\doiurl{10.1103/PhysRevC.88.044912}
\end{barticle}
\endbibitem

\bibitem{Russotto:2014EPJA}
\begin{barticle}
\bauthor{\bsnm{{Russotto, P.}}},
\bauthor{\bsnm{{Cozma, M. D.}}},
\bauthor{\bsnm{{Le F\`evre, A.}}},
\bauthor{\bsnm{{Leifels, Y.}}},
\bauthor{\bsnm{{Lemmon, R.}}},
\bauthor{\bsnm{{Li, Q.}}},
\bauthor{\bsnm{{Lukasik, J.}}},
\bauthor{\bsnm{{Trautmann, W.}}}:
\batitle{Flow probe of symmetry energy in relativistic heavy-ion reactions}.
\bjtitle{Eur. Phys. J. A}
\bvolume{50}(\bissue{2}),
\bfpage{38}
(\byear{2014}).
\doiurl{10.1140/epja/i2014-14038-5}
\end{barticle}
\endbibitem

\bibitem{MDCozEPJA18}
\begin{barticle}
\bauthor{\bsnm{{Cozma, M. D.}}}:
\batitle{Feasibility of constraining the curvature parameter of the symmetry
  energy using elliptic flow data}.
\bjtitle{Eur. Phys. J. A}
\bvolume{54}(\bissue{3}),
\bfpage{40}
(\byear{2018}).
\doiurl{10.1140/epja/i2018-12470-1}
\end{barticle}
\endbibitem

\bibitem{LAND:1991ffr}
\begin{barticle}
\bauthor{\bsnm{Blaich}, \binits{T.}}, \betal:
\batitle{{A Large area detector for high-energy neutrons}}.
\bjtitle{Nucl. Instrum. Meth. A}
\bvolume{314},
\bfpage{136}--\blpage{154}
(\byear{1992}).
\doiurl{10.1016/0168-9002(92)90507-Z}
\end{barticle}
\endbibitem

\bibitem{R:2021lxa}
\begin{barticle}
\bauthor{\bsnm{Boretzky}, \binits{K.}}, \betal:
\batitle{{NeuLAND: The high-resolution neutron time-of-flight spectrometer for
  R3B at FAIR}}.
\bjtitle{Nucl. Instrum. Meth. A}
\bvolume{1014},
\bfpage{165701}
(\byear{2021}).
\doiurl{10.1016/j.nima.2021.165701}
\end{barticle}
\endbibitem

\bibitem{Nakamura:2015phw}
\begin{barticle}
\bauthor{\bsnm{Nakamura}, \binits{T.}},
\bauthor{\bsnm{Kondo}, \binits{Y.}}:
\batitle{{Large acceptance spectrometers for invariant mass spectroscopy of
  exotic nuclei and future developments}}.
\bjtitle{Nucl. Instrum. Meth. B}
\bvolume{376},
\bfpage{156}--\blpage{161}
(\byear{2016})
{\href{https://arxiv.org/abs/1512.08380}{{arXiv:1512.08380}}}
{[physics.ins-det]}.
\doiurl{10.1016/j.nimb.2016.01.003}
\end{barticle}
\endbibitem

\bibitem{Bertsch:1988ik}
\begin{barticle}
\bauthor{\bsnm{Bertsch}, \binits{G.F.}},
\bauthor{\bsnm{Das~Gupta}, \binits{S.}}:
\batitle{{A Guide to microscopic models for intermediate-energy heavy ion
  collisions}}.
\bjtitle{Phys. Rept.}
\bvolume{160},
\bfpage{189}--\blpage{233}
(\byear{1988}).
\doiurl{10.1016/0370-1573(88)90170-6}
\end{barticle}
\endbibitem

\bibitem{Bonasera:1994zz}
\begin{barticle}
\bauthor{\bsnm{Bonasera}, \binits{A.}},
\bauthor{\bsnm{Gulminelli}, \binits{F.}},
\bauthor{\bsnm{Molitoris}, \binits{J.}}:
\batitle{{The Boltzmann equation at the borderline: a decade of Monte Carlo
  simulations of a quantum kinetic equation}}.
\bjtitle{Phys. Rept.}
\bvolume{243},
\bfpage{1}--\blpage{124}
(\byear{1994}).
\doiurl{10.1016/0370-1573(94)90108-2}
\end{barticle}
\endbibitem

\bibitem{Carruthers:1982fa}
\begin{barticle}
\bauthor{\bsnm{Carruthers}, \binits{P.}},
\bauthor{\bsnm{Zachariasen}, \binits{F.}}:
\batitle{{Quantum Collision Theory with Phase Space Distribution Functions}}.
\bjtitle{Rev. Mod. Phys.}
\bvolume{55},
\bfpage{245}
(\byear{1983}).
\doiurl{10.1103/RevModPhys.55.245}
\end{barticle}
\endbibitem

\bibitem{Abe:1995yw}
\begin{barticle}
\bauthor{\bsnm{Abe}, \binits{Y.}},
\bauthor{\bsnm{Ayik}, \binits{S.}},
\bauthor{\bsnm{Reinhard}, \binits{P.G.}},
\bauthor{\bsnm{Suraud}, \binits{E.}}:
\batitle{{On stochastic approaches of nuclear dynamics}}.
\bjtitle{Phys. Rept.}
\bvolume{275},
\bfpage{49}--\blpage{196}
(\byear{1996}).
\doiurl{10.1016/0370-1573(96)00003-8}
\end{barticle}
\endbibitem

\bibitem{Chomaz:2003dz}
\begin{barticle}
\bauthor{\bsnm{Chomaz}, \binits{P.}},
\bauthor{\bsnm{Colonna}, \binits{M.}},
\bauthor{\bsnm{Randrup}, \binits{J.}}:
\batitle{{Nuclear spinodal fragmentation}}.
\bjtitle{Phys. Rept.}
\bvolume{389},
\bfpage{263}--\blpage{440}
(\byear{2004}).
\doiurl{10.1016/j.physrep.2003.09.006}
\end{barticle}
\endbibitem

\bibitem{Aichelin:1991xy}
\begin{barticle}
\bauthor{\bsnm{Aichelin}, \binits{J.}}:
\batitle{{'Quantum' molecular dynamics: A Dynamical microscopic n body approach
  to investigate fragment formation and the nuclear equation of state in heavy
  ion collisions}}.
\bjtitle{Phys. Rept.}
\bvolume{202},
\bfpage{233}--\blpage{360}
(\byear{1991}).
\doiurl{10.1016/0370-1573(91)90094-3}
\end{barticle}
\endbibitem

\bibitem{Feldmeier:1989st}
\begin{barticle}
\bauthor{\bsnm{Feldmeier}, \binits{H.}}:
\batitle{{Fermionic molecular dynamcis}}.
\bjtitle{Nucl. Phys. A}
\bvolume{515},
\bfpage{147}--\blpage{172}
(\byear{1990}).
\doiurl{10.1016/0375-9474(90)90328-J}
\end{barticle}
\endbibitem

\bibitem{Ono:1998yd}
\begin{barticle}
\bauthor{\bsnm{Ono}, \binits{A.}}:
\batitle{{Antisymmetrized molecular dynamics with quantum branching processes
  for collisions of heavy nuclei}}.
\bjtitle{Phys. Rev. C}
\bvolume{59},
\bfpage{853}--\blpage{864}
(\byear{1999})
{\href{https://arxiv.org/abs/nucl-th/9809029}{{arXiv:nucl-th/9809029}}}.
\doiurl{10.1103/PhysRevC.59.853}
\end{barticle}
\endbibitem

\bibitem{Papa:2000ef}
\begin{barticle}
\bauthor{\bsnm{Papa}, \binits{M.}},
\bauthor{\bsnm{Maruyama}, \binits{T.}},
\bauthor{\bsnm{Bonasera}, \binits{A.}}:
\batitle{{Constraint molecular dynamics approach to fermionic systems}}.
\bjtitle{Phys. Rev. C}
\bvolume{64},
\bfpage{024612}
(\byear{2001})
{\href{https://arxiv.org/abs/nucl-th/0012083}{{arXiv:nucl-th/0012083}}}.
\doiurl{10.1103/PhysRevC.64.024612}
\end{barticle}
\endbibitem

\bibitem{Hartnack:1997ez}
\begin{barticle}
\bauthor{\bsnm{Hartnack}, \binits{C.}},
\bauthor{\bsnm{Puri}, \binits{R.K.}},
\bauthor{\bsnm{Aichelin}, \binits{J.}},
\bauthor{\bsnm{Konopka}, \binits{J.}},
\bauthor{\bsnm{Bass}, \binits{S.A.}},
\bauthor{\bsnm{Stoecker}, \binits{H.}},
\bauthor{\bsnm{Greiner}, \binits{W.}}:
\batitle{{Modeling the many body dynamics of heavy ion collisions: Present
  status and future perspective}}.
\bjtitle{Eur. Phys. J. A}
\bvolume{1},
\bfpage{151}--\blpage{169}
(\byear{1998}).
\doiurl{10.1007/s100500050045}
\end{barticle}
\endbibitem

\bibitem{Cassing:2009vt}
\begin{barticle}
\bauthor{\bsnm{Cassing}, \binits{W.}},
\bauthor{\bsnm{Bratkovskaya}, \binits{E.L.}}:
\batitle{{Parton-Hadron-String Dynamics: an off-shell transport approach for
  relativistic energies}}.
\bjtitle{Nucl. Phys. A}
\bvolume{831},
\bfpage{215}--\blpage{242}
(\byear{2009})
{\href{https://arxiv.org/abs/0907.5331}{{arXiv:0907.5331}}}
{[nucl-th]}.
\doiurl{10.1016/j.nuclphysa.2009.09.007}
\end{barticle}
\endbibitem

\bibitem{Buss:2011mx}
\begin{barticle}
\bauthor{\bsnm{Buss}, \binits{O.}},
\bauthor{\bsnm{Gaitanos}, \binits{T.}},
\bauthor{\bsnm{Gallmeister}, \binits{K.}},
\bauthor{\bparticle{van} \bsnm{Hees}, \binits{H.}},
\bauthor{\bsnm{Kaskulov}, \binits{M.}},
\bauthor{\bsnm{Lalakulich}, \binits{O.}},
\bauthor{\bsnm{Larionov}, \binits{A.B.}},
\bauthor{\bsnm{Leitner}, \binits{T.}},
\bauthor{\bsnm{Weil}, \binits{J.}},
\bauthor{\bsnm{Mosel}, \binits{U.}}:
\batitle{{Transport-theoretical Description of Nuclear Reactions}}.
\bjtitle{Phys. Rept.}
\bvolume{512},
\bfpage{1}--\blpage{124}
(\byear{2012})
{\href{https://arxiv.org/abs/1106.1344}{{arXiv:1106.1344}}}
{[hep-ph]}.
\doiurl{10.1016/j.physrep.2011.12.001}
\end{barticle}
\endbibitem

\bibitem{Hen:2014nza}
\begin{barticle}
\bauthor{\bsnm{Hen}, \binits{O.}}, \betal:
\batitle{{Momentum sharing in imbalanced Fermi systems}}.
\bjtitle{Science}
\bvolume{346},
\bfpage{614}--\blpage{617}
(\byear{2014})
{\href{https://arxiv.org/abs/1412.0138}{{arXiv:1412.0138}}}
{[nucl-ex]}.
\doiurl{10.1126/science.1256785}
\end{barticle}
\endbibitem

\bibitem{Hen:2016kwk}
\begin{barticle}
\bauthor{\bsnm{Hen}, \binits{O.}},
\bauthor{\bsnm{Miller}, \binits{G.A.}},
\bauthor{\bsnm{Piasetzky}, \binits{E.}},
\bauthor{\bsnm{Weinstein}, \binits{L.B.}}:
\batitle{{Nucleon-Nucleon Correlations, Short-lived Excitations, and the Quarks
  Within}}.
\bjtitle{Rev. Mod. Phys.}
\bvolume{89}(\bissue{4}),
\bfpage{045002}
(\byear{2017})
{\href{https://arxiv.org/abs/1611.09748}{{arXiv:1611.09748}}}
{[nucl-ex]}.
\doiurl{10.1103/RevModPhys.89.045002}
\end{barticle}
\endbibitem

\bibitem{Hen:2014yfa}
\begin{barticle}
\bauthor{\bsnm{Hen}, \binits{O.}},
\bauthor{\bsnm{Li}, \binits{B.-A.}},
\bauthor{\bsnm{Guo}, \binits{W.-J.}},
\bauthor{\bsnm{Weinstein}, \binits{L.B.}},
\bauthor{\bsnm{Piasetzky}, \binits{E.}}:
\batitle{{Symmetry Energy of Nucleonic Matter With Tensor Correlations}}.
\bjtitle{Phys. Rev. C}
\bvolume{91}(\bissue{2}),
\bfpage{025803}
(\byear{2015})
{\href{https://arxiv.org/abs/1408.0772}{{arXiv:1408.0772}}}
{[nucl-ex]}.
\doiurl{10.1103/PhysRevC.91.025803}
\end{barticle}
\endbibitem

\bibitem{Song:2015hua}
\begin{barticle}
\bauthor{\bsnm{Song}, \binits{T.}},
\bauthor{\bsnm{Ko}, \binits{C.M.}}:
\batitle{{Modifications of the pion-production threshold in the nuclear medium
  in heavy ion collisions and the nuclear symmetry energy}}.
\bjtitle{Phys. Rev.}
\bvolume{C91}(\bissue{1}),
\bfpage{014901}
(\byear{2015}).
\doiurl{10.1103/PhysRevC.91.014901}
\end{barticle}
\endbibitem

\bibitem{Kolomeitsev:2004np}
\begin{barticle}
\bauthor{\bsnm{Kolomeitsev}, \binits{E.E.}}, \betal:
\batitle{{Transport theories for heavy ion collisions in the 1-A-GeV regime}}.
\bjtitle{J. Phys. G}
\bvolume{31},
\bfpage{741}--\blpage{758}
(\byear{2005})
{\href{https://arxiv.org/abs/nucl-th/0412037}{{arXiv:nucl-th/0412037}}}.
\doiurl{10.1088/0954-3899/31/6/015}
\end{barticle}
\endbibitem

\bibitem{Reichert:2021ljd}
\begin{barticle}
\bauthor{\bsnm{Reichert}, \binits{T.}},
\bauthor{\bsnm{Elz}, \binits{A.}},
\bauthor{\bsnm{Song}, \binits{T.}},
\bauthor{\bsnm{Coci}, \binits{G.}},
\bauthor{\bsnm{Winn}, \binits{M.}},
\bauthor{\bsnm{Bratkovskaya}, \binits{E.}},
\bauthor{\bsnm{Aichelin}, \binits{J.}},
\bauthor{\bsnm{Steinheimer}, \binits{J.}},
\bauthor{\bsnm{Bleicher}, \binits{M.}}:
\batitle{{Comparison of heavy ion transport simulations: Ag + Ag collisions at
  $E_{lab}$ = 1.58A GeV}}.
\bjtitle{J. Phys. G}
\bvolume{49}(\bissue{5}),
\bfpage{055108}
(\byear{2022})
{\href{https://arxiv.org/abs/2111.07652}{{arXiv:2111.07652}}}
{[nucl-th]}.
\doiurl{10.1088/1361-6471/ac5dfe}
\end{barticle}
\endbibitem

\bibitem{LeFevre19}
\begin{barticle}
\bauthor{\bsnm{Le~F\`evre}, \binits{A.}},
\bauthor{\bsnm{Aichelin}, \binits{J.}},
\bauthor{\bsnm{Hartnack}, \binits{C.}},
\bauthor{\bsnm{Leifels}, \binits{Y.}}:
\batitle{Friga: A new approach to identify isotopes and hypernuclei in $n$-body
  transport models}.
\bjtitle{Phys. Rev. C}
\bvolume{100},
\bfpage{034904}
(\byear{2019}).
\doiurl{10.1103/PhysRevC.100.034904}
\end{barticle}
\endbibitem

\bibitem{Famiano:2006rb}
\begin{barticle}
\bauthor{\bsnm{Famiano}, \binits{M.A.}},
\bauthor{\bsnm{Liu}, \binits{T.}},
\bauthor{\bsnm{Lynch}, \binits{W.G.}},
\bauthor{\bsnm{Rogers}, \binits{A.M.}},
\bauthor{\bsnm{Tsang}, \binits{M.B.}},
\bauthor{\bsnm{Wallace}, \binits{M.S.}},
\bauthor{\bsnm{Charity}, \binits{R.J.}},
\bauthor{\bsnm{Komarov}, \binits{S.}},
\bauthor{\bsnm{Sarantites}, \binits{D.G.}},
\bauthor{\bsnm{Sobotka}, \binits{L.G.}}:
\batitle{{Neutron and Proton Transverse Emission Ratio Measurements and the
  Density Dependence of the Asymmetry Term of the Nuclear Equation of State}}.
\bjtitle{Phys. Rev. Lett.}
\bvolume{97},
\bfpage{052701}
(\byear{2006}).
\doiurl{10.1103/PhysRevLett.97.052701}
\end{barticle}
\endbibitem

\bibitem{Coupland-PhysRevC.94.011601}
\begin{barticle}
\bauthor{\bsnm{Coupland}, \binits{D.D.S.}}, \betal:
\batitle{Probing effective nucleon masses with heavy-ion collisions}.
\bjtitle{Phys. Rev. C}
\bvolume{94},
\bfpage{011601}
(\byear{2016}).
\doiurl{10.1103/PhysRevC.94.011601}
\end{barticle}
\endbibitem

\bibitem{morf19}
\begin{barticle}
\bauthor{\bsnm{Morfouace}, \binits{P.}},
\bauthor{\bsnm{Tsang}, \binits{C.Y.}},
\bauthor{\bsnm{Zhang}, \binits{Y.}},
\bauthor{\bsnm{Lynch}, \binits{W.G.}},
\bauthor{\bsnm{Tsang}, \binits{M.B.}},
\bauthor{\bsnm{Coupland}, \binits{D.D.S.}},
\bauthor{\bsnm{Youngs}, \binits{M.}},
\bauthor{\bsnm{Chajecki}, \binits{Z.}},
\bauthor{\bsnm{Famiano}, \binits{M.A.}},
\bauthor{\bsnm{Ghosh}, \binits{T.K.}},
\bauthor{\bsnm{Jhang}, \binits{G.}},
\bauthor{\bsnm{Lee}, \binits{J.}},
\bauthor{\bsnm{Liu}, \binits{H.}},
\bauthor{\bsnm{Sanetullaev}, \binits{A.}},
\bauthor{\bsnm{Showalter}, \binits{R.}},
\bauthor{\bsnm{Winkelbauer}, \binits{J.}}:
\batitle{{Constraining the symmetry energy with heavy-ion collisions and
  Bayesian analyses}}.
\bjtitle{Physics Letters B}
\bvolume{799},
\bfpage{135045}
(\byear{2019}).
\doiurl{10.1016/j.physletb.2019.135045}
\end{barticle}
\endbibitem

\bibitem{Danielewicz:1991dh}
\begin{barticle}
\bauthor{\bsnm{Danielewicz}, \binits{P.}},
\bauthor{\bsnm{Bertsch}, \binits{G.F.}}:
\batitle{{Production of deuterons and pions in a transport model of energetic
  heavy ion reactions}}.
\bjtitle{Nucl. Phys. A}
\bvolume{533},
\bfpage{712}--\blpage{748}
(\byear{1991}).
\doiurl{10.1016/0375-9474(91)90541-D}
\end{barticle}
\endbibitem

\bibitem{Ono:2013aaa}
\begin{barticle}
\bauthor{\bsnm{Ono}, \binits{A.}}:
\batitle{{Cluster correlations in multifragmentation}}.
\bjtitle{J. Phys. Conf. Ser.}
\bvolume{420},
\bfpage{012103}
(\byear{2013}).
\doiurl{10.1088/1742-6596/420/1/012103}
\end{barticle}
\endbibitem

\bibitem{Reisdorf:1997fx}
\begin{barticle}
\bauthor{\bsnm{Reisdorf}, \binits{W.}},
\bauthor{\bsnm{Ritter}, \binits{H.G.}}:
\batitle{{Collective flow in heavy-ion collisions}}.
\bjtitle{Ann. Rev. Nucl. Part. Sci.}
\bvolume{47},
\bfpage{663}--\blpage{709}
(\byear{1997}).
\doiurl{10.1146/annurev.nucl.47.1.663}
\end{barticle}
\endbibitem

\bibitem{Herrmann:1999axy}
\begin{barticle}
\bauthor{\bsnm{Herrmann}, \binits{N.}},
\bauthor{\bsnm{Wessels}, \binits{J.P.}},
\bauthor{\bsnm{Wienold}, \binits{T.}}:
\batitle{Collective flow in heavy-ion collisions}.
\bjtitle{Annual Review of Nuclear and Particle Science}
\bvolume{49}(\bissue{1}),
\bfpage{581}--\blpage{632}
(\byear{1999})
{\href{https://arxiv.org/abs/https://doi.org/10.1146/annurev.nucl.49.1.581}{{https://doi.org/10.1146/annurev.nucl.49.1.581}}}.
\doiurl{10.1146/annurev.nucl.49.1.581}
\end{barticle}
\endbibitem

\bibitem{Friman:2011zz}
\begin{botherref}
\oauthor{\bsnm{Friman}, \binits{B.}},
\oauthor{\bsnm{Hohne}, \binits{C.}},
\oauthor{\bsnm{Knoll}, \binits{J.}},
\oauthor{\bsnm{Leupold}, \binits{S.}},
\oauthor{\bsnm{Randrup}, \binits{J.}},
\oauthor{\bsnm{Rapp}, \binits{R.}},
\oauthor{\bsnm{Senger}, \binits{P.}}:
{The CBM physics book: Compressed baryonic matter in laboratory experiments}
\textbf{814}
(2011).
\doiurl{10.1007/978-3-642-13293-3}
\end{botherref}
\endbibitem

\bibitem{Gustafsson:1984ka}
\begin{barticle}
\bauthor{\bsnm{Gustafsson}, \binits{H.A.}}, \betal:
\batitle{{Collective Flow Observed in Relativistic Nuclear Collisions}}.
\bjtitle{Phys. Rev. Lett.}
\bvolume{52},
\bfpage{1590}--\blpage{1593}
(\byear{1984}).
\doiurl{10.1103/PhysRevLett.52.1590}
\end{barticle}
\endbibitem

\bibitem{Gutbrod:1989gh}
\begin{barticle}
\bauthor{\bsnm{Gutbrod}, \binits{H.H.}},
\bauthor{\bsnm{Kampert}, \binits{K.H.}},
\bauthor{\bsnm{Kolb}, \binits{B.}},
\bauthor{\bsnm{Poskanzer}, \binits{A.M.}},
\bauthor{\bsnm{Ritter}, \binits{H.G.}},
\bauthor{\bsnm{Schicker}, \binits{R.}},
\bauthor{\bsnm{Schmidt}, \binits{H.R.}}:
\batitle{{Squeezeout of Nuclear Matter as a Function of Projectile Energy and
  Mass}}.
\bjtitle{Phys. Rev. C}
\bvolume{42},
\bfpage{640}--\blpage{651}
(\byear{1990}).
\doiurl{10.1103/PhysRevC.42.640}
\end{barticle}
\endbibitem

\bibitem{Schnetzer:1982}
\begin{barticle}
\bauthor{\bsnm{Schnetzer}, \binits{S.}},
\bauthor{\bsnm{Lemaire}, \binits{M.-C.}},
\bauthor{\bsnm{Lombard}, \binits{R.}},
\bauthor{\bsnm{Moeller}, \binits{E.}},
\bauthor{\bsnm{Nagamiya}, \binits{S.}},
\bauthor{\bsnm{Shapiro}, \binits{G.}},
\bauthor{\bsnm{Steiner}, \binits{H.}},
\bauthor{\bsnm{Tanihata}, \binits{I.}}:
\batitle{{Production of ${K}^{+}$ Mesons in 2.1-GeV/Nucleon Nuclear
  Collisions}}.
\bjtitle{Phys. Rev. Lett.}
\bvolume{49},
\bfpage{989}--\blpage{992}
(\byear{1982}).
\doiurl{10.1103/PhysRevLett.49.989}
\end{barticle}
\endbibitem

\bibitem{Hubele:1991ss}
\begin{barticle}
\bauthor{\bsnm{Hubele}, \binits{J.}}, \betal:
\batitle{{Fragmentation of gold projectiles: From evaporation to total
  disassembly}}.
\bjtitle{Z. Phys. A}
\bvolume{340},
\bfpage{263}--\blpage{270}
(\byear{1991}).
\doiurl{10.1007/BF01294674}
\end{barticle}
\endbibitem

\bibitem{SENGER1993393}
\begin{barticle}
\bauthor{\bsnm{Senger}, \binits{P.}}, \betal:
\batitle{{The kaon spectrometer at SIS}}.
\bjtitle{Nuclear Instruments and Methods in Physics Research Section A:
  Accelerators, Spectrometers, Detectors and Associated Equipment}
\bvolume{327}(\bissue{2}),
\bfpage{393}--\blpage{411}
(\byear{1993}).
\doiurl{10.1016/0168-9002(93)90706-N}
\end{barticle}
\endbibitem

\bibitem{Novotny:1991}
\begin{barticle}
\bauthor{\bsnm{Novotny}, \binits{R.}}:
\batitle{{The BaF$_2$ photon spectrometer TAPS}}.
\bjtitle{IEEE Transactions on Nuclear Science}
\bvolume{38}(\bissue{2}),
\bfpage{379}--\blpage{385}
(\byear{1991}).
\doiurl{10.1109/23.289329}
\end{barticle}
\endbibitem

\bibitem{GOBBI1993156}
\begin{barticle}
\bauthor{\bsnm{Gobbi}, \binits{A.}}, \betal:
\batitle{{A highly-segmented $\Delta$E-time-of-flight wall as forward detector
  of the 4$\pi$-system for charged particles at the SIS/ESR accelerator}}.
\bjtitle{Nuclear Instruments and Methods A}
\bvolume{324}(\bissue{1}),
\bfpage{156}--\blpage{176}
(\byear{1993}).
\doiurl{10.1016/0168-9002(93)90974-M}
\end{barticle}
\endbibitem

\bibitem{Agakichiev_2009}
\begin{barticle}
\bauthor{\bsnm{Agakichiev}, \binits{G.}}, \betal:
\batitle{The high-acceptance dielectron spectrometer {HADES}}.
\bjtitle{The European Physical Journal A}
\bvolume{41}(\bissue{2}),
\bfpage{243}--\blpage{277}
(\byear{2009}).
\doiurl{10.1140/epja/i2009-10807-5}
\end{barticle}
\endbibitem

\bibitem{Inghirami:2022afu}
\begin{botherref}
\oauthor{\bsnm{Inghirami}, \binits{G.}},
\oauthor{\bsnm{Elfner}, \binits{H.}}:
{The applicability of hydrodynamics in heavy ion collisions at $\sqrt{s_{NN}}$=
  2.4-7.7 GeV}
(2022)
{\href{https://arxiv.org/abs/2201.05934}{{arXiv:2201.05934}}}
{[hep-ph]}
\end{botherref}
\endbibitem

\bibitem{Kaiser:2001bx}
\begin{barticle}
\bauthor{\bsnm{Kaiser}, \binits{N.}},
\bauthor{\bsnm{Weise}, \binits{W.}}:
\batitle{{Systematic calculation of S wave pion and kaon selfenergies in
  asymmetric nuclear matter}}.
\bjtitle{Phys. Lett. B}
\bvolume{512},
\bfpage{283}--\blpage{289}
(\byear{2001})
{\href{https://arxiv.org/abs/nucl-th/0102062}{{arXiv:nucl-th/0102062}}}.
\doiurl{10.1016/S0370-2693(01)00584-6}
\end{barticle}
\endbibitem

\bibitem{Korpa:2003bc}
\begin{barticle}
\bauthor{\bsnm{Korpa}, \binits{C.L.}},
\bauthor{\bsnm{Lutz}, \binits{M.F.M.}}:
\batitle{{Selfconsistent and covariant propagation of pions, nucleon and isobar
  resonances in cold nuclear matter}}.
\bjtitle{Nucl. Phys. A}
\bvolume{742},
\bfpage{305}--\blpage{321}
(\byear{2004})
{\href{https://arxiv.org/abs/nucl-th/0306063}{{arXiv:nucl-th/0306063}}}.
\doiurl{10.1016/j.nuclphysa.2004.06.031}
\end{barticle}
\endbibitem

\bibitem{Korpa:2004ae}
\begin{barticle}
\bauthor{\bsnm{Korpa}, \binits{C.L.}},
\bauthor{\bsnm{Lutz}, \binits{M.F.M.}}:
\batitle{{Kaon and antikaon properties in cold nuclear medium}}.
\bjtitle{Acta Phys. Hung. A}
\bvolume{22},
\bfpage{21}--\blpage{28}
(\byear{2005})
{\href{https://arxiv.org/abs/nucl-th/0404088}{{arXiv:nucl-th/0404088}}}.
\doiurl{10.1556/APH.22.2005.1-2.4}
\end{barticle}
\endbibitem

\bibitem{HARTNACK2012119}
\begin{barticle}
\bauthor{\bsnm{Hartnack}, \binits{C.}}, \betal:
\batitle{Strangeness production close to the threshold in proton–nucleus and
  heavy-ion collisions}.
\bjtitle{Physics Reports}
\bvolume{510}(\bissue{4}),
\bfpage{119}--\blpage{200}
(\byear{2012}).
\doiurl{10.1016/j.physrep.2011.08.004}.
\bcomment{Strangeness production close to the threshold in proton–nucleus and
  heavy-ion collisions}
\end{barticle}
\endbibitem

\bibitem{AICHKO}
\begin{barticle}
\bauthor{\bsnm{Aichelin}, \binits{J.}},
\bauthor{\bsnm{Ko}, \binits{C.M.}}:
\batitle{Subthreshold kaon production as a probe of the nuclear equation of
  state}.
\bjtitle{Phys. Rev. Lett.}
\bvolume{55},
\bfpage{2661}--\blpage{2663}
(\byear{1985}).
\doiurl{10.1103/PhysRevLett.55.2661}
\end{barticle}
\endbibitem

\bibitem{Fuchs:2001gv}
\begin{barticle}
\bauthor{\bsnm{Fuchs}, \binits{C.}},
\bauthor{\bsnm{Faessler}, \binits{A.}},
\bauthor{\bsnm{El-Basaouny}, \binits{S.}},
\bauthor{\bsnm{Shekhter}, \binits{K.M.}},
\bauthor{\bsnm{Zabrodin}, \binits{E.}},
\bauthor{\bsnm{Zheng}, \binits{Y.M.}}:
\batitle{{The Nuclear equation of state probed by K+ production in heavy ion
  collisions}}.
\bjtitle{J. Phys. G}
\bvolume{28},
\bfpage{1615}--\blpage{1622}
(\byear{2002}).
\doiurl{10.1088/0954-3899/28/7/313}
\end{barticle}
\endbibitem

\bibitem{Hartnack:2006}
\begin{barticle}
\bauthor{\bsnm{Hartnack}, \binits{C.}},
\bauthor{\bsnm{Oeschler}, \binits{H.}},
\bauthor{\bsnm{Aichelin}, \binits{J.}}:
\batitle{Hadronic matter is soft}.
\bjtitle{Phys. Rev. Lett.}
\bvolume{96},
\bfpage{012302}
(\byear{2006}).
\doiurl{10.1103/PhysRevLett.96.012302}
\end{barticle}
\endbibitem

\bibitem{Wagner:2000ak}
\begin{barticle}
\bauthor{\bsnm{Wagner}, \binits{A.}}, \betal:
\batitle{{The Emission pattern of high-energy pions: A New probe for the early
  phase of heavy ion collisions}}.
\bjtitle{Phys. Rev. Lett.}
\bvolume{85},
\bfpage{18}--\blpage{21}
(\byear{2000}).
\doiurl{10.1103/PhysRevLett.85.18}
\end{barticle}
\endbibitem

\bibitem{FOPI:2005tyo}
\begin{barticle}
\bauthor{\bsnm{Hong}, \binits{B.}}, \betal:
\batitle{{Charged pion production in
  ${}_{44}^{96}\mathrm{Ru}+{}_{44}^{96}\mathrm{Ru}$ collisions at $400A$ and
  $1528A\phantom{\rule{0.3em}{0ex}}\mathrm{MeV}$}}.
\bjtitle{Phys. Rev. C}
\bvolume{71},
\bfpage{034902}
(\byear{2005}).
\doiurl{10.1103/PhysRevC.71.034902}
\end{barticle}
\endbibitem

\bibitem{Li:2004cq}
\begin{barticle}
\bauthor{\bsnm{Li}, \binits{B.-A.}},
\bauthor{\bsnm{Yong}, \binits{G.-C.}},
\bauthor{\bsnm{Zuo}, \binits{W.}}:
\batitle{{Near-threshold pion production with radioactive beams at the rare
  isotope accelerator}}.
\bjtitle{Phys. Rev. C}
\bvolume{71},
\bfpage{014608}
(\byear{2005}).
\doiurl{10.1103/PhysRevC.71.014608}
\end{barticle}
\endbibitem

\bibitem{Stock:1985xe}
\begin{barticle}
\bauthor{\bsnm{Stock}, \binits{R.}}:
\batitle{Particle production in high energy nucleus-nucleus collisions}.
\bjtitle{Physics Report}
\bvolume{135}(\bissue{5}),
\bfpage{259}--\blpage{315}
(\byear{1986}).
\doiurl{10.1016/0370-1573(86)90134-1}
\end{barticle}
\endbibitem

\bibitem{Bass:1994af}
\begin{barticle}
\bauthor{\bsnm{Bass}, \binits{S.A.}},
\bauthor{\bsnm{Hartnack}, \binits{C.}},
\bauthor{\bsnm{Stoecker}, \binits{H.}},
\bauthor{\bsnm{Greiner}, \binits{W.}}:
\batitle{{High p(T) pions as probes of the early dense reaction phase in heavy
  ion collisions at 1-GeV/nucleon}}.
\bjtitle{Phys. Rev. C}
\bvolume{50},
\bfpage{2167}
(\byear{1994}).
\doiurl{10.1103/PhysRevC.50.2167}
\end{barticle}
\endbibitem

\bibitem{Schonhofen:1989pt}
\begin{barticle}
\bauthor{\bsnm{Schonhofen}, \binits{M.}},
\bauthor{\bsnm{Cubero}, \binits{M.}},
\bauthor{\bsnm{Gering}, \binits{M.}},
\bauthor{\bsnm{Sambataro}, \binits{M.}},
\bauthor{\bsnm{Feldmeier}, \binits{H.}},
\bauthor{\bsnm{Norenberg}, \binits{W.}}:
\batitle{{The Nuclear Equation of State in Effective Relativistic Field Thories
  and Pion Yields in Heavy Ion Collisions}}.
\bjtitle{Nucl. Phys. A}
\bvolume{504},
\bfpage{875}--\blpage{898}
(\byear{1989}).
\doiurl{10.1016/0375-9474(89)90015-8}
\end{barticle}
\endbibitem

\bibitem{Jaminon:1989wj}
\begin{barticle}
\bauthor{\bsnm{Jaminon}, \binits{M.}},
\bauthor{\bsnm{Mahaux}, \binits{C.}}:
\batitle{{Effective Masses in Relativistic Approaches to the Nucleon Nucleus
  Mean Field}}.
\bjtitle{Phys. Rev. C}
\bvolume{40},
\bfpage{354}--\blpage{367}
(\byear{1989}).
\doiurl{10.1103/PhysRevC.40.354}
\end{barticle}
\endbibitem

\bibitem{Schwalb:1994zz}
\begin{barticle}
\bauthor{\bsnm{Schwalb}, \binits{O.}}, \betal:
\batitle{{Mass dependence of pi0 production in heavy ion collisions at
  1-A/GeV}}.
\bjtitle{Phys. Lett. B}
\bvolume{321},
\bfpage{20}--\blpage{25}
(\byear{1994}).
\doiurl{10.1016/0370-2693(94)90322-0}
\end{barticle}
\endbibitem

\bibitem{Brill:1993xh}
\begin{barticle}
\bauthor{\bsnm{Brill}, \binits{D.}}, \betal:
\batitle{{Azimuthally anisotropic emission of pions in symmetric heavy ion
  collisions}}.
\bjtitle{Phys. Rev. Lett.}
\bvolume{71},
\bfpage{336}--\blpage{339}
(\byear{1993}).
\doiurl{10.1103/PhysRevLett.71.336}
\end{barticle}
\endbibitem

\bibitem{FOPI:2006ifg}
\begin{barticle}
\bauthor{\bsnm{Reisdorf}, \binits{W.}}, \betal:
\batitle{{Systematics of pion emission in heavy ion collisions in the 1A- GeV
  regime}}.
\bjtitle{Nucl. Phys. A}
\bvolume{781},
\bfpage{459}--\blpage{508}
(\byear{2007}).
\doiurl{10.1016/j.nuclphysa.2006.10.085}
\end{barticle}
\endbibitem

\bibitem{HADES:2009mtt}
\begin{botherref}
\oauthor{\bsnm{Tlusty}, \binits{P.}}, et al.:
{Charged pion production in C+C and Ar+KCl collisions measured with HADES}
(2009)
{\href{https://arxiv.org/abs/0906.2309}{{arXiv:0906.2309}}}
{[nucl-ex]}.
{Proceedings of the 47$^{th}$ International Winter Meeting on Nuclear Physics,
  Bormio (Italy)}
\end{botherref}
\endbibitem

\bibitem{HADES:2020ver}
\begin{barticle}
\bauthor{\bsnm{Adamczewski-Musch}, \binits{J.}}, \betal:
\batitle{{Charged-pion production in $\mathbf {Au+Au}$ collisions at
  $\sqrt{\mathbf {s}_{\mathbf {NN}}} = 2.4~{\mathbf {GeV}}$: HADES
  Collaboration}}.
\bjtitle{Eur. Phys. J. A}
\bvolume{56}(\bissue{10}),
\bfpage{259}
(\byear{2020}).
\doiurl{10.1140/epja/s10050-020-00237-2}
\end{barticle}
\endbibitem

\bibitem{Cozma:2021tfu}
\begin{barticle}
\bauthor{\bsnm{Cozma}, \binits{M.D.}},
\bauthor{\bsnm{Tsang}, \binits{M.B.}}:
\batitle{{In-medium $\Delta (1232)$ potential, pion production in heavy-ion
  collisions and the symmetry energy}}.
\bjtitle{Eur. Phys. J. A}
\bvolume{57}(\bissue{11}),
\bfpage{309}
(\byear{2021}).
\doiurl{10.1140/epja/s10050-021-00616-3}
\end{barticle}
\endbibitem

\bibitem{Zhang:2020dvn}
\begin{barticle}
\bauthor{\bsnm{Zhang}, \binits{Y.}},
\bauthor{\bsnm{Wang}, \binits{N.}},
\bauthor{\bsnm{Li}, \binits{Q.-F.}},
\bauthor{\bsnm{Ou}, \binits{L.}},
\bauthor{\bsnm{Tian}, \binits{J.-L.}},
\bauthor{\bsnm{Liu}, \binits{M.}},
\bauthor{\bsnm{Zhao}, \binits{K.}},
\bauthor{\bsnm{Wu}, \binits{X.-Z.}},
\bauthor{\bsnm{Li}, \binits{Z.-X.}}:
\batitle{Progress of quantum molecular dynamics model and its applications in
  heavy ion collisions}.
\bjtitle{Frontiers of Physics}
\bvolume{15}(\bissue{5}),
\bfpage{54301}
(\byear{2020}).
\doiurl{10.1007/s11467-020-0961-9}
\end{barticle}
\endbibitem

\bibitem{Andronic:2006ra}
\begin{barticle}
\bauthor{\bsnm{Andronic}, \binits{A.}},
\bauthor{\bsnm{{\L}ukasik}, \binits{J.}},
\bauthor{\bsnm{Reisdorf}, \binits{W.}},
\bauthor{\bsnm{Trautmann}, \binits{W.}}:
\batitle{{Systematics of Stopping and Flow in Au+Au Collisions}}.
\bjtitle{Eur. Phys. J. A}
\bvolume{30},
\bfpage{31}--\blpage{46}
(\byear{2006}).
\doiurl{10.1140/epja/i2006-10101-2}
\end{barticle}
\endbibitem

\bibitem{LeFevre:2015paj}
\begin{barticle}
\bauthor{\bsnm{Le~F\`evre}, \binits{A.}},
\bauthor{\bsnm{Leifels}, \binits{Y.}},
\bauthor{\bsnm{Reisdorf}, \binits{W.}},
\bauthor{\bsnm{Aichelin}, \binits{J.}},
\bauthor{\bsnm{Hartnack}, \binits{C.}}:
\batitle{{Constraining the nuclear matter equation of state around twice
  saturation density}}.
\bjtitle{Nucl. Phys. A}
\bvolume{945},
\bfpage{112}--\blpage{133}
(\byear{2016}).
\doiurl{10.1016/j.nuclphysa.2015.09.015}
\end{barticle}
\endbibitem

\bibitem{FOPI:2010xrt}
\begin{barticle}
\bauthor{\bsnm{Reisdorf}, \binits{W.}}, \betal:
\batitle{{Systematics of central heavy ion collisions in the 1A GeV regime}}.
\bjtitle{Nucl. Phys. A}
\bvolume{848},
\bfpage{366}--\blpage{427}
(\byear{2010}).
\doiurl{10.1016/j.nuclphysa.2010.09.008}
\end{barticle}
\endbibitem

\bibitem{FOPI:2011aa}
\begin{barticle}
\bauthor{\bsnm{Reisdorf}, \binits{W.}}, \betal:
\batitle{{Systematics of azimuthal asymmetries in heavy ion collisions in the 1
  A GeV regime}}.
\bjtitle{Nucl. Phys. A}
\bvolume{876},
\bfpage{1}--\blpage{60}
(\byear{2012}).
\doiurl{10.1016/j.nuclphysa.2011.12.006}
\end{barticle}
\endbibitem

\bibitem{Wang:2018hsw}
\begin{barticle}
\bauthor{\bsnm{Wang}, \binits{Y.}},
\bauthor{\bsnm{Guo}, \binits{C.}},
\bauthor{\bsnm{Li}, \binits{Q.}},
\bauthor{\bsnm{Le~F\`evre}, \binits{A.}},
\bauthor{\bsnm{Leifels}, \binits{Y.}},
\bauthor{\bsnm{Trautmann}, \binits{W.}}:
\batitle{{Determination of the nuclear incompressibility from the
  rapidity-dependent elliptic flow in heavy-ion collisions at beam energies 0.4
  A \textendash{}1.0 A GeV}}.
\bjtitle{Phys. Lett. B}
\bvolume{778},
\bfpage{207}--\blpage{212}
(\byear{2018}).
\doiurl{10.1016/j.physletb.2018.01.035}
\end{barticle}
\endbibitem

\bibitem{LeFevre:2016vpp}
\begin{barticle}
\bauthor{\bsnm{Le~F\`evre}, \binits{A.}},
\bauthor{\bsnm{Leifels}, \binits{Y.}},
\bauthor{\bsnm{Hartnack}, \binits{C.}},
\bauthor{\bsnm{Aichelin}, \binits{J.}}:
\batitle{Origin of elliptic flow and its dependence on the equation of state in
  heavy ion reactions at intermediate energies}.
\bjtitle{Phys. Rev. C}
\bvolume{98},
\bfpage{034901}
(\byear{2018}).
\doiurl{10.1103/PhysRevC.98.034901}
\end{barticle}
\endbibitem

\bibitem{Stoecker:1981pg}
\begin{barticle}
\bauthor{\bsnm{Stoecker}, \binits{H.}},
\bauthor{\bsnm{Csernai}, \binits{L.P.}},
\bauthor{\bsnm{Graebner}, \binits{G.}},
\bauthor{\bsnm{Buchwald}, \binits{G.}},
\bauthor{\bsnm{Kruse}, \binits{H.}},
\bauthor{\bsnm{Cusson}, \binits{R.Y.}},
\bauthor{\bsnm{Maruhn}, \binits{J.A.}},
\bauthor{\bsnm{Greiner}, \binits{W.}}:
\batitle{{Jets of Nuclear Matter From High-energy Heavy Ion Collisions}}.
\bjtitle{Phys. Rev. C}
\bvolume{25},
\bfpage{1873}--\blpage{1876}
(\byear{1982}).
\doiurl{10.1103/PhysRevC.25.1873}
\end{barticle}
\endbibitem

\bibitem{FOPI:2004hyz}
\begin{barticle}
\bauthor{\bsnm{Stoicea}, \binits{G.}}, \betal:
\batitle{{Azimuthal dependence of collective expansion for symmetric heavy ion
  collisions}}.
\bjtitle{Phys. Rev. Lett.}
\bvolume{92},
\bfpage{072303}
(\byear{2004}).
\doiurl{10.1103/PhysRevLett.92.072303}
\end{barticle}
\endbibitem

\bibitem{Sturm:2000dm}
\begin{barticle}
\bauthor{\bsnm{Sturm}, \binits{C.T.}}, \betal:
\batitle{{Evidence for a soft nuclear equation of state from kaon production in
  heavy ion collisions}}.
\bjtitle{Phys. Rev. Lett.}
\bvolume{86},
\bfpage{39}--\blpage{42}
(\byear{2001}).
\doiurl{10.1103/PhysRevLett.86.39}
\end{barticle}
\endbibitem

\bibitem{bor08}
\begin{barticle}
\bauthor{\bsnm{Borderie}, \binits{B.}},
\bauthor{\bsnm{Rivet}, \binits{M.F.}}:
\batitle{{Nuclear multifragmentation and phase transition for hot nuclei}}.
\bjtitle{Progress in Particle and Nuclear Physics}
\bvolume{61}(\bissue{2}),
\bfpage{551}--\blpage{601}
(\byear{2008}).
\doiurl{10.1016/j.ppnp.2008.01.003}
\end{barticle}
\endbibitem

\bibitem{poc95}
\begin{barticle}
\bauthor{\bsnm{Pochodzalla}, \binits{J.}}, \betal:
\batitle{{Probing the Nuclear Liquid-Gas Phase Transition}}.
\bjtitle{Phys. Rev. Lett.}
\bvolume{75},
\bfpage{1040}--\blpage{1043}
(\byear{1995}).
\doiurl{10.1103/PhysRevLett.75.1040}
\end{barticle}
\endbibitem

\bibitem{nat02}
\begin{barticle}
\bauthor{\bsnm{Natowitz}, \binits{J.B.}},
\bauthor{\bsnm{Wada}, \binits{R.}},
\bauthor{\bsnm{Hagel}, \binits{K.}},
\bauthor{\bsnm{Keutgen}, \binits{T.}},
\bauthor{\bsnm{Murray}, \binits{M.}},
\bauthor{\bsnm{Makeev}, \binits{A.}},
\bauthor{\bsnm{Qin}, \binits{L.}},
\bauthor{\bsnm{Smith}, \binits{P.}},
\bauthor{\bsnm{Hamilton}, \binits{C.}}:
\batitle{{Caloric curves and critical behavior in nuclei}}.
\bjtitle{Phys. Rev. C}
\bvolume{65},
\bfpage{034618}
(\byear{2002}).
\doiurl{10.1103/PhysRevC.65.034618}
\end{barticle}
\endbibitem

\bibitem{wad12}
\begin{barticle}
\bauthor{\bsnm{Wada}, \binits{R.}},
\bauthor{\bsnm{Hagel}, \binits{K.}},
\bauthor{\bsnm{Qin}, \binits{L.}},
\bauthor{\bsnm{Natowitz}, \binits{J.B.}},
\bauthor{\bsnm{Ma}, \binits{Y.G.}},
\bauthor{\bsnm{R\"opke}, \binits{G.}},
\bauthor{\bsnm{Shlomo}, \binits{S.}},
\bauthor{\bsnm{Bonasera}, \binits{A.}},
\bauthor{\bsnm{Typel}, \binits{S.}},
\bauthor{\bsnm{Chen}, \binits{Z.}},
\bauthor{\bsnm{Huang}, \binits{M.}},
\bauthor{\bsnm{Wang}, \binits{J.}},
\bauthor{\bsnm{Zheng}, \binits{H.}},
\bauthor{\bsnm{Kowalski}, \binits{S.}},
\bauthor{\bsnm{Bottosso}, \binits{C.}},
\bauthor{\bsnm{Barbui}, \binits{M.}},
\bauthor{\bsnm{Rodrigues}, \binits{M.R.D.}},
\bauthor{\bsnm{Schmidt}, \binits{K.}},
\bauthor{\bsnm{Fabris}, \binits{D.}},
\bauthor{\bsnm{Lunardon}, \binits{M.}},
\bauthor{\bsnm{Moretto}, \binits{S.}},
\bauthor{\bsnm{Nebbia}, \binits{G.}},
\bauthor{\bsnm{Pesente}, \binits{S.}},
\bauthor{\bsnm{Rizzi}, \binits{V.}},
\bauthor{\bsnm{Viesti}, \binits{G.}},
\bauthor{\bsnm{Cinausero}, \binits{M.}},
\bauthor{\bsnm{Prete}, \binits{G.}},
\bauthor{\bsnm{Keutgen}, \binits{T.}},
\bauthor{\bsnm{El~Masri}, \binits{Y.}},
\bauthor{\bsnm{Majka}, \binits{Z.}}:
\batitle{{Nuclear matter symmetry energy at $0.03 \leq\rho/\rho_0\leq 0.2$}}.
\bjtitle{Phys. Rev. C}
\bvolume{85},
\bfpage{064618}
(\byear{2012}).
\doiurl{10.1103/PhysRevC.85.064618}
\end{barticle}
\endbibitem

\bibitem{qin12}
\begin{barticle}
\bauthor{\bsnm{Qin}, \binits{L.}},
\bauthor{\bsnm{Hagel}, \binits{K.}},
\bauthor{\bsnm{Wada}, \binits{R.}},
\bauthor{\bsnm{Natowitz}, \binits{J.B.}},
\bauthor{\bsnm{Shlomo}, \binits{S.}},
\bauthor{\bsnm{Bonasera}, \binits{A.}},
\bauthor{\bsnm{R\"opke}, \binits{G.}},
\bauthor{\bsnm{Typel}, \binits{S.}}, \betal:
\batitle{{Laboratory Tests of Low Density Astrophysical Nuclear Equations of
  State}}.
\bjtitle{Phys. Rev. Lett.}
\bvolume{108},
\bfpage{172701}
(\byear{2012}).
\doiurl{10.1103/PhysRevLett.108.172701}
\end{barticle}
\endbibitem

\bibitem{rob12}
\begin{barticle}
\bauthor{\bsnm{Roberts}, \binits{L.F.}},
\bauthor{\bsnm{Reddy}, \binits{S.}},
\bauthor{\bsnm{Shen}, \binits{G.}}:
\batitle{Medium modification of the charged-current neutrino opacity and its
  implications}.
\bjtitle{Phys. Rev. C}
\bvolume{86},
\bfpage{065803}
(\byear{2012}).
\doiurl{10.1103/PhysRevC.86.065803}
\end{barticle}
\endbibitem

\bibitem{hag16}
\begin{barticle}
\bauthor{\bsnm{{Hagel, K.}}},
\bauthor{\bsnm{{Hempel, M.}}},
\bauthor{\bsnm{{Natowitz, J. B.}}},
\bauthor{\bsnm{{R\"opke, G.}}},
\bauthor{\bsnm{{Typel, S.}}},
\bauthor{\bsnm{{Wuenschel, S.}}},
\bauthor{\bsnm{{Wada, R.}}},
\bauthor{\bsnm{{Barbui, M.}}},
\bauthor{\bsnm{{Schmidt, K.}}}:
\batitle{From femtonova to supernova: Heavy-ion collisions and the supernova
  equation of state}.
\bjtitle{EPJ Web of Conferences}
\bvolume{117},
\bfpage{07018}
(\byear{2016}).
\doiurl{10.1051/epjconf/201611707018}
\end{barticle}
\endbibitem

\bibitem{carl17}
\begin{barticle}
\bauthor{\bsnm{Carlson}, \binits{J.}}, \betal:
\batitle{{White paper on nuclear astrophysics and low-energy nuclear physics,
  Part 2: Low-energy nuclear physics}}.
\bjtitle{Progress in Particle and Nuclear Physics}
\bvolume{94},
\bfpage{68}--\blpage{124}
(\byear{2017}).
\doiurl{10.1016/j.ppnp.2016.11.002}
\end{barticle}
\endbibitem

\bibitem{hor14}
\begin{barticle}
\bauthor{\bsnm{Horowitz}, \binits{C.J.}},
\bauthor{\bsnm{Brown}, \binits{E.F.}},
\bauthor{\bsnm{Kim}, \binits{Y.}},
\bauthor{\bsnm{Lynch}, \binits{W.G.}},
\bauthor{\bsnm{Michaels}, \binits{R.}},
\bauthor{\bsnm{Ono}, \binits{A.}},
\bauthor{\bsnm{Piekarewicz}, \binits{J.}},
\bauthor{\bsnm{Tsang}, \binits{M.B.}},
\bauthor{\bsnm{Wolter}, \binits{H.H.}}:
\batitle{A way forward in the study of the symmetry energy: experiment, theory,
  and observation}.
\bjtitle{Journal of Physics G: Nuclear and Particle Physics}
\bvolume{41}(\bissue{9}),
\bfpage{093001}
(\byear{2014}).
\doiurl{10.1088/0954-3899/41/9/093001}
\end{barticle}
\endbibitem

\bibitem{tsa12}
\begin{barticle}
\bauthor{\bsnm{Tsang}, \binits{M.B.}},
\bauthor{\bsnm{Stone}, \binits{J.R.}},
\bauthor{\bsnm{Camera}, \binits{F.}},
\bauthor{\bsnm{Danielewicz}, \binits{P.}},
\bauthor{\bsnm{Gandolfi}, \binits{S.}},
\bauthor{\bsnm{Hebeler}, \binits{K.}},
\bauthor{\bsnm{Horowitz}, \binits{C.J.}},
\bauthor{\bsnm{Lee}, \binits{J.}},
\bauthor{\bsnm{Lynch}, \binits{W.G.}},
\bauthor{\bsnm{Kohley}, \binits{Z.}},
\bauthor{\bsnm{Lemmon}, \binits{R.}},
\bauthor{\bsnm{M\"oller}, \binits{P.}},
\bauthor{\bsnm{Murakami}, \binits{T.}},
\bauthor{\bsnm{Riordan}, \binits{S.}},
\bauthor{\bsnm{Roca-Maza}, \binits{X.}},
\bauthor{\bsnm{Sammarruca}, \binits{F.}},
\bauthor{\bsnm{Steiner}, \binits{A.W.}},
\bauthor{\bsnm{Vida\~na}, \binits{I.}},
\bauthor{\bsnm{Yennello}, \binits{S.J.}}:
\batitle{Constraints on the symmetry energy and neutron skins from experiments
  and theory}.
\bjtitle{Phys. Rev. C}
\bvolume{86},
\bfpage{015803}
(\byear{2012}).
\doiurl{10.1103/PhysRevC.86.015803}
\end{barticle}
\endbibitem

\bibitem{mcin19}
\begin{barticle}
\bauthor{\bsnm{McIntosh}, \binits{A.B.}},
\bauthor{\bsnm{Yennello}, \binits{S.J.}}:
\batitle{Interplay of neutron–proton equilibration and nuclear dynamics}.
\bjtitle{Progress in Particle and Nuclear Physics}
\bvolume{108},
\bfpage{103707}
(\byear{2019}).
\doiurl{10.1016/j.ppnp.2019.06.001}
\end{barticle}
\endbibitem

\bibitem{col20}
\begin{barticle}
\bauthor{\bsnm{Colonna}, \binits{M.}}:
\batitle{Collision dynamics at medium and relativistic energies}.
\bjtitle{Progress in Particle and Nuclear Physics}
\bvolume{113},
\bfpage{103775}
(\byear{2020}).
\doiurl{10.1016/j.ppnp.2020.103775}
\end{barticle}
\endbibitem

\bibitem{LeFevre22}
\begin{botherref}
\oauthor{\bsnm{Le~F\`evre}, \binits{A.}}, et al.:
{Result of the ALADiN experiment at GSI: The asymmetry energy at sub-saturation
  density}.
to be submitted
(2022)
\end{botherref}
\endbibitem

\bibitem{reed21}
\begin{barticle}
\bauthor{\bsnm{Reed}, \binits{B.T.}},
\bauthor{\bsnm{Fattoyev}, \binits{F.J.}},
\bauthor{\bsnm{Horowitz}, \binits{C.J.}},
\bauthor{\bsnm{Piekarewicz}, \binits{J.}}:
\batitle{{Implications of PREX-2 on the Equation of State of Neutron-Rich
  Matter}}.
\bjtitle{Phys. Rev. Lett.}
\bvolume{126},
\bfpage{172503}
(\byear{2021}).
\doiurl{10.1103/PhysRevLett.126.172503}
\end{barticle}
\endbibitem

\bibitem{she07}
\begin{barticle}
\bauthor{\bsnm{Shetty}, \binits{D.V.}},
\bauthor{\bsnm{Yennello}, \binits{S.J.}},
\bauthor{\bsnm{Souliotis}, \binits{G.A.}}:
\batitle{{Density dependence of the symmetry energy and the nuclear equation of
  state: A dynamical and statistical model perspective}}.
\bjtitle{Phys. Rev. C}
\bvolume{76},
\bfpage{024606}
(\byear{2007}).
\doiurl{10.1103/PhysRevC.76.024606}
\end{barticle}
\endbibitem

\bibitem{sun10}
\begin{barticle}
\bauthor{\bsnm{Sun}, \binits{Z.Y.}}, \betal:
\batitle{{Isospin diffusion and equilibration for $\mathrm{Sn}+\mathrm{Sn}$
  collisions at $E/A=35$ MeV}}.
\bjtitle{Phys. Rev. C}
\bvolume{82},
\bfpage{051603}
(\byear{2010}).
\doiurl{10.1103/PhysRevC.82.051603}
\end{barticle}
\endbibitem

\bibitem{cam21}
\begin{barticle}
\bauthor{\bsnm{Camaiani}, \binits{A.}}, \betal:
\batitle{{Isospin diffusion measurement from the direct detection of a
  quasiprojectile remnant}}.
\bjtitle{Phys. Rev. C}
\bvolume{103},
\bfpage{014605}
(\byear{2021}).
\doiurl{10.1103/PhysRevC.103.014605}
\end{barticle}
\endbibitem

\bibitem{hud12}
\begin{barticle}
\bauthor{\bsnm{Hudan}, \binits{S.}},
\bauthor{\bsnm{McIntosh}, \binits{A.B.}},
\bauthor{\bparticle{de} \bsnm{Souza}, \binits{R.T.}},
\bauthor{\bsnm{Bianchin}, \binits{S.}},
\bauthor{\bsnm{Black}, \binits{J.}},
\bauthor{\bsnm{Chbihi}, \binits{A.}},
\bauthor{\bsnm{Famiano}, \binits{M.}},
\bauthor{\bsnm{Fr\'egeau}, \binits{M.O.}},
\bauthor{\bsnm{Gauthier}, \binits{J.}},
\bauthor{\bsnm{Mercier}, \binits{D.}},
\bauthor{\bsnm{Moisan}, \binits{J.}},
\bauthor{\bsnm{Metelko}, \binits{C.J.}},
\bauthor{\bsnm{Roy}, \binits{R.}},
\bauthor{\bsnm{Schwarz}, \binits{C.}},
\bauthor{\bsnm{Trautmann}, \binits{W.}},
\bauthor{\bsnm{Yanez}, \binits{R.}}:
\batitle{{Tracking saddle-to-scission dynamics using $N/Z$ in projectile
  breakup reactions}}.
\bjtitle{Phys. Rev. C}
\bvolume{86},
\bfpage{021603}
(\byear{2012}).
\doiurl{10.1103/PhysRevC.86.021603}
\end{barticle}
\endbibitem

\bibitem{pia21}
\begin{barticle}
\bauthor{\bsnm{Piantelli}, \binits{S.}},
\bauthor{\bsnm{Casini}, \binits{G.}},
\bauthor{\bsnm{Ono}, \binits{A.}},
\bauthor{\bsnm{Poggi}, \binits{G.}},
\bauthor{\bsnm{Pastore}, \binits{G.}}, \betal:
\batitle{{Isospin transport phenomena for the systems
  $^{80}\mathrm{Kr}+^{40,48}\mathrm{Ca}$ at 35 MeV/nucleon}}.
\bjtitle{Phys. Rev. C}
\bvolume{103},
\bfpage{014603}
(\byear{2021}).
\doiurl{10.1103/PhysRevC.103.014603}
\end{barticle}
\endbibitem

\bibitem{riz08}
\begin{barticle}
\bauthor{\bsnm{Rizzo}, \binits{J.}},
\bauthor{\bsnm{Chomaz}, \binits{P.}},
\bauthor{\bsnm{Colonna}, \binits{M.}}:
\batitle{{A new approach to solve the Boltzmann–Langevin equation for
  fermionic systems}}.
\bjtitle{Nuclear Physics A}
\bvolume{806}(\bissue{1}),
\bfpage{40}--\blpage{64}
(\byear{2008}).
\doiurl{10.1016/j.nuclphysa.2008.02.304}
\end{barticle}
\endbibitem

\bibitem{DeFilippo:2013ipa}
\begin{barticle}
\bauthor{\bsnm{De~Filippo}, \binits{E.}},
\bauthor{\bsnm{Pagano}, \binits{A.}}:
\batitle{{Experimental effects on dynamics and thermodynamics in nuclear
  reactions on the symmetry energy as seen by the CHIMERA 4$\pi$ detector}}.
\bjtitle{Eur. Phys. J. A}
\bvolume{50},
\bfpage{32}
(\byear{2014}).
\doiurl{10.1140/epja/i2014-14032-y}
\end{barticle}
\endbibitem

\bibitem{pag20}
\begin{barticle}
\bauthor{\bsnm{Pagano}, \binits{A.}}, \betal:
\batitle{{Nuclear neck-density determination at Fermi energy with CHIMERA
  detector}}.
\bjtitle{The European Physical Journal A}
\bvolume{56},
\bfpage{102}
(\byear{2020}).
\doiurl{10.1140/epja/s10050-020-00105-z}
\end{barticle}
\endbibitem

\bibitem{rusfis20}
\begin{barticle}
\bauthor{\bsnm{Russotto}, \binits{P.}}, \betal:
\batitle{{Dynamical versus statistical production of Intermediate Mass
  Fragments at Fermi Energies}}.
\bjtitle{Eur. Phys. J. A}
\bvolume{56}(\bissue{1}),
\bfpage{12}
(\byear{2020}).
\doiurl{10.1140/epja/s10050-019-00011-z}
\end{barticle}
\endbibitem

\bibitem{jedel17}
\begin{barticle}
\bauthor{\bsnm{Jedele}, \binits{A.}},
\bauthor{\bsnm{McIntosh}, \binits{A.B.}},
\bauthor{\bsnm{Hagel}, \binits{K.}},
\bauthor{\bsnm{Huang}, \binits{M.}},
\bauthor{\bsnm{Heilborn}, \binits{L.}},
\bauthor{\bsnm{Kohley}, \binits{Z.}},
\bauthor{\bsnm{May}, \binits{L.W.}},
\bauthor{\bsnm{McCleskey}, \binits{E.}},
\bauthor{\bsnm{Youngs}, \binits{M.}},
\bauthor{\bsnm{Zarrella}, \binits{A.}},
\bauthor{\bsnm{Yennello}, \binits{S.J.}}:
\batitle{{Characterizing Neutron-Proton Equilibration in Nuclear Reactions with
  Subzeptosecond Resolution}}.
\bjtitle{Phys. Rev. Lett.}
\bvolume{118},
\bfpage{062501}
(\byear{2017}).
\doiurl{10.1103/PhysRevLett.118.062501}
\end{barticle}
\endbibitem

\bibitem{zhaplb08}
\begin{barticle}
\bauthor{\bsnm{Zhang}, \binits{Y.}},
\bauthor{\bsnm{Danielewicz}, \binits{P.}},
\bauthor{\bsnm{Famiano}, \binits{M.}},
\bauthor{\bsnm{Li}, \binits{Z.}},
\bauthor{\bsnm{Lynch}, \binits{W.G.}},
\bauthor{\bsnm{Tsang}, \binits{M.B.}}:
\batitle{The influence of cluster emission and the symmetry energy on
  neutron–proton spectral double ratios}.
\bjtitle{Physics Letters B}
\bvolume{664}(\bissue{1}),
\bfpage{145}--\blpage{148}
(\byear{2008}).
\doiurl{10.1016/j.physletb.2008.03.075}
\end{barticle}
\endbibitem

\bibitem{sfi09}
\begin{barticle}
\bauthor{\bsnm{Sfienti}, \binits{C.}}, \betal:
\batitle{{Isotopic Dependence of the Nuclear Caloric Curve}}.
\bjtitle{Phys. Rev. Lett.}
\bvolume{102},
\bfpage{152701}
(\byear{2009}).
\doiurl{10.1103/PhysRevLett.102.152701}
\end{barticle}
\endbibitem

\bibitem{zha15}
\begin{barticle}
\bauthor{\bsnm{Zhang}, \binits{Z.}},
\bauthor{\bsnm{Chen}, \binits{L.-W.}}:
\batitle{{Electric dipole polarizability in $^{208}\mathbf{Pb}$ as a probe of
  the symmetry energy and neutron matter around ${\ensuremath{\rho}}_{0}/3$}}.
\bjtitle{Phys. Rev. C}
\bvolume{92},
\bfpage{031301}
(\byear{2015}).
\doiurl{10.1103/PhysRevC.92.031301}
\end{barticle}
\endbibitem

\bibitem{roc13}
\begin{barticle}
\bauthor{\bsnm{Roca-Maza}, \binits{X.}},
\bauthor{\bsnm{Brenna}, \binits{M.}},
\bauthor{\bsnm{Agrawal}, \binits{B.K.}},
\bauthor{\bsnm{Bortignon}, \binits{P.F.}},
\bauthor{\bsnm{Col\`o}, \binits{G.}},
\bauthor{\bsnm{Cao}, \binits{L.-G.}},
\bauthor{\bsnm{Paar}, \binits{N.}},
\bauthor{\bsnm{Vretenar}, \binits{D.}}:
\batitle{{Giant quadrupole resonances in ${}^{208}$Pb, the nuclear symmetry
  energy, and the neutron skin thickness}}.
\bjtitle{Phys. Rev. C}
\bvolume{87},
\bfpage{034301}
(\byear{2013}).
\doiurl{10.1103/PhysRevC.87.034301}
\end{barticle}
\endbibitem

\bibitem{fan14}
\begin{barticle}
\bauthor{\bsnm{Fan}, \binits{X.}},
\bauthor{\bsnm{Dong}, \binits{J.}},
\bauthor{\bsnm{Zuo}, \binits{W.}}:
\batitle{Density-dependent symmetry energy at subsaturation densities from
  nuclear mass differences}.
\bjtitle{Phys. Rev. C}
\bvolume{89},
\bfpage{017305}
(\byear{2014}).
\doiurl{10.1103/PhysRevC.89.017305}
\end{barticle}
\endbibitem

\bibitem{adh22}
\begin{barticle}
\bauthor{\bsnm{Adhikari}, \binits{D.}}, \betal:
\batitle{Precision determination of the neutral weak form factor of
  $^{48}\mathrm{Ca}$}.
\bjtitle{Phys. Rev. Lett.}
\bvolume{129},
\bfpage{042501}
(\byear{2022}).
\doiurl{10.1103/PhysRevLett.129.042501}
\end{barticle}
\endbibitem

\bibitem{heb13}
\begin{barticle}
\bauthor{\bsnm{Hebeler}, \binits{K.}},
\bauthor{\bsnm{Lattimer}, \binits{J.M.}},
\bauthor{\bsnm{Pethick}, \binits{C.J.}},
\bauthor{\bsnm{Schwenk}, \binits{A.}}:
\batitle{{Equation of state and neutron star properties constrained by nuclear
  physics and observations}}.
\bjtitle{The Astrophysical Journal}
\bvolume{773}(\bissue{1}),
\bfpage{11}
(\byear{2013}).
\doiurl{10.1088/0004-637x/773/1/11}
\end{barticle}
\endbibitem

\bibitem{duc11}
\begin{barticle}
\bauthor{\bsnm{Ducoin}, \binits{C.}},
\bauthor{\bsnm{Margueron}, \binits{J.}},
\bauthor{\bsnm{Provid\^encia}, \binits{C.}},
\bauthor{\bsnm{Vida\~na}, \binits{I.}}:
\batitle{{Core-crust transition in neutron stars: Predictivity of density
  developments}}.
\bjtitle{Phys. Rev. C}
\bvolume{83},
\bfpage{045810}
(\byear{2011}).
\doiurl{10.1103/PhysRevC.83.045810}
\end{barticle}
\endbibitem

\bibitem{carb10}
\begin{barticle}
\bauthor{\bsnm{Carbone}, \binits{A.}},
\bauthor{\bsnm{Col\`o}, \binits{G.}},
\bauthor{\bsnm{Bracco}, \binits{A.}},
\bauthor{\bsnm{Cao}, \binits{L.-G.}},
\bauthor{\bsnm{Bortignon}, \binits{P.F.}},
\bauthor{\bsnm{Camera}, \binits{F.}},
\bauthor{\bsnm{Wieland}, \binits{O.}}:
\batitle{{Constraints on the symmetry energy and neutron skins from pygmy
  resonances in $^{68}\mathrm{Ni}$ and $^{132}\mathrm{Sn}$}}.
\bjtitle{Phys. Rev. C}
\bvolume{81},
\bfpage{041301}
(\byear{2010}).
\doiurl{10.1103/PhysRevC.81.041301}
\end{barticle}
\endbibitem

\bibitem{ono19}
\begin{barticle}
\bauthor{\bsnm{Ono}, \binits{A.}}:
\batitle{Dynamics of clusters and fragments in heavy-ion collisions}.
\bjtitle{Progress in Particle and Nuclear Physics}
\bvolume{105},
\bfpage{139}--\blpage{179}
(\byear{2019}).
\doiurl{10.1016/j.ppnp.2018.11.001}
\end{barticle}
\endbibitem

\bibitem{Ferrini:2005jw}
\begin{barticle}
\bauthor{\bsnm{Ferini}, \binits{G.}},
\bauthor{\bsnm{Colonna}, \binits{M.}},
\bauthor{\bsnm{Gaitanos}, \binits{T.}},
\bauthor{\bsnm{Di~Toro}, \binits{M.}}:
\batitle{{Aspects of particle production in charge asymmetric matter}}.
\bjtitle{Nucl. Phys. A}
\bvolume{762},
\bfpage{147}--\blpage{166}
(\byear{2005}).
\doiurl{10.1016/j.nuclphysa.2005.08.007}
\end{barticle}
\endbibitem

\bibitem{Cozma:2014yna}
\begin{barticle}
\bauthor{\bsnm{Cozma}, \binits{M.D.}}:
\batitle{{The impact of energy conservation in transport models on the
  $\pi^-/\pi^+$ multiplicity ratio in heavy-ion collisions and the symmetry
  energy}}.
\bjtitle{Phys. Lett. B}
\bvolume{753},
\bfpage{166}--\blpage{172}
(\byear{2016}).
\doiurl{10.1016/j.physletb.2015.12.015}
\end{barticle}
\endbibitem

\bibitem{Zhang:2017nck}
\begin{barticle}
\bauthor{\bsnm{Zhang}, \binits{Z.}},
\bauthor{\bsnm{Ko}, \binits{C.M.}}:
\batitle{{Effects of energy conservation on equilibrium properties of hot
  asymmetric nuclear matter}}.
\bjtitle{Phys. Rev. C}
\bvolume{97}(\bissue{1}),
\bfpage{014610}
(\byear{2018}).
\doiurl{10.1103/PhysRevC.97.014610}
\end{barticle}
\endbibitem

\bibitem{Zhang:2017mps}
\begin{barticle}
\bauthor{\bsnm{Zhang}, \binits{Z.}},
\bauthor{\bsnm{Ko}, \binits{C.M.}}:
\batitle{{Medium effects on pion production in heavy ion collisions}}.
\bjtitle{Phys. Rev. C}
\bvolume{95}(\bissue{6}),
\bfpage{064604}
(\byear{2017}).
\doiurl{10.1103/PhysRevC.95.064604}
\end{barticle}
\endbibitem

\bibitem{Zhang:2018ool}
\begin{barticle}
\bauthor{\bsnm{Zhang}, \binits{Z.}},
\bauthor{\bsnm{Ko}, \binits{C.M.}}:
\batitle{{Pion production in a transport model based on mean fields from chiral
  effective field theory}}.
\bjtitle{Phys. Rev. C}
\bvolume{98}(\bissue{5}),
\bfpage{054614}
(\byear{2018}).
\doiurl{10.1103/PhysRevC.98.054614}
\end{barticle}
\endbibitem

\bibitem{Cozma:2016qej}
\begin{barticle}
\bauthor{\bsnm{Cozma}, \binits{M.D.}}:
\batitle{{Constraining the density dependence of the symmetry energy using the
  multiplicity and average $p_T$ ratios of charged pions}}.
\bjtitle{Phys. Rev. C}
\bvolume{95}(\bissue{1}),
\bfpage{014601}
(\byear{2017}).
\doiurl{10.1103/PhysRevC.95.014601}
\end{barticle}
\endbibitem

\bibitem{Shane:2014tsa}
\begin{barticle}
\bauthor{\bsnm{Shane}, \binits{R.}}, \betal:
\batitle{{S$\pi$RIT: A time-projection chamber for symmetry-energy studies}}.
\bjtitle{Nucl. Instrum. Meth. A}
\bvolume{784},
\bfpage{513}
(\byear{2015}).
\doiurl{10.1016/j.nima.2015.01.026}
\end{barticle}
\endbibitem

\bibitem{SpRIT:2016aqk}
\begin{barticle}
\bauthor{\bsnm{Tangwancharoen}, \binits{S.}}, \betal:
\batitle{{A Gating Grid Driver for Time Projection Chambers}}.
\bjtitle{Nucl. Instrum. Meth. A}
\bvolume{853},
\bfpage{44}--\blpage{52}
(\byear{2017}).
\doiurl{10.1016/j.nima.2017.02.001}
\end{barticle}
\endbibitem

\bibitem{Barney:2020mxk}
\begin{barticle}
\bauthor{\bsnm{Barney}, \binits{J.}}, \betal:
\batitle{{The S\ensuremath{\pi}RIT time projection chamber}}.
\bjtitle{Rev. Sci. Instrum.}
\bvolume{92}(\bissue{6}),
\bfpage{063302}
(\byear{2021}).
\doiurl{10.1063/5.0041191}
\end{barticle}
\endbibitem

\bibitem{Yong:2021nwn}
\begin{barticle}
\bauthor{\bsnm{Yong}, \binits{G.-C.}}:
\batitle{{Symmetry energy extracted from the S\ensuremath{\pi}RIT pion data in
  Sn+Sn systems}}.
\bjtitle{Phys. Rev. C}
\bvolume{104}(\bissue{1}),
\bfpage{014613}
(\byear{2021}).
\doiurl{10.1103/PhysRevC.104.014613}
\end{barticle}
\endbibitem

\bibitem{Wei:2021arw}
\begin{botherref}
\oauthor{\bsnm{Wei}, \binits{G.-F.}},
\oauthor{\bsnm{Huang}, \binits{X.}},
\oauthor{\bsnm{Zhi}, \binits{Q.-J.}},
\oauthor{\bsnm{Dong}, \binits{A.-J.}},
\oauthor{\bsnm{Peng}, \binits{C.-G.}},
\oauthor{\bsnm{Long}, \binits{Z.-W.}}:
{Constraints on the momentum dependence of nuclear symmetry potential from Sn +
  Sn collisions at 270 MeV/nucleon}
(2021)
{\href{https://arxiv.org/abs/2112.13518}{{arXiv:2112.13518}}}
{[nucl-th]}
\end{botherref}
\endbibitem

\bibitem{Teis:1996kx}
\begin{barticle}
\bauthor{\bsnm{Teis}, \binits{S.}},
\bauthor{\bsnm{Cassing}, \binits{W.}},
\bauthor{\bsnm{Effenberger}, \binits{M.}},
\bauthor{\bsnm{Hombach}, \binits{A.}},
\bauthor{\bsnm{Mosel}, \binits{U.}},
\bauthor{\bsnm{Wolf}, \binits{G.}}:
\batitle{{Pion production in heavy ion collisions at SIS energies}}.
\bjtitle{Z. Physik A}
\bvolume{356},
\bfpage{421}--\blpage{435}
(\byear{1997}).
\doiurl{10.1007/s002180050198}
\end{barticle}
\endbibitem

\bibitem{Li:2005zza}
\begin{barticle}
\bauthor{\bsnm{Li}, \binits{Q.}},
\bauthor{\bsnm{Li}, \binits{Z.}},
\bauthor{\bsnm{Soff}, \binits{S.}},
\bauthor{\bsnm{Gupta}, \binits{R.K.}},
\bauthor{\bsnm{Bleicher}, \binits{M.}},
\bauthor{\bsnm{Stoecker}, \binits{H.}}:
\batitle{{Probing the density dependence of the symmetry potential in
  intermediate energy heavy ion collisions}}.
\bjtitle{J. Phys. G}
\bvolume{31},
\bfpage{1359}--\blpage{1374}
(\byear{2005}).
\doiurl{10.1088/0954-3899/31/11/016}
\end{barticle}
\endbibitem

\bibitem{Li:2005gfa}
\begin{barticle}
\bauthor{\bsnm{Li}, \binits{Q.-f.}},
\bauthor{\bsnm{Li}, \binits{Z.-x.}},
\bauthor{\bsnm{Soff}, \binits{S.}},
\bauthor{\bsnm{Bleicher}, \binits{M.}},
\bauthor{\bsnm{Stoecker}, \binits{H.}}:
\batitle{{Probing the equation of state with pions}}.
\bjtitle{J. Phys. G}
\bvolume{32}(\bissue{2}),
\bfpage{151}--\blpage{164}
(\byear{2006})
{\href{https://arxiv.org/abs/nucl-th/0509070}{{arXiv:nucl-th/0509070}}}.
\doiurl{10.1088/0954-3899/32/2/007}
\end{barticle}
\endbibitem

\bibitem{Li:2006ez}
\begin{barticle}
\bauthor{\bsnm{Li}, \binits{Q.-f.}},
\bauthor{\bsnm{Li}, \binits{Z.-x.}},
\bauthor{\bsnm{Soff}, \binits{S.}},
\bauthor{\bsnm{Bleicher}, \binits{M.}},
\bauthor{\bsnm{Stoecker}, \binits{H.}}:
\batitle{{Medium modifications of the nucleon-nucleon elastic cross section in
  neutron-rich intermediate energy HICs}}.
\bjtitle{J. Phys. G}
\bvolume{32},
\bfpage{407}--\blpage{416}
(\byear{2006})
{\href{https://arxiv.org/abs/nucl-th/0601047}{{arXiv:nucl-th/0601047}}}.
\doiurl{10.1088/0954-3899/32/4/001}
\end{barticle}
\endbibitem

\bibitem{FOPI:2004bfz}
\begin{barticle}
\bauthor{\bsnm{Andronic}, \binits{A.}}, \betal:
\batitle{{Excitation function of elliptic flow in Au+Au collisions and the
  nuclear matter equation of state}}.
\bjtitle{Phys. Lett. B}
\bvolume{612},
\bfpage{173}--\blpage{180}
(\byear{2005}).
\doiurl{10.1016/j.physletb.2005.02.060}
\end{barticle}
\endbibitem

\bibitem{Cozma:2011nr}
\begin{barticle}
\bauthor{\bsnm{Cozma}, \binits{M.D.}}:
\batitle{{Neutron-proton elliptic flow difference as a probe for the high
  density dependence of the symmetry energy}}.
\bjtitle{Phys. Lett. B}
\bvolume{700},
\bfpage{139}--\blpage{144}
(\byear{2011}).
\doiurl{10.1016/j.physletb.2011.05.002}
\end{barticle}
\endbibitem

\bibitem{Giordano:2010pv}
\begin{barticle}
\bauthor{\bsnm{Giordano}, \binits{V.}},
\bauthor{\bsnm{Colonna}, \binits{M.}},
\bauthor{\bsnm{Di~Toro}, \binits{M.}},
\bauthor{\bsnm{Greco}, \binits{V.}},
\bauthor{\bsnm{Rizzo}, \binits{J.}}:
\batitle{{Isospin emission and flows at high baryon density: a test of the
  symmetry potential}}.
\bjtitle{Phys. Rev. C}
\bvolume{81},
\bfpage{044611}
(\byear{2010}).
\doiurl{10.1103/PhysRevC.81.044611}
\end{barticle}
\endbibitem

\bibitem{FOPI:1993wdf}
\begin{barticle}
\bauthor{\bsnm{Leifels}, \binits{Y.}}, \betal:
\batitle{{Exclusive studies of neutron and charged particle emission in
  collisions of $^{197}$Au + $^{197}$Au at 400-MeV/nucleon}}.
\bjtitle{Phys. Rev. Lett.}
\bvolume{71},
\bfpage{963}--\blpage{966}
(\byear{1993}).
\doiurl{10.1103/PhysRevLett.71.963}
\end{barticle}
\endbibitem

\bibitem{FOPI:1994xpn}
\begin{barticle}
\bauthor{\bsnm{Lambrecht}, \binits{D.}}, \betal:
\batitle{{Energy dependence of collective flow of neutrons and protons in
  $^{197}$Au + $^{197}$Au collisions}}.
\bjtitle{Z. Phys. A}
\bvolume{350},
\bfpage{115}--\blpage{120}
(\byear{1994}).
\doiurl{10.1007/BF01290679}
\end{barticle}
\endbibitem

\bibitem{Wang2014-PhysRevC.89.044603}
\begin{barticle}
\bauthor{\bsnm{Wang}, \binits{Y.}},
\bauthor{\bsnm{Guo}, \binits{C.}},
\bauthor{\bsnm{Li}, \binits{Q.}},
\bauthor{\bsnm{Zhang}, \binits{H.}},
\bauthor{\bsnm{Leifels}, \binits{Y.}},
\bauthor{\bsnm{Trautmann}, \binits{W.}}:
\batitle{Constraining the high-density nuclear symmetry energy with the
  transverse-momentum-dependent elliptic flow}.
\bjtitle{Phys. Rev. C}
\bvolume{89},
\bfpage{044603}
(\byear{2014}).
\doiurl{10.1103/PhysRevC.89.044603}
\end{barticle}
\endbibitem

\bibitem{Li:2013ola}
\begin{barticle}
\bauthor{\bsnm{Li}, \binits{B.-A.}},
\bauthor{\bsnm{Han}, \binits{X.}}:
\batitle{{Constraining the neutron-proton effective mass splitting using
  empirical constraints on the density dependence of nuclear symmetry energy
  around normal density}}.
\bjtitle{Phys. Lett. B}
\bvolume{727},
\bfpage{276}--\blpage{281}
(\byear{2013}).
\doiurl{10.1016/j.physletb.2013.10.006}
\end{barticle}
\endbibitem

\bibitem{Oertel:2016bki}
\begin{barticle}
\bauthor{\bsnm{Oertel}, \binits{M.}},
\bauthor{\bsnm{Hempel}, \binits{M.}},
\bauthor{\bsnm{Kl\"ahn}, \binits{T.}},
\bauthor{\bsnm{Typel}, \binits{S.}}:
\batitle{{Equations of state for supernovae and compact stars}}.
\bjtitle{Rev. Mod. Phys.}
\bvolume{89}(\bissue{1}),
\bfpage{015007}
(\byear{2017}).
\doiurl{10.1103/RevModPhys.89.015007}
\end{barticle}
\endbibitem

\bibitem{Das:2002fr}
\begin{barticle}
\bauthor{\bsnm{Das}, \binits{C.B.}},
\bauthor{\bsnm{Gupta}, \binits{S.D.}},
\bauthor{\bsnm{Gale}, \binits{C.}},
\bauthor{\bsnm{Li}, \binits{B.-A.}}:
\batitle{{Momentum dependence of symmetry potential in asymmetric nuclear
  matter for transport model calculations}}.
\bjtitle{Phys. Rev. C}
\bvolume{67},
\bfpage{034611}
(\byear{2003}).
\doiurl{10.1103/PhysRevC.67.034611}
\end{barticle}
\endbibitem

\bibitem{Ikeno:2016xpr}
\begin{barticle}
\bauthor{\bsnm{Ikeno}, \binits{N.}},
\bauthor{\bsnm{Ono}, \binits{A.}},
\bauthor{\bsnm{Nara}, \binits{Y.}},
\bauthor{\bsnm{Ohnishi}, \binits{A.}}:
\batitle{{Probing neutron-proton dynamics by pions}}.
\bjtitle{Phys. Rev.}
\bvolume{C93}(\bissue{4}),
\bfpage{044612}
(\byear{2016}).
\doiurl{10.1103/PhysRevC.93.044612}
\end{barticle}
\endbibitem

\bibitem{Liu:2020jbg}
\begin{barticle}
\bauthor{\bsnm{Liu}, \binits{Y.}},
\bauthor{\bsnm{Wang}, \binits{Y.}},
\bauthor{\bsnm{Cui}, \binits{Y.}},
\bauthor{\bsnm{Xia}, \binits{C.-J.}},
\bauthor{\bsnm{Li}, \binits{Z.}},
\bauthor{\bsnm{Chen}, \binits{Y.}},
\bauthor{\bsnm{Li}, \binits{Q.}},
\bauthor{\bsnm{Zhang}, \binits{Y.}}:
\batitle{{Insights into the pion production mechanism and the symmetry energy
  at high density}}.
\bjtitle{Phys. Rev. C}
\bvolume{103}(\bissue{1}),
\bfpage{014616}
(\byear{2021}).
\doiurl{10.1103/PhysRevC.103.014616}
\end{barticle}
\endbibitem

\bibitem{ozel16a}
\begin{barticle}
\bauthor{\bsnm{\"{O}zel}, \binits{F.}},
\bauthor{\bsnm{Freire}, \binits{P.}}:
\batitle{{Masses, Radii, and the Equation of State of Neutron Stars}}.
\bjtitle{Annual Review of Astronomy and Astrophysics}
\bvolume{54}(\bissue{1}),
\bfpage{401}--\blpage{440}
(\byear{2016}).
\doiurl{10.1146/annurev-astro-081915-023322}
\end{barticle}
\endbibitem

\bibitem{dege18}
\begin{botherref}
\oauthor{\bsnm{Degenaar}, \binits{N.}},
\oauthor{\bsnm{Suleimanov}, \binits{V.F.}}:
{Testing the Equation of State with Electromagnetic Observations}
\textbf{457},
185--253
(2018)
{\href{https://arxiv.org/abs/1806.02833}{{arXiv:1806.02833}}}.
\doiurl{10.1007/978-3-319-97616-7_5}.
{In: Rezzolla, L., Pizzochero, P., Jones, D., Rea, N., Vida\~na, I. (eds) The
  Physics and Astrophysics of Neutron Stars. Astrophysics and Space Science
  Library}
\end{botherref}
\endbibitem

\bibitem{latt21}
\begin{barticle}
\bauthor{\bsnm{Lattimer}, \binits{J.M.}}:
\batitle{Neutron stars and the nuclear matter equation of state}.
\bjtitle{Annual Review of Nuclear and Particle Science}
\bvolume{71}(\bissue{1}),
\bfpage{433}--\blpage{464}
(\byear{2021})
{\href{https://arxiv.org/abs/https://doi.org/10.1146/annurev-nucl-102419-124827}{{https://doi.org/10.1146/annurev-nucl-102419-124827}}}
\end{barticle}
\endbibitem

\bibitem{lind92}
\begin{barticle}
\bauthor{\bsnm{Lindblom}, \binits{L.}}:
\batitle{{Determining the Nuclear Equation of State from Neutron-Star Masses
  and Radii }}.
\bjtitle{Astrophys. J.}
\bvolume{398},
\bfpage{569}
(\byear{1992})
{\href{https://arxiv.org/abs/https://ui.adsabs.harvard.edu/abs/1992ApJ...398..569L}{{https://ui.adsabs.harvard.edu/abs/1992ApJ...398..569L}}}
\end{barticle}
\endbibitem

\bibitem{LAT07}
\begin{barticle}
\bauthor{\bsnm{{James M. Lattimer and Madappa Prakash}}}:
\batitle{{Neutron star observations: Prognosis for equation of state
  constraints}}.
\bjtitle{Physics Reports}
\bvolume{442}(\bissue{1}),
\bfpage{109}--\blpage{165}
(\byear{2007}).
\doiurl{10.1016/j.physrep.2007.02.003}
\end{barticle}
\endbibitem

\bibitem{Gal08}
\begin{barticle}
\bauthor{\bsnm{Galloway}, \binits{D.K.}},
\bauthor{\bsnm{\"{O}zel}, \binits{F.}},
\bauthor{\bsnm{Psaltis}, \binits{D.}}:
\batitle{{Biases for neutron star mass, radius and distance measurements from
  Eddington-limited X-ray bursts}}.
\bjtitle{MNRAS}
\bvolume{387}(\bissue{1}),
\bfpage{268}--\blpage{272}
(\byear{2008})
{\href{https://arxiv.org/abs/https://academic.oup.com/mnras/article-pdf/387/1/268/3202628/mnras0387-0268.pdf}{{https://academic.oup.com/mnras/article-pdf/387/1/268/3202628/mnras0387-0268.pdf}}}
\end{barticle}
\endbibitem

\bibitem{bali21}
\begin{botherref}
\oauthor{\bsnm{Li}, \binits{B.-A.}}, et al.:
{Progress in constraining nuclear symmetry energy using neutron star
  observables since GW170817}.
Universe
\textbf{7}(6)
(2021).
\doiurl{10.3390/universe7060182}
\end{botherref}
\endbibitem

\bibitem{luk18}
\begin{barticle}
\bauthor{\bsnm{\L{}ukasik}, \binits{J.}}:
\batitle{{Constraints on the density dependence of the symmetry energy}}.
\bjtitle{Il Nuovo Cimento C}
\bvolume{41},
\bfpage{182}
(\byear{2018}).
\doiurl{10.1393/ncc/i2018-18182-8}
\end{barticle}
\endbibitem

\bibitem{gui13}
\begin{barticle}
\bauthor{\bsnm{Guillot}, \binits{S.}},
\bauthor{\bsnm{Servillat}, \binits{M.}},
\bauthor{\bsnm{Webb}, \binits{N.A.}},
\bauthor{\bsnm{Rutledge}, \binits{R.E.}}:
\batitle{Measurement of the radius of neutron stars with high signal-to-noise
  quiescent low-mass {X}-ray binaries in globular clusters}.
\bjtitle{The Astrophysical Journal}
\bvolume{772}(\bissue{1}),
\bfpage{7}
(\byear{2013}).
\doiurl{10.1088/0004-637x/772/1/7}
\end{barticle}
\endbibitem

\bibitem{Steiner:2012xt}
\begin{barticle}
\bauthor{\bsnm{Steiner}, \binits{A.W.}},
\bauthor{\bsnm{Lattimer}, \binits{J.M.}},
\bauthor{\bsnm{Brown}, \binits{E.F.}}:
\batitle{{{The Neutron Star Mass-Radius Relation and the Equation of State of
  Dense Matter}}}.
\bjtitle{Astrophys. J. Lett.}
\bvolume{765},
\bfpage{5}
(\byear{2013}).
\doiurl{10.1088/2041-8205/765/1/L5}
\end{barticle}
\endbibitem

\bibitem{gui16}
\begin{barticle}
\bauthor{\bsnm{Guillot}, \binits{S.}}:
\batitle{Neutron stars in globular clusters as tests of nuclear physics}.
\bjtitle{Memorie della Societa Astronomica Italiana}
\bvolume{87},
\bfpage{521}
(\byear{2016})
{\href{https://arxiv.org/abs/https://ui.adsabs.harvard.edu/abs/2016MmSAI..87..521G/abstract}{{https://ui.adsabs.harvard.edu/abs/2016MmSAI..87..521G/abstract}}}
\end{barticle}
\endbibitem

\bibitem{ozel16}
\begin{barticle}
\bauthor{\bsnm{\"{O}zel}, \binits{F.}}, \betal:
\batitle{The dense matter equation of state from neutron star radius and mass
  measurements}.
\bjtitle{The Astrophysical Journal}
\bvolume{820}(\bissue{1}),
\bfpage{28}
(\byear{2016}).
\doiurl{10.3847/0004-637x/820/1/28}
\end{barticle}
\endbibitem

\bibitem{bog16}
\begin{barticle}
\bauthor{\bsnm{Bogdanov}, \binits{S.}}, \betal:
\batitle{Neutron star mass\textendash radius constraints of the quiescent
  low-mass {X}-ray binaries {X7} and {X5} in the globular cluster 47 {Tuc}}.
\bjtitle{The Astrophysical Journal}
\bvolume{831}(\bissue{2}),
\bfpage{184}
(\byear{2016}).
\doiurl{10.3847/0004-637x/831/2/184}
\end{barticle}
\endbibitem

\bibitem{shaw18}
\begin{barticle}
\bauthor{\bsnm{Shaw}, \binits{A.W.}}, \betal:
\batitle{{The radius of the quiescent neutron star in the globular cluster
  M13}}.
\bjtitle{MNRAS}
\bvolume{476}(\bissue{4}),
\bfpage{4713}--\blpage{4718}
(\byear{2018}).
\doiurl{10.1093/mnras/sty582}
\end{barticle}
\endbibitem

\bibitem{ste18}
\begin{barticle}
\bauthor{\bsnm{Steiner}, \binits{A.W.}}, \betal:
\batitle{{Constraining the mass and radius of neutron stars in globular
  clusters}}.
\bjtitle{MNRAS}
\bvolume{476}(\bissue{1}),
\bfpage{421}--\blpage{435}
(\byear{2018}).
\doiurl{10.1093/mnras/sty215}
\end{barticle}
\endbibitem

\bibitem{kim21}
\begin{barticle}
\bauthor{\bsnm{Kim}, \binits{M.}}, \betal:
\batitle{{Measuring the masses and radii of neutron stars in low-mass X-ray
  binaries: Effects of the atmospheric composition and touchdown radius}}.
\bjtitle{A\&A}
\bvolume{650},
\bfpage{139}
(\byear{2021}).
\doiurl{10.1051/0004-6361/202038126}
\end{barticle}
\endbibitem

\bibitem{nat17}
\begin{barticle}
\bauthor{\bsnm{{N\"attil\"a, J.}}}, \betal:
\batitle{{Neutron star mass and radius measurements from atmospheric model fits
  to X-ray burst cooling tail spectra}}.
\bjtitle{A\&A}
\bvolume{608},
\bfpage{31}
(\byear{2017}).
\doiurl{10.1051/0004-6361/201731082}
\end{barticle}
\endbibitem

\bibitem{bai19}
\begin{barticle}
\bauthor{\bsnm{{Baillot d'Etivaux}}, \binits{N.}}, \betal:
\batitle{{New Constraints on the Nuclear Equation of State from the Thermal
  Emission of Neutron Stars in Quiescent Low-mass X-Ray Binaries}}.
\bjtitle{The Astrophysical Journal}
\bvolume{887}(\bissue{1}),
\bfpage{48}
(\byear{2019}).
\doiurl{10.3847/1538-4357/ab4f6c}
\end{barticle}
\endbibitem

\bibitem{mos18}
\begin{barticle}
\bauthor{\bsnm{Most}, \binits{E.R.}}, \betal:
\batitle{{New Constraints on Radii and Tidal Deformabilities of Neutron Stars
  from GW170817}}.
\bjtitle{Phys. Rev. Lett.}
\bvolume{120},
\bfpage{261103}
(\byear{2018}).
\doiurl{10.1103/PhysRevLett.120.261103}
\end{barticle}
\endbibitem

\bibitem{lim18}
\begin{barticle}
\bauthor{\bsnm{Lim}, \binits{Y.}},
\bauthor{\bsnm{Holt}, \binits{J.W.}}:
\batitle{{Neutron Star Tidal Deformabilities Constrained by Nuclear Theory and
  Experiment}}.
\bjtitle{Phys. Rev. Lett.}
\bvolume{121},
\bfpage{062701}
(\byear{2018}).
\doiurl{10.1103/PhysRevLett.121.062701}
\end{barticle}
\endbibitem

\bibitem{LIGOScientific:2018cki}
\begin{barticle}
\bauthor{\bsnm{Abbott}, \binits{B.P.}}, \betal:
\batitle{{GW170817: Measurements of neutron star radii and equation of state}}.
\bjtitle{Phys. Rev. Lett.}
\bvolume{121}(\bissue{16}),
\bfpage{161101}
(\byear{2018})
{\href{https://arxiv.org/abs/1805.11581}{{arXiv:1805.11581}}}
{[gr-qc]}.
\doiurl{10.1103/PhysRevLett.121.161101}
\end{barticle}
\endbibitem

\bibitem{ann18}
\begin{barticle}
\bauthor{\bsnm{Annala}, \binits{E.}},
\bauthor{\bsnm{Gorda}, \binits{T.}},
\bauthor{\bsnm{Kurkela}, \binits{A.}},
\bauthor{\bsnm{Vuorinen}, \binits{A.}}:
\batitle{{Gravitational-Wave Constraints on the Neutron-Star-Matter Equation of
  State}}.
\bjtitle{Phys. Rev. Lett.}
\bvolume{120},
\bfpage{172703}
(\byear{2018}).
\doiurl{10.1103/PhysRevLett.120.172703}
\end{barticle}
\endbibitem

\bibitem{tews18}
\begin{barticle}
\bauthor{\bsnm{Tews}, \binits{I.}},
\bauthor{\bsnm{Margueron}, \binits{J.}},
\bauthor{\bsnm{Reddy}, \binits{S.}}:
\batitle{{Critical examination of constraints on the equation of state of dense
  matter obtained from GW170817}}.
\bjtitle{Phys. Rev. C}
\bvolume{98},
\bfpage{045804}
(\byear{2018}).
\doiurl{10.1103/PhysRevC.98.045804}
\end{barticle}
\endbibitem

\bibitem{de18}
\begin{barticle}
\bauthor{\bsnm{De}, \binits{S.}},
\bauthor{\bsnm{Finstad}, \binits{D.}},
\bauthor{\bsnm{Lattimer}, \binits{J.M.}},
\bauthor{\bsnm{Brown}, \binits{D.A.}},
\bauthor{\bsnm{Berger}, \binits{E.}},
\bauthor{\bsnm{Biwer}, \binits{C.M.}}:
\batitle{{Tidal Deformabilities and Radii of Neutron Stars from the Observation
  of GW170817}}.
\bjtitle{Phys. Rev. Lett.}
\bvolume{121},
\bfpage{091102}
(\byear{2018}).
\doiurl{10.1103/PhysRevLett.121.091102}
\end{barticle}
\endbibitem

\bibitem{rad19}
\begin{barticle}
\bauthor{\bsnm{Radice}, \binits{D.}},
\bauthor{\bsnm{Dai}, \binits{L.}}:
\batitle{{Multimessenger parameter estimation of GW170817}}.
\bjtitle{Eur. Phys. J. A}
\bvolume{55},
\bfpage{50}
(\byear{2019}).
\doiurl{10.1140/epja/i2019-12716-4}
\end{barticle}
\endbibitem

\bibitem{cap20}
\begin{barticle}
\bauthor{\bsnm{Capano}, \binits{C.D.}}, \betal:
\batitle{Stringent constraints on neutron-star radii from multimessenger
  observations and nuclear theory}.
\bjtitle{Nature Astronomy}
\bvolume{4},
\bfpage{625}
(\byear{2020}).
\doiurl{10.1038/s41550-020-1014-6}
\end{barticle}
\endbibitem

\bibitem{ess20}
\begin{barticle}
\bauthor{\bsnm{Essick}, \binits{R.}},
\bauthor{\bsnm{Tews}, \binits{I.}},
\bauthor{\bsnm{Landry}, \binits{P.}},
\bauthor{\bsnm{Reddy}, \binits{S.}},
\bauthor{\bsnm{Holz}, \binits{D.E.}}:
\batitle{{Direct astrophysical tests of chiral effective field theory at
  supranuclear densities}}.
\bjtitle{Phys. Rev. C}
\bvolume{102},
\bfpage{055803}
(\byear{2020}).
\doiurl{10.1103/PhysRevC.102.055803}
\end{barticle}
\endbibitem

\bibitem{gon19}
\begin{barticle}
\bauthor{\bsnm{D.~Gonz\'{a}lez-Caniulef}, \binits{D.}},
\bauthor{\bsnm{Guillot}, \binits{S.}},
\bauthor{\bsnm{Reisenegger}, \binits{A.}}:
\batitle{{Neutron star radius measurement from the ultraviolet and soft X-ray
  thermal emission of PSR J0437-4715}}.
\bjtitle{MNRAS}
\bvolume{490}(\bissue{4}),
\bfpage{5848}--\blpage{5859}
(\byear{2019}).
\doiurl{10.1093/mnras/stz2941}
\end{barticle}
\endbibitem

\bibitem{ril21}
\begin{barticle}
\bauthor{\bsnm{Riley}, \binits{T.E.}}, \betal:
\batitle{{A NICER View of the Massive Pulsar PSR J0740+6620 Informed by Radio
  Timing and XMM-Newton Spectroscopy}}.
\bjtitle{The Astrophysical Journal Letters}
\bvolume{918}(\bissue{2}),
\bfpage{27}
(\byear{2021}).
\doiurl{10.3847/2041-8213/ac0a81}
\end{barticle}
\endbibitem

\bibitem{mil21}
\begin{barticle}
\bauthor{\bsnm{Miller}, \binits{M.C.}}, \betal:
\batitle{{The Radius of PSR J0740+6620 from NICER and XMM-Newton Data}}.
\bjtitle{The Astrophysical Journal Letters}
\bvolume{918}(\bissue{2}),
\bfpage{28}
(\byear{2021}).
\doiurl{10.3847/2041-8213/ac089b}
\end{barticle}
\endbibitem

\bibitem{jiang20}
\begin{barticle}
\bauthor{\bsnm{Jiang}, \binits{J.-L.}}, \betal:
\batitle{{PSR J0030+0451, {GW}170817, and the Nuclear Data: Joint Constraints
  on Equation of State and Bulk Properties of Neutron Stars}}.
\bjtitle{The Astrophysical Journal}
\bvolume{892}(\bissue{1}),
\bfpage{55}
(\byear{2020}).
\doiurl{10.3847/1538-4357/ab77cf}
\end{barticle}
\endbibitem

\bibitem{lan20}
\begin{barticle}
\bauthor{\bsnm{Landry}, \binits{P.}},
\bauthor{\bsnm{Essick}, \binits{R.}},
\bauthor{\bsnm{Chatziioannou}, \binits{K.}}:
\batitle{{Nonparametric constraints on neutron star matter with existing and
  upcoming gravitational wave and pulsar observations}}.
\bjtitle{Phys. Rev. D}
\bvolume{101},
\bfpage{123007}
(\byear{2020}).
\doiurl{10.1103/PhysRevD.101.123007}
\end{barticle}
\endbibitem

\bibitem{die20}
\begin{barticle}
\bauthor{\bsnm{Dietrich}, \binits{T.}}, \betal:
\batitle{{Multimessenger constraints on the neutron-star equation of state and
  the Hubble constant}}.
\bjtitle{Science}
\bvolume{370}(\bissue{6523}),
\bfpage{1450}--\blpage{1453}
(\byear{2020}).
\doiurl{10.1126/science.abb4317}
\end{barticle}
\endbibitem

\bibitem{raa21}
\begin{barticle}
\bauthor{\bsnm{Raaijmakers}, \binits{G.}}, \betal:
\batitle{{Constraints on the Dense Matter Equation of State and Neutron Star
  Properties from {NICER}'s Mass{\textendash}Radius Estimate of {PSR}
  J0740+6620 and Multimessenger Observations}}.
\bjtitle{The Astrophysical Journal Letters}
\bvolume{918}(\bissue{2}),
\bfpage{29}
(\byear{2021}).
\doiurl{10.3847/2041-8213/ac089a}
\end{barticle}
\endbibitem

\bibitem{tan21}
\begin{barticle}
\bauthor{\bsnm{Tang}, \binits{S.-P.}}, \betal:
\batitle{Constraint on phase transition with the multimessenger data of neutron
  stars}.
\bjtitle{Phys. Rev. D}
\bvolume{103},
\bfpage{063026}
(\byear{2021}).
\doiurl{10.1103/PhysRevD.103.063026}
\end{barticle}
\endbibitem

\bibitem{bis21}
\begin{barticle}
\bauthor{\bsnm{Biswas}, \binits{B.}}:
\batitle{{Impact of {PREX}-{II} and Combined Radio/{NICER}/{XMM}-Newton's
  Mass{\textendash}radius Measurement of {PSR} J0740+6620 on the Dense-matter
  Equation of State}}.
\bjtitle{The Astrophysical Journal}
\bvolume{921}(\bissue{1}),
\bfpage{63}
(\byear{2021}).
\doiurl{10.3847/1538-4357/ac1c72}
\end{barticle}
\endbibitem

\bibitem{alm21}
\begin{barticle}
\bauthor{\bsnm{Al-Mamun}, \binits{M.}}, \betal:
\batitle{{Combining Electromagnetic and Gravitational-Wave Constraints on
  Neutron-Star Masses and Radii}}.
\bjtitle{Phys. Rev. Lett.}
\bvolume{126},
\bfpage{061101}
(\byear{2021}).
\doiurl{10.1103/PhysRevLett.126.061101}
\end{barticle}
\endbibitem

\bibitem{rai21}
\begin{barticle}
\bauthor{\bsnm{Raithel}, \binits{C.A.}},
\bauthor{\bsnm{\"{O}zel}, \binits{F.}},
\bauthor{\bsnm{Psaltis}, \binits{D.}}:
\batitle{{Optimized Statistical Approach for Comparing Multi-messenger Neutron
  Star Data}}.
\bjtitle{The Astrophysical Journal}
\bvolume{908}(\bissue{1}),
\bfpage{103}
(\byear{2021}).
\doiurl{10.3847/1538-4357/abd3a4}
\end{barticle}
\endbibitem

\bibitem{Trautmann:2019gmh}
\begin{barticle}
\bauthor{\bsnm{Trautmann}, \binits{W.}}:
\batitle{{High Density with Elliptic Flows}}.
\bjtitle{AIP Conf. Proc.}
\bvolume{2127}(\bissue{1}),
\bfpage{020003}
(\byear{2019})
{\href{https://arxiv.org/abs/1903.12543}{{arXiv:1903.12543}}}
{[nucl-ex]}.
\doiurl{10.1063/1.5117793}
\end{barticle}
\endbibitem

\bibitem{gui14}
\begin{barticle}
\bauthor{\bsnm{Guillot}, \binits{S.}},
\bauthor{\bsnm{Rutledge}, \binits{R.E.}}:
\batitle{Rejecting proposed dense matter equations of state with quiescent
  low-mass {X}-ray binaries}.
\bjtitle{The Astrophysical Journal}
\bvolume{796}(\bissue{1}),
\bfpage{3}
(\byear{2014}).
\doiurl{10.1088/2041-8205/796/1/l3}
\end{barticle}
\endbibitem

\bibitem{gwem17}
\begin{barticle}
\bauthor{\bsnm{Abbott}, \binits{B.P.}}, \betal:
\batitle{{Multi-messenger Observations of a Binary Neutron Star Merger}}.
\bjtitle{The Astrophysical Journal}
\bvolume{848}(\bissue{2}),
\bfpage{12}
(\byear{2017}).
\doiurl{10.3847/2041-8213/aa91c9}
\end{barticle}
\endbibitem

\bibitem{Dem10}
\begin{barticle}
\bauthor{\bsnm{Demorest}, \binits{P.B.}},
\bauthor{\bsnm{Pennucci}, \binits{T.}},
\bauthor{\bsnm{Ransom}, \binits{S.M.}},
\bauthor{\bsnm{Roberts}, \binits{M.S.E.}},
\bauthor{\bsnm{Hessels}, \binits{J.W.T.}}:
\batitle{{A two-solar-mass neutron star measured using Shapiro delay}}.
\bjtitle{Nature}
\bvolume{467},
\bfpage{1081}
(\byear{2010}).
\doiurl{10.1038/nature09466}
\end{barticle}
\endbibitem

\bibitem{Krastev:2018nwr}
\begin{barticle}
\bauthor{\bsnm{Krastev}, \binits{P.G.}},
\bauthor{\bsnm{Li}, \binits{B.-A.}}:
\batitle{{Imprints of the nuclear symmetry energy on the tidal deformability of
  neutron stars}}.
\bjtitle{J. Phys. G}
\bvolume{46}(\bissue{7}),
\bfpage{074001}
(\byear{2019}).
\doiurl{10.1088/1361-6471/ab1a7a}
\end{barticle}
\endbibitem

\bibitem{Zhang:2018vbw}
\begin{barticle}
\bauthor{\bsnm{Zhang}, \binits{N.-B.}},
\bauthor{\bsnm{Li}, \binits{B.-A.}}:
\batitle{{Delineating Effects of Nuclear Symmetry Energy on the Radii and Tidal
  Polarizabilities of Neutron Stars}}.
\bjtitle{J. Phys. G}
\bvolume{46}(\bissue{1}),
\bfpage{014002}
(\byear{2019}).
\doiurl{10.1088/1361-6471/aaef54}
\end{barticle}
\endbibitem

\bibitem{Forbes-PhysRevD.100.083010}
\begin{barticle}
\bauthor{\bsnm{Forbes}, \binits{M.M.}},
\bauthor{\bsnm{Bose}, \binits{S.}},
\bauthor{\bsnm{Reddy}, \binits{S.}},
\bauthor{\bsnm{Zhou}, \binits{D.}},
\bauthor{\bsnm{Mukherjee}, \binits{A.}},
\bauthor{\bsnm{De}, \binits{S.}}:
\batitle{Constraining the neutron-matter equation of state with gravitational
  waves}.
\bjtitle{Phys. Rev. D}
\bvolume{100},
\bfpage{083010}
(\byear{2019}).
\doiurl{10.1103/PhysRevD.100.083010}
\end{barticle}
\endbibitem

\bibitem{Lattimer:2015nhk}
\begin{barticle}
\bauthor{\bsnm{Lattimer}, \binits{J.M.}},
\bauthor{\bsnm{Prakash}, \binits{M.}}:
\batitle{{The Equation of State of Hot, Dense Matter and Neutron Stars}}.
\bjtitle{Phys. Rept.}
\bvolume{621},
\bfpage{127}--\blpage{164}
(\byear{2016}).
\doiurl{10.1016/j.physrep.2015.12.005}
\end{barticle}
\endbibitem

\bibitem{tews17}
\begin{barticle}
\bauthor{\bsnm{Tews}, \binits{I.}},
\bauthor{\bsnm{Lattimer}, \binits{J.M.}},
\bauthor{\bsnm{Ohnishi}, \binits{A.}},
\bauthor{\bsnm{Kolomeitsev}, \binits{E.E.}}:
\batitle{{Symmetry Parameter Constraints from a Lower Bound on Neutron-matter
  Energy}}.
\bjtitle{The Astrophysical Journal}
\bvolume{848}(\bissue{2}),
\bfpage{105}
(\byear{2017}).
\doiurl{10.3847/1538-4357/aa8db9}
\end{barticle}
\endbibitem

\bibitem{sot22}
\begin{botherref}
\oauthor{\bsnm{Sotani}, \binits{H.}},
\oauthor{\bsnm{Nishimura}, \binits{N.}},
\oauthor{\bsnm{Naito}, \binits{T.}}:
{New constraints on the neutron-star mass and radius relation from terrestrial
  nuclear experiments}.
Progress of Theoretical and Experimental Physics
\textbf{2022}(4)
(2022)
{\href{https://arxiv.org/abs/arXiv:2203.05410 [nucl-th]}{{arXiv:2203.05410
  [nucl-th]}}}.
\doiurl{10.1093/ptep/ptac055}.
041D01
\end{botherref}
\endbibitem

\bibitem{sot22d}
\begin{barticle}
\bauthor{\bsnm{Sotani}, \binits{H.}},
\bauthor{\bsnm{Togashi}, \binits{H.}}:
\batitle{Neutron star mass formula with nuclear saturation parameters}.
\bjtitle{Phys. Rev. D}
\bvolume{105},
\bfpage{063010}
(\byear{2022}).
\doiurl{10.1103/PhysRevD.105.063010}
\end{barticle}
\endbibitem

\bibitem{Zhang:2019fog}
\begin{barticle}
\bauthor{\bsnm{Zhang}, \binits{N.-B.}},
\bauthor{\bsnm{Li}, \binits{B.-A.}}:
\batitle{{Implications of the Mass $M=2.17^{+0.11}_{-0.10}$M$_\odot$ of PSR
  J0740+6620 on the Equation of State of Super-dense Neutron-rich Nuclear
  Matter}}.
\bjtitle{Astrophys. J.}
\bvolume{879}(\bissue{2}),
\bfpage{99}
(\byear{2019}).
\doiurl{10.3847/1538-4357/ab24cb}
\end{barticle}
\endbibitem

\bibitem{Xie:2019sqb}
\begin{barticle}
\bauthor{\bsnm{Xie}, \binits{W.-J.}},
\bauthor{\bsnm{Li}, \binits{B.-A.}}:
\batitle{{Bayesian Inference of High-density Nuclear Symmetry Energy from Radii
  of Canonical Neutron Stars}}.
\bjtitle{Astrophys. J.}
\bvolume{883},
\bfpage{174}
(\byear{2019}).
\doiurl{10.3847/1538-4357/ab3f37}
\end{barticle}
\endbibitem

\bibitem{Wiringa:1988tp}
\begin{barticle}
\bauthor{\bsnm{Wiringa}, \binits{R.B.}},
\bauthor{\bsnm{Fiks}, \binits{V.}},
\bauthor{\bsnm{Fabrocini}, \binits{A.}}:
\batitle{{Equation of state for dense nucleon matter}}.
\bjtitle{Phys. Rev. C}
\bvolume{38},
\bfpage{1010}--\blpage{1037}
(\byear{1988}).
\doiurl{10.1103/PhysRevC.38.1010}
\end{barticle}
\endbibitem

\bibitem{Akmal:1998cf}
\begin{barticle}
\bauthor{\bsnm{Akmal}, \binits{A.}},
\bauthor{\bsnm{Pandharipande}, \binits{V.R.}},
\bauthor{\bsnm{Ravenhall}, \binits{D.G.}}:
\batitle{{The Equation of state of nucleon matter and neutron star structure}}.
\bjtitle{Phys. Rev. C}
\bvolume{58},
\bfpage{1804}--\blpage{1828}
(\byear{1998})
{\href{https://arxiv.org/abs/nucl-th/9804027}{{arXiv:nucl-th/9804027}}}.
\doiurl{10.1103/PhysRevC.58.1804}
\end{barticle}
\endbibitem

\bibitem{Lackey:2005tk}
\begin{barticle}
\bauthor{\bsnm{Lackey}, \binits{B.D.}},
\bauthor{\bsnm{Nayyar}, \binits{M.}},
\bauthor{\bsnm{Owen}, \binits{B.J.}}:
\batitle{{Observational constraints on hyperons in neutron stars}}.
\bjtitle{Phys. Rev. D}
\bvolume{73},
\bfpage{024021}
(\byear{2006})
{\href{https://arxiv.org/abs/astro-ph/0507312}{{arXiv:astro-ph/0507312}}}.
\doiurl{10.1103/PhysRevD.73.024021}
\end{barticle}
\endbibitem

\bibitem{Engel:1996ic}
\begin{barticle}
\bauthor{\bsnm{Engel}, \binits{A.}},
\bauthor{\bsnm{Dutt-Mazumder}, \binits{A.K.}},
\bauthor{\bsnm{Shyam}, \binits{R.}},
\bauthor{\bsnm{Mosel}, \binits{U.}}:
\batitle{{Pion production in proton proton collisions in a covariant one boson
  exchange model}}.
\bjtitle{Nucl. Phys. A}
\bvolume{603},
\bfpage{387}--\blpage{414}
(\byear{1996}).
\doiurl{10.1016/0375-9474(96)80008-F}
\end{barticle}
\endbibitem

\bibitem{Shyam:1996id}
\begin{barticle}
\bauthor{\bsnm{Shyam}, \binits{R.}},
\bauthor{\bsnm{Mosel}, \binits{U.}}:
\batitle{{NN $\rightarrow$ NN$\pi$ reaction near threshold in a covariant
  one-boson-exchange model}}.
\bjtitle{Physics Letters B}
\bvolume{426}(\bissue{1}),
\bfpage{1}--\blpage{6}
(\byear{1998}).
\doiurl{10.1016/S0370-2693(98)00297-4}
\end{barticle}
\endbibitem

\bibitem{dcQMD22}
\begin{botherref}
\oauthor{\bsnm{Cozma}, \binits{M.D.}}:
private communication
(2022)
\end{botherref}
\endbibitem

\bibitem{Russotto:2021mpu}
\begin{botherref}
\oauthor{\bsnm{Russotto}, \binits{P.}}, et al.:
{Symmetry energy at high densities from neutron/proton flow excitation
  functions}
(2021)
{\href{https://arxiv.org/abs/2105.09233}{{arXiv:2105.09233}}}
{[nucl-ex]}
\end{botherref}
\endbibitem

\bibitem{Li:2002yda}
\begin{barticle}
\bauthor{\bsnm{Li}, \binits{B.-A.}}:
\batitle{{High density behavior of nuclear symmetry energy and high-energy
  heavy ion collisions}}.
\bjtitle{Nucl. Phys. A}
\bvolume{708},
\bfpage{365}--\blpage{390}
(\byear{2002}).
\doiurl{10.1016/S0375-9474(02)01018-7}
\end{barticle}
\endbibitem

\bibitem{PhysRevC.47.R2467}
\begin{barticle}
\bauthor{\bsnm{Ehehalt}, \binits{W.}},
\bauthor{\bsnm{Cassing}, \binits{W.}},
\bauthor{\bsnm{Engel}, \binits{A.}},
\bauthor{\bsnm{Mosel}, \binits{U.}},
\bauthor{\bsnm{Wolf}, \binits{G.}}:
\batitle{Resonance properties in nuclear matter}.
\bjtitle{Phys. Rev. C}
\bvolume{47},
\bfpage{2467}--\blpage{2469}
(\byear{1993}).
\doiurl{10.1103/PhysRevC.47.R2467}
\end{barticle}
\endbibitem

\bibitem{Zhang:2014sva}
\begin{barticle}
\bauthor{\bsnm{Zhang}, \binits{Y.}},
\bauthor{\bsnm{Tsang}, \binits{M.B.}},
\bauthor{\bsnm{Li}, \binits{Z.}},
\bauthor{\bsnm{Liu}, \binits{H.}}:
\batitle{{Constraints on nucleon effective mass splitting with heavy ion
  collisions}}.
\bjtitle{Phys. Lett. B}
\bvolume{732},
\bfpage{186}--\blpage{190}
(\byear{2014}).
\doiurl{10.1016/j.physletb.2014.03.030}
\end{barticle}
\endbibitem

\bibitem{WANG2020135249}
\begin{barticle}
\bauthor{\bsnm{Wang}, \binits{Y.}},
\bauthor{\bsnm{Li}, \binits{Q.}},
\bauthor{\bsnm{Leifels}, \binits{Y.}},
\bauthor{\bsnm{{Le Fèvre}}, \binits{A.}}:
\batitle{{Study of the nuclear symmetry energy from the rapidity-dependent
  elliptic flow in heavy-ion collisions around 1 GeV/nucleon regime}}.
\bjtitle{Physics Letters B}
\bvolume{802},
\bfpage{135249}
(\byear{2020}).
\doiurl{10.1016/j.physletb.2020.135249}
\end{barticle}
\endbibitem

\end{thebibliography}

\end{document}